# Electrically and Magnetically Charged States and Particles in the 2+1-Dimensional $\mathbb{Z}_N$-Higgs Gauge Model.[1]

João C. A. Barata[2] and Florian Nill[3].
Institut für Theoretische Physik der Freien Universität Berlin.
Arnimallee 14, 14195 Berlin, Germany.

**Abstract:** Electrically as well as magnetically charged states are constructed in the 2+1-dimensional Euclidean $\mathbb{Z}_N$-Higgs lattice gauge model, the former following ideas of Fredenhagen and Marcu [1] and the latter using duality transformations on the algebra of observables. The existence of electrically and of magnetically charged particles is also established. With this work we prepare the ground for the constructive study of anyonic statistics of multiparticle scattering states of electrically and magnetically charged particles in this model.

## Contents

---

[1]Work supported by the Deutsche Forschungsgemeinschaft (SFB 288 "Differentialgeometrie und Quantenphysik").

[2]Supported by the Deutsche Forschungsgemeinschaft (SFB 288 "Differentialgeometrie und Quantenphysik") and with a travel grant from Fapesp. On leave from the Instituto de Física da Universidade de São Paulo. E-mail: barata@omega.physik.fu-berlin.de or jbarata@snfma1.if.usp.br

[3]E-mail: nill@vax1.physik.fu-berlin.d400.de

# 1 Introduction.

The study of the statistics of particles and fields in low dimensional Quantum Field Theory became one of the most fruitful lines of research of the last few years, in part due to some physically and mathematically appealing connections, like those to the Quantum Hall effect, to the so called exactly integrable models, to the theory of the "Quantum Groups" and to the so called "Topological Quantum Field Theories". The emergence of non-trivial, i.e., non-bosonic and non-fermionic, statistics in two and three space-time dimensions has been already described in the general framework of the Algebraic Quantum Field Theory [2].

In the work started with the present paper we intend to exhibit in a purely constructive way the emergence of non-trivial statistics in a (2+1)-dimensional quantum spin system, namely in a self-dual $\mathbb{Z}_N$-Higgs lattice gauge field theory with a dynamics defined through the transfer matrix formalism. The vacuum expectations in this theory are given by classical expectations of an (Euclidean) statistical mechanics model with an action given by a generalized Wilson action:

$$-\frac{1}{\sqrt{N}}\sum_{p}\sum_{n=0}^{N-1}\beta_g(n)\cos\left(\frac{2\pi n}{N}dA(p)\right)-\frac{1}{\sqrt{N}}\sum_{b}\sum_{n=0}^{N-1}\beta_h(n)\cos\left(\frac{2\pi n}{N}(d\varphi(b)-A(b))\right), \tag{1.1}$$

where $\varphi$ and $A$ are $\mathbb{Z}_N$-valued Higgs fields, respectively gauge fields, taking values in $\{0,\ldots,N-1\}$. $\beta_g$ and $\beta_h$ are the gauge and Higgs coupling constants. They will be chosen to satisfy $\beta_g(n)=\beta_g(N-n)$, $\beta_h(n)=\beta_h(N-n)$. We will be mostly interested in the so called "free charges phase" of this model, corresponding to "large" positive values of all the $\{\beta_g(1),\ldots,\beta_g(N-1)\}$ and "small" positive values of all the $\{\beta_h(1),\ldots,\beta_h(N-1)\}$[4]. In this region convergent polymer and cluster expansions are available and are analyzed in detail here. Other phases like the "confinement" and the "Higgs" phase may also exist, but we will not consider them since no charged states are expected there. See [1] for a study of these phases in the $\mathbb{Z}_2$ case.

In [1] Fredenhagen and Marcu have shown the existence of electrically charged states in the phase of "free charges" of the $\mathbb{Z}_2$-Higgs gauge model in three or more space time dimensions. Subsequently, the existence of electrically charged particles in this model was established in [3]. Multiparticle scattering states were constructed in [4], combining the methods of [1] with a general analysis of particle scattering in Euclidean lattice field theories given in [5].

In three space-time dimensions $\mathbb{Z}_N$-Higgs models are well known to enjoy self-duality properties [6, 7]. This suggests that there should exist magnetically charged states as well. For the $\mathbb{Z}_2$-model, this has been established in [8], where also a dyonic state (i.e. electrically *and* magnetically charged) has been constructed.

In the present work we extend these results to general $\mathbb{Z}_N$-Higgs models. In the above mentioned region of parameter space we construct electrically charged states with $\mathbb{Z}_N$-charges $n=1,\ldots,N-1$ in three or more space-time dimensions, following the lines of [1]. For the $(2+1)$-dimensional case we then give a detailed discussion of algebraic duality transformations, which in many aspects appear to be more subtle than the corresponding Euclidean Lattice formulation [6, 7]. The reason for this lies on the fact that there are two choices for defining transfer matrices from the Euclidean actions. This corresponds to the fact that in the Osterwalder-Schrader reconstruction one may use reflection positivity for time reflections across spatial planes of either the original lattice or of the dual lattice. Correspondingly, the algebraic version of duality transformations maps one choice of the Euclidean dynamics onto the other.

[4] The values of $\beta_g(0)$ and $\beta_h(0)$ only determine additive constants to the action, which will be fixed by convenient normalization conditions.

Among other results we are able to show that the global transfer matrices associated to dual states are unitarily equivalent. We then use these results to prove that there is an isomorphism between the model at parameters $(\beta_g, \beta_h)$ and the dual model at the dual parameters $(\beta_g', \beta_h')$, mapping electric states onto magnetic ones and vice-versa. We finally construct unitary generators of the translation group and prove the existence of electrically (and therefore, in $2+1$ space-time dimensions, also of magnetically) charged particles.

In a forthcoming paper [9] we will construct the dyonic sectors of this model, i.e., sectors where electric and magnetic charge distributions are simultaneously present. There translation invariant global transfer matrices will be constructed and the existence of a unitary representation of the translation group, a question which is much more subtle in the dyonic case, will be analyzed. No trace of the existence of dyonic particles, i.e., of particles carrying simultaneously electric and magnetic charges, was found and it remains an open problem to prove or to disprove their existence.

Our ultimate goal is to show the emergence of anyonic statistics in this model. Since no method is available for constructing fields generating the various sectors mentioned above, in [9] we will address the question of the statistics through the analysis of multiparticle scattering states of electrically and magnetically charged particles, whose existence will be proven in the present article.

For the orientation of the reader we describe now the organization of the work. Section 2 is devoted to the description of the $\mathbb{Z}_N$-Higgs gauge model as a quantum spin system. The algebras of fields and of observables are defined, local transfer matrices are introduced as well as the important concept of a ground state. In section 3 we describe how to use the transfer matrix formalism in order to reconstruct the classical (Euclidean) vacuum expectations. In section 4 we develop polymer and cluster expansions for classical expectations in the so called free charges phase, combining high and low temperature expansions. A full convergence proof is given in appendix A. In section 5 we resume our algebraic analysis and consider duality transformations from the algebraic point of view. Four different types of ground states related by duality transformations are presented. Global transfer matrices are defined for each of these states and their interplay is discussed. One shows in a precise sense that duality transformations keep the joint spectrum of the transfer matrix and momentum operator invariant. Section 6 is devoted to the construction of electrically and of magnetically charged states. The first following [1] the latter using duality transformations. This method provides a natural frame for further analyses, specially concerning the relationship between transfer matrices in magnetically and electrically charged sectors and concerning the structure of charged particles. In this section we also establish that our charged states are ground states with respect to certain automorphisms generated by modified transfer matrices, a fact which will be of relevance for our future construction of dyonic states. In section 7 we define global transfer matrices for the charged sectors and analyze the relationship between then, and in section 8 the translation invariance of these global transfer matrices is established. We prove that, due to the duality transformations, the existence of electric particles implies the existence of magnetic ones with masses related by dually transformed couplings. In section 9 the existence of electrically and of magnetically charged particles is directly established through the analysis of suitably defined two point functions. In the other appendices we complete some important proofs.

*Remarks on the notation.* The symbol $\square$ indicates end of statement and $\blacksquare$ end of proof. Products of operators run from the left to the right, i.e., $\prod_{a=1}^{n} A_a$ means $A_1 \cdots A_n$. For an invertible operator $B$, $\mathrm{ad}_B$ is the automorphism $B \cdot B^{-1}$. $\mathcal{Z}$ denotes the set of all functions $\{0, \ldots, N-1\} \to \mathbb{C}$ $\square$

*Acknowledgments.* We which to express our gratitude to Klaus Fredenhagen and Frank Gaebler, whom we are indebted for many valuable and stimulating discussions.

# 2 Basic Setting and Algebraic Formulation.

In this paper we use the well known transfer matrix formalism to represent the Euclidean lattice $\mathbb{Z}_N$-Higgs model ($N \in \mathbb{N}$, $N \geq 2$) as a d-dimensional *quantum* spin system, rather than a (d+1)-dimensional *classical* statistical system. This brings us closer to quantum field theories where relevant quantities are described in terms of operators and states. Here the dynamics is described by a transfer matrix, which has to be interpreted as the generator of time translations by one unit in *imaginary* time direction. In a first step transfer matrices can only be properly defined at finite volume. The associated ground state is a ground state for a finite volume system but, if its thermodynamic limit exists, one can use the GNS construction associated to the limit state to define a global (infinite volume) transfer matrix, in a fashion first proposed in [1]. Translation invariance of the limit state also provides a way for defining generators of space translations in this GNS Hilbert space, thus making the analogy with quantum field theoretical systems even more appealing. In this work we build up these structures for neutral and charged states associated to the $\mathbb{Z}_N$-Higgs model and study some of their properties. We now start by describing our quantum spin system.

We will consider the hypercubic $\mathbb{Z}^{d+1}$ Euclidean space-time lattice with $d \geq 2$ and particularly the case $d = 2$, where our main results hold. For simplicity we will fix the lattice spacing as $a = 1$. We denote by $l_i$ the set of $i$-cells of $\mathbb{Z}^{d+1}$ with the identification $l_0 = \mathbb{Z}^{d+1}$. $l_1$ is the set of oriented bonds associated to $\mathbb{Z}^{d+1}$, $l_2$ the set of oriented plaquettes, etc. The quantum spin system is defined on a $\mathbb{Z}^d$ lattice, the time-zero hyperplane. In our notation the elements of $\mathbb{Z}^d$, its cells or subsets are usually underlined.

We introduce the local algebra of time-zero Higgs and gauge fields in the following way. To each $\underline{x} \in \underline{l}_0$ we associate the unitary $\mathbb{Z}_N$-fields $P_H(\underline{x})$ and $Q_H(\underline{x})$ and to each $\underline{b} \in \underline{l}_1$ we associate the unitary $\mathbb{Z}_N$-fields $P_G(\underline{b})$ and $Q_G(\underline{b})$ (the subscripts $G$ and $H$ stand for "gauge" and "Higgs" respectively) satisfying the relations:

$$P_H(\underline{x})^* = P_H(\underline{x})^{-1} = P_H(\underline{x})^{N-1}, \tag{2.1}$$

$$Q_H(\underline{x})^* = Q_H(\underline{x})^{-1} = Q_H(\underline{x})^{N-1}, \tag{2.2}$$

$$P_G(\underline{b})^* = P_G(\underline{b})^{-1} = P_G(\underline{b})^{N-1}, \tag{2.3}$$

$$Q_G(\underline{b})^* = Q_G(\underline{b})^{-1} = Q_G(\underline{b})^{N-1}, \tag{2.4}$$

and the $\mathbb{Z}_N$ Weyl-algebra relations

$$P_H(\alpha)Q_H(\beta) = e^{-\frac{2\pi i}{N}\langle\alpha,\beta\rangle_{\underline{l}^0}} Q_H(\beta)P_H(\alpha), \tag{2.5}$$

$$P_G(\gamma)Q_G(\delta) = e^{-\frac{2\pi i}{N}\langle\gamma,\delta\rangle_{\underline{l}^1}} Q_G(\delta)P_G(\gamma), \tag{2.6}$$

where $\alpha$, $\beta \in \underline{l}^0$: $\underline{l}_0 \to \{0, \ldots N-1\}$ are 0-forms with finite support; $\gamma$, $\delta \in \underline{l}^1$: $\underline{l}_1 \to \{0, \ldots N-1\}$ are 1-forms with finite support and $P_H(\alpha) := \prod_{\underline{x}\in\underline{l}_0} P_H(\underline{x})^{\alpha(\underline{x})}$, etc. The brackets $\langle\cdot,\cdot\rangle$ indicate the scalar product of forms. We also convention that $P_H(-\underline{x}) = P_H(\underline{x})^{-1}$, etc, where here $-\underline{x}$ indicates the cell $\underline{x}$ with reverse orientation. Operators at different sites and bonds commute. Finally the $G$ operators commute with the $H$ operators.

We will generally define $[\delta Q_H](\alpha) := Q_H(d\alpha)$, $[\delta^* P_G](\beta) := P_G(d^*\beta)$, etc, where $d$ is the exterior derivative on forms.

We will realize these operators by attaching to each lattice point and to each lattice bond a Hilbert space $\mathcal{H}_{\underline{x}} = \mathcal{H}_{\underline{b}} = \mathbb{C}^N$ with each $Q_H(\underline{x})$, $P_H(\underline{x})$, $Q_G(\underline{b})$ and $P_G(\underline{b})$ acting on $\mathcal{H}_{\underline{x}}$, respectively on $\mathcal{H}_{\underline{b}}$ as matrices with matrix elements:

$$P_H(\underline{x})_{a,b} = P_G(\underline{b})_{a,b} = \delta_{a,\, b+1 (\mathrm{mod} N)} \quad \text{and} \tag{2.7}$$

$$Q_H(\underline{x})_{a,b} = Q_G(\underline{b})_{a,b} = \delta_{a,b} e^{\frac{2\pi i}{N} a}, \tag{2.8}$$

for $a,\ b \in \{0, \ldots, N-1\}$.

One should interpret the operators $Q_H$ and $Q_G$ as the $\mathbb{Z}_N$ versions of the Higgs field and gauge field, respectively: $Q_H(\underline{x}) = e^{\frac{2\pi i}{N}\varphi(\underline{x})}$, $Q_G(\underline{x}) = e^{\frac{2\pi i}{N}A(\underline{b})}$, with $\varphi$ and $A$ taking values in $\{0, \ldots, N-1\}$. The operators $P_H$ and $P_G$ are their respective canonically conjugated momenta, in $\mathbb{Z}_N$ version.

We denote by $\mathfrak{F}_0$ the $*$-algebra generated by these operators together with a unit $\mathbb{1}$. Denoting by $\mathfrak{F}(\underline{V})$ the C$^*$-subalgebra generated by $\mathbb{1}$, $Q_H(\underline{x})$, $P_H(\underline{x})$, $Q_G(\underline{b})$ and $P_G(\underline{b})$ for $\underline{x},\ \underline{b} \in \underline{V} \subset \mathbb{Z}^d$, a finite set, one has $\mathfrak{F}_0 = \cup_{|\Lambda|<\infty}\mathfrak{F}(\underline{\Lambda})$. The algebra $\mathfrak{F}(\underline{V})$ acts on $\mathcal{H}_{\underline{V}} := \otimes_{\underline{x}\in\underline{V}_0}\mathcal{H}_{\underline{x}} \otimes_{\underline{b}\in\underline{V}_1} \mathcal{H}_{\underline{b}}$. We will denote by $\mathfrak{F}$ the unique C$^*$-algebra generated by $\mathfrak{F}_0$. By $\tau_{\underline{x}}$ we denote the $*$-automorphism of $\mathfrak{F}$ implementing translations by $\underline{x} \in \mathbb{Z}^d$.

The dynamics we will consider is invariant under the $*$-automorphism implemented by the unitaries

$$\mathcal{Q}(\underline{x}) := P_H(\underline{x})\left[\delta^* P_G\right](\underline{x})^*. \tag{2.9}$$

Note that $\mathcal{Q}(\underline{x})^* = \mathcal{Q}(\underline{x})^{-1} = \mathcal{Q}(\underline{x})^{N-1}$. The operator $\mathcal{Q}(\underline{x})$ is to be interpreted as the generator of a $\mathbb{Z}_N$ gauge transformation at the point $\underline{x}$, as one can easily checks, since it can be interpreted as $\exp\left(-2\pi i(\mathrm{div}\ \mathbb{E} - \rho)/N\right)$.

The algebra of observables $\mathfrak{A}$ is defined as the set of fixed points of $\mathfrak{F}$ by $\mathrm{ad}_{\mathcal{Q}(\underline{x})}$ for all $\underline{x}$:

$$\mathfrak{A} := \{A \in \mathfrak{F} :\ \mathcal{Q}(\underline{x})A\mathcal{Q}(\underline{x})^* = A \text{ for all } \underline{x} \in \mathbb{Z}^d\}. \tag{2.10}$$

The norm dense sub-algebra $\mathfrak{A}_0$ is generated by $\mathbb{1}$, $P_G(\underline{b})$, $\left[\delta Q_H\right](\underline{b})Q_G(\underline{b})^*$ and $P_H(\underline{x})$, $\underline{x},\ \underline{b} \in \mathbb{Z}^d$. We call $\mathfrak{A}(\underline{\Lambda}) := \mathfrak{F}(\underline{\Lambda}) \cap \mathfrak{A}$.

The algebra $\mathfrak{A}$ contains a non-trivial two-sided norm closed ideal $J$ generated by $\mathcal{Q}(\underline{x})^a - \mathbb{1}$, $\underline{x} \in \mathbb{Z}^d$, $a = 1, \ldots, N-1$. Since the operator $\mathcal{Q}(\underline{x})$ can be interpreted as an operator measuring a external $\mathbb{Z}_N$-electric charge at $\underline{x}$ the relevant algebra of observables is actually $\mathfrak{B} := \mathfrak{A}/J$. The elements of $\mathfrak{B}$ are equivalence classes $[A] := \{B \in \mathfrak{A},\ \text{so that } A - B \in J\}$. Below we will mostly prefer the notation $A + J$ for $[A]$, $A \in \mathfrak{A}$. We call $\mathfrak{B}(\underline{\Lambda}) := \{A + J,\ A \in \mathfrak{A}(\underline{\Lambda})\}$ the $*$-algebra which is generated by $\mathbb{1} + \mathrm{J}$, $P_G(\underline{b}) + J$ and $\left[\delta Q_H\right](\underline{b})Q_G(\underline{b})^* + J$, $\underline{b} \in \underline{\Lambda}$. Finally we call $\mathfrak{B}_0 = \cup_{|\Lambda|<\infty}\mathfrak{B}(\underline{\Lambda})$. The algebras $\mathfrak{B}$ and $\mathfrak{B}(\underline{\Lambda})$ can be regarded as C$^*$-algebras with norm $\|[A]\| := \inf\{\|A + j\|,\ j \in J\}$ (for a proof see [10], Proposition 2.2.19). We will denote by

$$U_1(\underline{b}) := P_G(\underline{b}) + J, \tag{2.11}$$

$$U_3(\underline{b}) := \left[\delta Q_H\right](\underline{b})Q_G(\underline{b})^* + J \tag{2.12}$$

the generators of $\mathfrak{B}_0$.

Let us now introduce the dynamics we are interested in by defining suitable finite volume transfer matrices. The form of the transfer matrix is justified by the finite volume ground state we will associate to it, which is identical to the classical expectation associated to the Euclidean $\mathbb{Z}_N$-Higgs model we are considering.

We will consider local transfer matrices $T_{\underline{V}} \in \mathfrak{A}_0$, $\underline{V} \subset \mathbb{Z}^d$ defined by:

$$T_{\underline{V}} = e^{A_{\underline{V}}/2} e^{B_{\underline{V}}} e^{A_{\underline{V}}/2}, \tag{2.13}$$

with

$$A_{\underline{V}} := \frac{1}{\sqrt{N}} \sum_{\underline{p}\in\underline{V}_2^+} \sum_{n=0}^{N-1} \beta_g(n)\ [\delta Q_G]\,(\underline{p})^n + \frac{1}{\sqrt{N}} \sum_{\underline{b}\in\underline{V}_1^+} \sum_{n=0}^{N-1} \beta_h(n)\,([\delta Q_H]\,(\underline{b})Q_G(\underline{b})^*)^n, \tag{2.14}$$

and

$$B_{\underline{V}} := \frac{1}{\sqrt{N}} \sum_{\underline{b}\in\underline{V}_1^+} \sum_{n=0}^{N-1} \gamma_g(n)\, P_G(\underline{b})^n + \frac{1}{\sqrt{N}} \sum_{\underline{x}\in\underline{V}_0^+} \sum_{n=0}^{N-1} \gamma_h(n)\, P_H(\underline{x})^n \tag{2.15}$$

where $\underline{V}_i^+$ are the positively oriented elements of the set of $i$-cells of $\underline{V}$. The relation of these definitions with a formulation in terms of an Euclidean action will be given below (eqs. (3.16)).

Above $\beta_g$, $\beta_h$, $\gamma_g$ and $\gamma_h \in \mathcal{Z}$, the set of functions $\{0, \ldots, N-1\} \to \mathbb{C}$, and satisfy the condition

$$\overline{\beta_{g,h}(n)} = \beta_{g,h}(N-n), \qquad \overline{\gamma_{g,h}(n)} = \gamma_{g,h}(N-n) \tag{2.16}$$

for all $n \in \{0, \ldots, N-1\}$, with $\beta_{g,h}(N) := \beta_{g,h}(0)$, $\gamma_{g,h}(N) := \gamma_{g,h}(0)$ in order to assure self-adjointness of $T_{\underline{V}}$. They are the coupling constants of the model. Later when we treat the duality transformations we will have to restrict ourself to real couplings. For further purposes we also define

$$T_{\underline{V}}^T := e^{B_{\underline{V}}/2} e^{A_{\underline{V}}} e^{B_{\underline{V}}/2}. \tag{2.17}$$

Notice that transfer matrices of a classical statistical mechanics spin system are usually defined in order to provide a way of expressing the partition function as $Z = \mathrm{Tr}(T_{\underline{V}}^n)$, for a system in a volume $\underline{V} \times [0, \ldots, n]$ and periodic boundary conditions in "time" direction. In this particular sense it should be immaterial to use $T_{\underline{V}}$ or $T_{\underline{V}}^T$, as defined above. However, from the point of view of the quantum spin system we are constructing, both transfer matrices provide different quantum dynamics. Interestingly, the interplay between both will be of relevance for the study of duality transformations in this model.

We can also write

$$e^{A_{\underline{V}}} = \prod_{\underline{p}\in\underline{V}_2^+} \varrho_G(\underline{p}) \prod_{\underline{b}\in\underline{V}_1^+} \varrho_H(\underline{b}), \tag{2.18}$$

$$e^{B_{\underline{V}}} = \prod_{\underline{b}\in\underline{V}_1^+} \zeta_G(\underline{b}) \prod_{\underline{x}\in\underline{V}_0^+} \zeta_H(\underline{x}), \tag{2.19}$$

where,

$$\varrho_G(\underline{p}) := \exp\left(\frac{1}{\sqrt{N}} \sum_{n=0}^{N-1} \beta_g(n)\ [\delta Q_G]\,(\underline{p})^n\right) = \frac{1}{\sqrt{N}} \sum_{m=0}^{N-1} \mathcal{E}[\beta_g](m)\,[\delta Q_G]\,(\underline{p})^m, \tag{2.20}$$

$$\varrho_H(\underline{b}) := \exp\left(\frac{1}{\sqrt{N}} \sum_{n=0}^{N-1} \beta_h(n)\,([\delta Q_H]\,(\underline{b})Q_G(\underline{b})^*)^n\right) = \frac{1}{\sqrt{N}} \sum_{m=0}^{N-1} \mathcal{E}[\beta_h](m)\,([\delta Q_H]\,(\underline{b})Q_G(\underline{b})^*)^m, \tag{2.21}$$

$$\zeta_G(\underline{b}) := \exp\left(\frac{1}{\sqrt{N}} \sum_{n=0}^{N-1} \gamma_g(n)\, P_G(\underline{b})^n\right) = \frac{1}{\sqrt{N}} \sum_{m=0}^{N-1} \mathcal{E}[\gamma_g](m) P_G(\underline{b})^m, \tag{2.22}$$

$$\zeta_H(\underline{x}) := \exp\left(\frac{1}{\sqrt{N}} \sum_{n=0}^{N-1} \gamma_h(n)\, P_H(\underline{x})^n\right) = \frac{1}{\sqrt{N}} \sum_{m=0}^{N-1} \mathcal{E}[\gamma_h](m) P_H(\underline{x})^m, \tag{2.23}$$

where, for functions $\alpha \in \mathcal{Z}$ we define the transformations $\mathcal{E}$: $\mathcal{Z} \to \mathcal{Z}$ by

$$\mathcal{E}[\alpha] = \mathcal{F}^{-1}\left[\exp\left(\mathcal{F}[\alpha]\right)\right], \tag{2.24}$$

where the Fourier transform $\mathcal{F}$ and its inverse $\mathcal{F}^{-1}$ are defined on functions $\rho \in \mathcal{Z}$ by

$$\mathcal{F}[\rho](m) := N^{-1/2} \sum_{v\in\{0,\ldots,N-1\}} \rho(v) e^{+\frac{2\pi i}{N} mv}, \tag{2.25}$$

$$\mathcal{F}^{-1}[\rho](m) := N^{-1/2} \sum_{v\in\{0,\ldots,N-1\}} \rho(v) e^{-\frac{2\pi i}{N} mv} \tag{2.26}$$

and satisfy

$$(\mathcal{F}[a])^{\#} = \mathcal{F}^{-1}[a] = \mathcal{F}[a^{\#}], \tag{2.27}$$

where, by definition, $a^{\#}(n) := a(N-n)$, $n \in \{0, \ldots, N-1\}$ with $a(N) := a(0)$.

Note that in a C$^*$-algebra, by the Gelfand transform [11], $\varrho = e^{\sum_{n=0}^{N-1} a(n)\mathcal{P}^n}$ with $\overline{a(n)} = a(N-n)$, $a(N) := a(0)$ is the most general way of writing a positive self-adjoint element $\varrho$ of the abelian sub-algebra generated by $\mathbb{1}$ and an element $\mathcal{P}$ satisfying $\mathcal{P}^* = \mathcal{P}^{-1} = \mathcal{P}^{N-1}$. In this way the generalized $\mathbb{Z}_N$-Higgs model above represents the most general class of models with "nearest neighbors" interactions of the sort considered.

At this algebraic level the Euclidean dynamics is given by the strong limit of local (non-$*$)-automorphisms of $\mathfrak{F}_0$ generated by local transfer matrices :

$$\alpha_i(A) := \lim_{\underline{V}\uparrow\mathbb{Z}^d} \alpha_i(A)_{\underline{V}}, \qquad A \in \mathfrak{F}_0 \tag{2.28}$$

where $\alpha_i(\cdot)_{\underline{V}}$ is the automorphism of $\mathfrak{F}$ defined through

$$\alpha_i(A)_{\underline{V}} := T_{\underline{V}} A T_{\underline{V}}^{-1}, \qquad A \in \mathfrak{F}. \tag{2.29}$$

The limit in (2.28) clearly exists and defines a (non $*$)-automorphism of $\mathfrak{F}_0$. Following the notation introduced in [1] the subindex $i$ in $\alpha_i$ is due to the interpretation of $\alpha_i$ as the generator of translations of one unit in imaginary (Euclidean) time direction. Frequently we will use the notation $\alpha_{ir}$ for $(\alpha_i)^r$, for $r \in \mathbb{Z}$.

Since $\alpha_i$ keeps the ideal $J$ invariant one naturally defines the action of $\alpha_i$ on $\mathfrak{B}_0$, which we will denote by the same symbol $\alpha_i$, by $\alpha_i(A+J) = \alpha_i(A) + J$, $A \in \mathfrak{A}_0$, as a non-$*$ automorphism.

For further purposes we introduce the important concept of a ground state. According to the definition introduced in [1], a state $\omega$ of $\mathfrak{F}$ is called a "ground state" with respect to the dynamics defined by $\alpha_i$ if it is $\alpha_i$-invariant and if

$$0 \le \omega(A^*\alpha_i(A)) \le \omega(A^*A), \qquad A \in \mathfrak{F}_0. \tag{2.30}$$

Let us generalize this concept and derive some general results from it. Let $\gamma$ be an automorphism of a unital $*$-algebra $\mathfrak{C}$. A state $\omega$ on $\mathfrak{C}$ is called a "ground state" with respect to $\gamma$ and $\mathfrak{C}$ if it is $\gamma$-invariant and if

$$0 \le \omega(A^*\gamma(A)) \le \omega(A^*A), \qquad \forall A \in \mathfrak{C}. \tag{2.31}$$

Actually the $\gamma$-invariance of $\omega$ follows from (2.31). If one has $0 \le \omega(A^*\gamma(A))$, $\forall A \in \mathfrak{C}$, then the sesquilinear form on $\mathfrak{C}$, $(X, Y) := \omega(X^*\gamma(Y))$, $X, Y \in \mathfrak{C}$ is positive and so $(X, Y) = \overline{(Y, X)}$ [10].

Therefore $\omega(X^*\gamma(Y)) = \omega(\gamma(X)^*Y)$. Taking $X = 1$ the invariance of $\omega$ under $\gamma$ follows. Let us now introduce two important concepts.

First, the adjoint $\gamma^*$ of an automorphism $\gamma$ of a unital $*$-algebra $\mathfrak{C}$ is defined through $\gamma^*(A) := (\gamma(A^*))^*$, $A \in \mathfrak{C}$. Note that $\gamma$ is a $*$-automorphism iff $\gamma = \gamma^*$. In general, $\gamma^{**} = \gamma$ and note also that for the composition of automorphisms one has $(\alpha\circ\beta)^* = \alpha^*\circ\beta^*$ and consequently $\alpha^{*-1} = \alpha^{-1*}$. For an invertible element $A \in \mathfrak{C}$ one also has $(\mathrm{ad}_A)^* = \mathrm{ad}_{A^{*-1}}$. Finally, note that if $\omega$ is a $\gamma$-invariant state on $\mathfrak{C}$ then it is also $\gamma^*$-invariant.

Second, we say that a state $\omega$ on a $*$-algebra $\mathfrak{C}$ has the cluster property for the automorphism $\gamma$ if, for all $A$, $B \in \mathfrak{C}$, one has

$$\lim_{n\to\infty} \omega(A\gamma^n(B)) = \omega(A)\omega(B). \tag{2.32}$$

Using these definitions we are prepared to formulate the following Lemma, which will be very useful for proving that certain states are ground states with respect to a given automorphism. This Lemma was already implicitly used in [1].

**Lemma 2.1** *Let $\gamma$ be an automorphism on a $*$-algebra $\mathfrak{C}$ satisfying $\gamma^* = \gamma^{-1}$ and let $\omega$ be a $\gamma$-invariant state on $\mathfrak{C}$ which has the cluster property for $\gamma$. Actually one just needs that, for each $A \in \mathfrak{C}$, the sequence $\omega(A^*\gamma^a(A))$, $a \in \mathbb{N}$, is bounded, what follows from the cluster property. Then, for all $A \in \mathfrak{C}$,*

$$|\omega(A^*\gamma(A))| \le \omega(A^*A) \qquad \square \tag{2.33}$$

**Proof.** If we use $a$-times the Cauchy-Schwarz inequality and the invariance of $\omega$ we get

$$|\omega\left(A^*\gamma(A)\right)| \le \omega(A^*A)^{1-2^{-a}} \left|\omega\left(A^*\gamma^{2a}(A)\right)\right|^{2^{-a}}. \tag{2.34}$$

By the cluster property, the factor $\omega\left(A^*\gamma^{2a}(A)\right)$ on the right hand side is bounded on $a$ and taking $a \to \infty$ we complete the proof ■

# 3 The Ground State, the Construction of the Path Space and the Associated Classical Spin System.

This section is basically devoted to the construction of a ground state w.r.t. $\alpha_i$ by means of the transfer matrix formalism as described in [1]. The ground state we will find is given by the expectation of a classical statistical mechanics spin system.

Define for $n$, $n' \in \{0, \ldots, N-1\}$

$$E^H_{n,\,n'}(\underline{x}) = P_H(\underline{x})^n \left[N^{-1}\sum_{m=0}^{N-1} Q_H(\underline{x})^m\right] P_H(\underline{x})^{N-n'}, \tag{3.1}$$

$$E^G_{n,\,n'}(\underline{b}) = P_G(\underline{b})^n \left[N^{-1}\sum_{m=0}^{N-1} Q_G(\underline{b})^m\right] P_G(\underline{b})^{N-n'}. \tag{3.2}$$

$E^H_{n,\,n'}(\underline{x})$ and $E^G_{n,\,n'}(\underline{b})$ are unit matrices with matrix elements

$$\left(E^H_{n,\,n'}(\underline{x})\right)_{a,\,b} = \delta_{a,\,n}\;\delta_{b,\,n'} = \left(E^G_{n,\,n'}(\underline{b})\right)_{a,\,b}, \tag{3.3}$$

$a,\, b \in \{0, \ldots, N-1\}$.

Let be the functions $\varphi$: $\underline{V}_0 \to \{0, \ldots, N-1\}$, $A$: $\underline{V}_1 \to \{0, \ldots, N-1\}$. Here $\underline{V} \subset \mathbb{Z}^2$ is a finite set, for instance a square centered at the origin. Define

$$E_{(\varphi, A), (\varphi', A')} := \prod_{\underline{x} \in \underline{V}_0^+} E^H_{\varphi(\underline{x}), \varphi'(\underline{x})}(\underline{x}) \prod_{\underline{b} \in \underline{V}_1^+} E^G_{A(\underline{b}), A'(\underline{b})}(\underline{b}). \tag{3.4}$$

Since the $E_{(\varphi,A),(\varphi',A')}$ form a basis of unit matrices we can write

$$T_{\underline{V}} = \sum_{(\varphi, A), (\varphi', A')} T_{\underline{V}}(\varphi, A; \varphi', A') E_{(\varphi, A), (\varphi', A')}. \tag{3.5}$$

The $E_{(\varphi,A),(\varphi',A')}$ are partial isometries with one dimensional range and one finds for the expansion coefficients of $T_{\underline{V}}$

$$T_{\underline{V}}(\varphi, A; \varphi', A') = Tr_{\mathcal{H}_{\underline{V}}} \left( E^*_{(\varphi, A), (\varphi', A')} T_{\underline{V}} \right). \tag{3.6}$$

Using the last equalities in (2.20)-(2.23), the relation

$$Tr \left( P^{n'} \left[ N^{-1} \sum_{m=0}^{N-1} Q^m \right] P^{N-n} Q^a P^b Q^c \right) = \left( \delta_{b,\, n-n' \bmod N} \right) e^{\frac{2\pi i}{N}(na + n'c)}, \tag{3.7}$$

and expression (2.24) we get:

$$\begin{aligned} T_{\underline{V}}(\varphi, A; \varphi', A') \;&= \exp \left( \frac{1}{2} \sum_{\underline{p} \in \underline{V}_2^+} \left( \mathcal{F}[\beta_g](dA(\underline{p})) + \mathcal{F}[\beta_g](dA'(\underline{p})) \right) \right) \times \\ &\exp \left( \frac{1}{2} \sum_{\underline{b} \in \underline{V}_1^+} \left( \mathcal{F}[\beta_h](d\varphi(\underline{b}) - A(\underline{b})) + \mathcal{F}[\beta_h](d\varphi'(\underline{b}) - A'(\underline{b})) \right) \right) \times \\ &\prod_{\underline{x} \in \underline{V}_0^+} \frac{\mathcal{E}[\gamma_h](\varphi(\underline{x}) - \varphi'(\underline{x}))}{\sqrt{N}} \prod_{\underline{b} \in \underline{V}_1^+} \frac{\mathcal{E}[\gamma_g](A(\underline{b}) - A'(\underline{b}))}{\sqrt{N}}. \end{aligned} \tag{3.8}$$

We will assume $\gamma_g$ and $\gamma_h$ to be such that

$$\mathcal{E}[\gamma_g](m) = \exp(\mathcal{F}[\beta_g^0](m)), \tag{3.9}$$

$$\mathcal{E}[\gamma_h](m) = \exp(\mathcal{F}^{-1}[\beta_h^0](m)), \tag{3.10}$$

for some $\beta^0_{g,\,h} \in \mathcal{Z}$. Since in general $\mathcal{E}[\gamma_{g,\,h}] = \mathcal{F}^{-1}\left[\exp(\mathcal{F}[\gamma_{g,\,h}])\right]$ this assumption means that

$$\begin{aligned} \gamma_g \;&=\; \mathcal{F}^{-1}\left[\ln\left(\mathcal{F}\left[\exp(\mathcal{F}[\beta_g^0])\right]\right)\right], & (3.11)\\ \gamma_h \;&=\; \mathcal{F}^{-1}\left[\ln\left(\mathcal{F}\left[\exp(\mathcal{F}^{-1}[\beta_h^0])\right]\right)\right] & (3.12)\\ &=\; \mathcal{F}^{-1}\left[\ln\left(\mathcal{F}^{-1}\left[\exp(\mathcal{F}[\beta_h^0])\right]\right)\right], & (3.13) \end{aligned}$$

the last equality coming from (2.27). Note that, for $a \geq 0$, the Fourier transform and the inverse Fourier transform of $\exp(\mathcal{F}^{\pm 1}[a])$ are strictly positive since, for instance,

$$\exp(\mathcal{F}[a]) = \mathcal{F}\left[\sum_{k=0}^{\infty} \frac{1}{k!} \underbrace{a * \cdots * a}_{k}\right], \tag{3.14}$$

with the convolution product defined by

$$a * b(n) = \frac{1}{\sqrt{N}} \sum_{i=0}^{N-1} a(n-i)b(i). \tag{3.15}$$

The same can be proven for functions $a$ with $a(n) \geq 0$ for $n \neq 0$, but with $a(0)$ being eventually negative, which is our case of interest. For, define $b \in \mathcal{Z}$ with $b(m) = (-a(0)+\epsilon)\delta_{m,0}$, $\epsilon \geq 0$. One has $\mathcal{F}^{-1}\left[\exp(\mathcal{F}[a])\right] = e^{-(-a(0)+\epsilon)/\sqrt{N}} \mathcal{F}^{-1}\left[\exp(\mathcal{F}[a+b])\right] > 0$ by the previous argument since $a + b \geq 0$.

In order to have isotropic couplings in the classical expectations to be defined below we will take $\beta^0_{g,h} = \beta_{g,h}$. Expression (3.8) becomes

$$\begin{aligned} T_{\underline{V}}(\varphi, A; \varphi', A') = \exp\left(\frac{1}{2}\sum_{\underline{p}\in\underline{V}_2^+}\left(\mathcal{F}[\beta_g](dA(\underline{p})) + \mathcal{F}[\beta_g](dA'(\underline{p}))\right)\right) \times \\ \exp\left(\frac{1}{2}\sum_{\underline{b}\in\underline{V}_1^+}\left(\mathcal{F}[\beta_h](d\varphi(\underline{b}) - A(\underline{b})) + \mathcal{F}[\beta_h](d\varphi'(\underline{b}) - A'(\underline{b}))\right)\right) \times \\ \exp\left(\sum_{\underline{b}\in\underline{V}_1^+}\mathcal{F}[\beta_g](A(\underline{b}) - A'(\underline{b})) + \sum_{\underline{x}\in\underline{V}_0^+}\mathcal{F}[\beta_h](\varphi'(\underline{x}) - \varphi(\underline{x}))\right) N^{-1/2(|\underline{V}_0^+|+|\underline{V}_1^+|)}, \end{aligned} \tag{3.16}$$

where we used $\mathcal{F}^{-1}[\beta_h](n) = \mathcal{F}[\beta_h](-n)$ on the last factor.

The correspondence to the $\mathbb{Z}_2$ case of Fredenhagen and Marcu, whose couplings we call $\beta^{FM}_{g,h}$, is found by taking $\beta_g(1) = -\beta_g(0) = \sqrt{2}\beta^{FM}_g$, with $\beta_h(0) = -2^{-1/2}\ln(2(\cosh\beta^{FM}_h)^2)$ and $\beta_h(1) = \sqrt{2}\beta^{FM}_h$.

Following the argumentation of [1], since the expansion coefficients (3.16) of $T_{\underline{V}}$ are strictly positive, we conclude by the Perron-Frobenius Theorem that there exists in $\mathcal{H}_{\underline{V}}$ a unique eigenvector $\Omega_{\underline{V}}$ of $T_{\underline{V}}$ corresponding to eigenvalue $\|T_{\underline{V}}\|_{\mathcal{H}_{\underline{V}}}$. The associated vector state $\omega_{\underline{V}}$ can be obtained by

$$\omega_{\underline{V}}(A) = \lim_{n\to\infty} \frac{Tr_{\mathcal{H}_{\underline{V}}}\left(T^n_{\underline{V}} A T^n_{\underline{V}} E_{\underline{V}}\right)}{Tr_{\mathcal{H}_{\underline{V}}}\left(T^{2n}_{\underline{V}} E_{\underline{V}}\right)}, \quad A \in \mathfrak{F}(\underline{V}), \tag{3.17}$$

where $E_{\underline{V}}$ is any matrix with strictly positive expansion coefficients in the basis $E_{(\varphi,A)\,(\varphi',A')}$. Again by the Perron-Frobenius Theorem the spectral projection associated with $\Omega_{\underline{V}}$ has also strictly positive expansion coefficients and therefore one has $(\Omega_{\underline{V}},\, E_{\underline{V}}\Omega_{\underline{V}}) \neq 0$. In order to obtain for $\omega_{\underline{V}}$ a classical expectation with free boundary conditions in Euclidean time direction the choice for $E_{\underline{V}}$ is [1]:

$$E_{\underline{V}} := \sum_{(\varphi,A),(\varphi',A')} e_{\underline{V}}(\varphi,\, A) e_{\underline{V}}(\varphi',\, A') E_{(\varphi,A),(\varphi',A')} \tag{3.18}$$

where

$$e_{\underline{V}}(\varphi,\, A) := \exp\left[\frac{1}{2}\sum_{\underline{p}\in\underline{V}_2^+}\mathcal{F}[\beta_g](dA(\underline{p})) + \frac{1}{2}\sum_{\underline{b}\in\underline{V}_1^+}\mathcal{F}[\beta_h](d\varphi(\underline{b}) - A(\underline{b}))\right]. \tag{3.19}$$

Periodic boundary conditions can be obtained with the choice $E_{\underline{V}} = 1\!\!1$. We have, for $V^{(n)} := \underline{V} \times \{-n+1, \ldots, n\} \subset \mathbb{Z}^3$,

$$\omega_{\underline{V}}(B) = \lim_{n\to\infty} \langle B^{cl} \rangle_{V(n)} = \frac{\sum_{(\varphi,A)} B^{cl}(\varphi,\, A) e^{-H_{V(n)}(\varphi, A)}}{Z_{V(n)}}, \tag{3.20}$$

where,

$$Z_{V(n)} = \sum_{(\varphi,A)} e^{-H_{V(n)}(\varphi,A)}, \tag{3.21}$$

with the generalized Wilson action

$$-H_{V(n)}(\varphi,\, A) := \sum_{a=-n+1}^{n} \left[ \sum_{\underline{p}\in \underline{V}_2^+} \mathcal{F}[\beta_g](dA((\underline{p},\, a))) + \sum_{\underline{b}\in \underline{V}_1^+} \mathcal{F}[\beta_h](d\varphi((\underline{b},\, a)) - A((\underline{b}\, a))) \right] +$$

$$+ \sum_{a=-n+1}^{n-1} \left[ \sum_{\underline{b}\in \underline{V}_1^+} \mathcal{F}[\beta_g](A((\underline{b},\, a)) - A((\underline{b},\, a+1)) + \sum_{\underline{x}\in \underline{V}_0^+} \mathcal{F}[\beta_h](\varphi((\underline{x},\, a+1)) - \varphi((\underline{x},\, a))) \right], \tag{3.22}$$

with free boundary conditions. Here $(\underline{x},\, a) \in \mathbb{Z}^{d+1}$, etc. Above $B^{cl}$ is a classical function of the classical fields $\varphi$ and $A$ associated to the operator $B$. The choice of $B^{cl}$ is generally non-unique and some prescriptions for determining it from a given operator can be found in [1]. We will not enter into details here. The important fact is that one can choose $B^{cl}$ as

$$B^{cl}(\varphi,\, A) = \frac{Tr_{\mathcal{H}_{\underline{V}}} \left( E^*_{(\varphi(0),A(0)),(\varphi(1),A(1))} B\, T_{\underline{V}} \right)}{T_{\underline{V}}(\varphi(0), A(0);\, \varphi(1), A(1))}, \tag{3.23}$$

where $\varphi(k)$, $A(k)$ refers to the variables in the $k$-th Euclidean time hyperplane. One can also use the following useful rules. If $B \in \mathfrak{F}(\underline{V}_1)$ and $C \in \mathfrak{F}(\underline{V}_2)$ with dist $(\underline{V}_1,\, \underline{V}_2) \geq 2$, then one can choose $(BC)^{cl} = B^{cl}C^{cl}$. Beyond this, if $B$ is of the form $B = \alpha_{ia_1}(C^1) \cdots \alpha_{ia_k}(C^k)$ with $a_1 < \ldots < a_k$, then one can choose

$$B^{cl} = \prod_{j=1}^{k} (C^j)^{cl}(\varphi(a_j), A(a_j);\, \varphi(a_j+1), A(a_j+1)). \tag{3.24}$$

Finally we note that $J^{cl} = 0$ and so the classical function above is constant on the equivalence classes defining the elements of $\mathfrak{B}_0$.

It is useful to change to the unitary gauge by defining the new function $u$, $V_1^{(n)} \to \{0, \ldots, N-1\}$: $u(b) = d\varphi - A(b) \bmod N$, if $b$ is a space-like bond, $u(b) = \varphi(y_b) - \varphi(x_b) \bmod N$; for $b$ time-like, where the $x_b$ and $y_b = x_b + (1,\, \underline{0})$ are the boundaries of the bond $b$. We get for the partition function

$$Z_{V(n)} = \sum_{u} \prod_{b\in V_1^{(n)}} \mathcal{F}[h](u(b)) \prod_{p\in V_2^{(n)}} g(du(p)), \tag{3.25}$$

with the following important definitions:

$$g(n) := e^{\mathcal{F}[\beta_g](n)} \quad \text{and} \quad h(n) := \mathcal{F}^{-1}[e^{\mathcal{F}[\beta_h]}](n) = e^{\mathcal{F}[\gamma_h](n)}. \tag{3.26}$$

For gauge invariant classical observables one has,

$$\langle B\rangle_{V^{(n)}} = Z_{V^{(n)}}^{-1}\sum_u B(u)\prod_{b\in V_1^{(n)+}}\mathcal{F}[h](u(b))\prod_{p\in V_2^{(n)+}} g(du(p)). \tag{3.27}$$

The existence of the thermodynamic limit of the last classical expectation can be established by standard techniques, for instance, using the polymer expansion introduced below or Griffiths inequalities. We will not enter into details here. This limit defines a translation invariant state $\omega_0$ of $\mathfrak{A}_0$, which we call the vacuum state. By construction it is a ground state with respect to $\alpha_i$ (see [1]).

An important observation is the fact that, for $\mathcal{Q}(\lambda) := \prod_{\underline{x}}\mathcal{Q}(\underline{x})^{\lambda(\underline{x})}$, where $\lambda$ is a 0-form with finite support in $\mathbb{Z}^2$, and for all $F\in\mathfrak{F}_0$, one has

$$\omega_0(F(\mathcal{Q}(\lambda)-1)) = 0. \tag{3.28}$$

This follows from (3.18)-(3.17) and from the fact that $\mathcal{Q}(\lambda)E_{(\varphi,A),(\varphi',A')} = E_{(\varphi+\lambda,A+d\lambda),(\varphi',A')}$ which by (3.19) implies $\mathcal{Q}(\lambda)E_{\underline{V}} = E_{\underline{V}}$. Therefore, for the previously defined two-sided ideal $J$ one has $\omega_0(J)=0$ and so we are allowed to define $\omega_0$ on $\mathfrak{B}$ by $\omega_0(A+J) := \omega_0(A)$, $A\in\mathfrak{A}$.

# 4 The Polymer Expansion.

In this section we develop polymer and cluster expansions for the classical expectations found in the previous section in their "free charges" phase and show their convergence regions. Cluster expansions are the technically most important tool of this work (see also [1]) because they provide a method for rigorously extracting informations from the classical expectations, and therefore, from the various states we will consider on the quantum spin system.

Let $V\subset\mathbb{Z}^{d+1}$ be a cubic box of the form $V=\underline{V}\times\{-n+1,\ldots,n\}$. To simplify matters we can consider periodic boundary conditions here. Free boundary conditions can also be treated with the polymer expansions below. Call $\mathcal{D}_V$ the set of all defect-networks of $V$, i.e., a function $D$: $V_2\to\{0,\ldots,N-1\}$ belongs to $\mathcal{D}_V$ iff $dD=0$ mod $N$. We write the partition function as

$$Z_V = \sum_{D\in\mathcal{D}_V}\left[\prod_{p\in\mathrm{supp}D} g(D(p))\right]\sum_{u:\,du=D\,\mathrm{mod}N}\ \prod_{b\in V_1^+}\mathcal{F}[h](u(b)). \tag{4.1}$$

Above we have chosen $\beta_g(0)$ so that $g(0)=1$.

Let us associate to each $D\in\mathcal{D}_V$ a configuration $u^D\in V^1$ so that $du^D=D$. Then we can write:

$$\sum_{u:\,du=D}\prod_{b\in V_1^+}\mathcal{F}[h](u(b)) = \sum_{\chi\in V^0}\prod_{b\in V_1^+}\mathcal{F}[h](u^D(b)+d\chi(b)) =$$

$$N^{-|V_1^+|/2}\sum_{E\in V^1}\left(\prod_{b\in V_1^+} h(E(b))\sum_{\chi\in V^0}\exp\left(\frac{2\pi i}{N}\left(\langle u^D,\,E\rangle_{V^1}+\langle d\chi,\,E\rangle_{V^1}\right)\right)\right). \tag{4.2}$$

Since the sum over $\chi$ above equals $N^{|V_0^+|}\delta_{d^*E,0\mathrm{mod}N}$ we get:

$$Z_V^1 := N^{-|V_0^+|+|V_1^+|/2}Z_V =$$

$$\sum_{\substack{D\in V^2\\ dD=0\,\mathrm{mod}\,N}} \sum_{\substack{E\in V^1\\ d^*E=0\,\mathrm{mod}\,N}} \exp\left(\frac{2\pi i}{N}\langle u^D,\, E\rangle_{V^1}\right)\left[\prod_{p\in \mathrm{supp} D} g(D(p))\right]\left[\prod_{b\in \mathrm{supp} E} h(E(b))\right], \tag{4.3}$$

where we have chosen $h(0) = 1$.

We will use the following

**Notation 4.1** *For 1-forms $E$ with $d^*E = 0$ and for 2-forms $D$ with $dD = 0$, both with finite support we define*

$$[D:\, E] := \exp\left(\frac{2\pi i}{N}\langle u^D,\, E\rangle\right) \qquad \Box \tag{4.4}$$

Define the sets

$$\mathcal{P}(V) = \Big\{P \in V_2^+ :\ P \text{ is co-connected and } P = \text{ supp}\, D, \text{ for some } D \in V^2,\ dD = 0,\ D \neq 0\Big\}, \tag{4.5}$$

$$\mathcal{B}(V) = \Big\{M \in V_1^+ :\ M \text{ is connected and } M = \text{ supp}\, E, \text{ for some } E \in V^1,\ d^*E = 0,\ E \neq 0\Big\}, \tag{4.6}$$

and

$$\mathcal{P}_{total}(V) \ = \ \Big\{P \in V_2^+ \text{ so that } P = \text{ supp}\, D, \text{ for some } D \in V^2,\ dD = 0\Big\}, \tag{4.7}$$

$$\mathcal{B}_{total}(V) \ = \ \Big\{M \in V_1^+ \text{ so that } M = \text{ supp}\, E, \text{ for some } E \in V^1,\ d^*E = 0\Big\}. \tag{4.8}$$

Above and below relations like $dD = 0$ mean, more precisely, that $dD = 0 \ \mathrm{mod} N$.

Note that the sets $\mathcal{P}_{total}(V)$ and $\mathcal{B}_{total}(V)$ contain the empty set and that the non-empty elements of $\mathcal{P}_{total}(V)$ and of $\mathcal{B}_{total}(V)$ are build up by unions of co-disjoint elements of $\mathcal{P}(V)$, respectively, by unions of disjoint elements of $\mathcal{B}(V)$. One has naturally $\mathcal{P}(V) \subset \mathcal{P}_{total}(V)$ and $\mathcal{B}(V) \subset \mathcal{B}_{total}(V)$.

We get

$$Z_V^1 = \sum_{\substack{P\in\mathcal{P}_{total}(V)\\ M\in\mathcal{B}_{total}(V)}} \sum_{\substack{D\in V^2\\ \mathrm{supp} D=P\\ dD=0}} \sum_{\substack{E\in V^1\\ \mathrm{supp} E=M\\ d^*E=0}} [D:\, E]\left[\prod_{p\in P} g(D(p))\right]\left[\prod_{b\in M} h(E(b))\right]. \tag{4.9}$$

We recall the assumptions that $g(0) = 1$ and $h(0) = 1$.

Each non-empty set $P \in \mathcal{P}_{total}(V)$ and $M \in \mathcal{B}_{total}(V)$ can be uniquely decomposed into disjoint unions $P = P_1 + \cdots + P_{A_P}$, $M = M_1 + \cdots + M_{B_M}$ where $P_i \in \mathcal{P}(V)$ and $M_j \in \mathcal{B}(V)$. Then, if $D \in V^2$ is such that $\text{supp}\, D = P$, there is a unique decomposition $D = D_1 + \cdots + D_{A_P}$ with $D_i \in V^2$, $\text{supp}\, D_i = P_i$ and respectively, if $E \in V^1$ is such that $\text{supp}\, E = M$ then there is a unique decomposition $E = E_1 + \cdots + E_{B_M}$ with $E_i \in V^1$, $\text{supp}\, E_i = M_i$. One can also decompose $u = u^{D_1} + \cdots + u^{D_{A_P}}$ with $u^{D_i} \in V^1$, $du^{D_i} = D_i$.

We get

$$Z_V^1 = \sum_{\substack{P\in\mathcal{P}_{total}(V)\\ M\in\mathcal{B}_{total}(V)}} \prod_{i=1}^{A_P}\prod_{j=1}^{B_M} \sum_{\substack{D_i\in V^2\\ \mathrm{supp} D_i=P_i\\ dD_i=0}} \sum_{\substack{E_j\in V^1\\ \mathrm{supp} E_j=M_j\\ d^*E_j=0}} [D_i:\, E_j]\left[\prod_{p\in P_i} g(D_i(p))\right]\left[\prod_{b\in M_j} h(E_j(b))\right], \tag{4.10}$$

with the conventions $A_\emptyset = 0$, $B_\emptyset = 0$. For $P \in \mathcal{P}_{total}(V)$ and $M \in \mathcal{B}_{total}(V)$ we define the sets

$$\mathcal{D}(P) \ := \ \{D \in V^2 \text{ so that } \text{supp}\, D = P \text{ and } dD = 0\} \tag{4.11}$$

$$\mathcal{E}(M) \ := \ \{E \in V^1 \text{ so that } \text{supp}\, E = M \text{ and } d^*E = 0\}. \tag{4.12}$$

We consider now pairs $(P,\ D)$ with $P \in \mathcal{P}_{total}(V)$ and $D \in \mathcal{D}(P)$ and pairs $(M,\ E)$ with $M \in \mathcal{B}_{total}(V)$ and $E \in \mathcal{D}(M)$ and define $w((P,\ D),(M,\ E)) = w((M,\ E),(P,\ D)) \in \{0, \ldots, N-1\}$ as the "$\mathbb{Z}_N$-winding number" of $(M,\ E)$ around $(P,\ D)$:

$$w((P,\ D),(M,\ E)) = w((M,\ E),(P,\ D)) := \langle u^D,\ E\rangle \mod N. \tag{4.13}$$

The pairs with $P \in \mathcal{P}(V)$ and $M \in \mathcal{B}(V)$ will be the building blocks of our polymers.

With the help of $w$ we can establish a connectivity relation between pairs $(P,\ D)$ with $P \in \mathcal{P}(V)$, $D \in \mathcal{D}(P)$ and pairs $(M,\ E)$ with $M \in \mathcal{B}(V)$, $E \in \mathcal{E}(M)$: we say that $(P,\ D)$ and $(M,\ E)$ are "$w$-connected" if $w((P,\ D),(M,\ E)) \neq 0$ and "$w$-disconnected" otherwise.

Now we are able to define our polymer model. A polymer $\gamma$ is formed by two pairs

$$\{(P^\gamma,\ D^\gamma),\ (M^\gamma,\ E^\gamma)\},$$

with $P^\gamma \in \mathcal{P}_{total}(V)$, $M^\gamma \in \mathcal{B}_{total}(V)$ and $D^\gamma \in \mathcal{D}(P^\gamma)$, $E^\gamma \in \mathcal{E}(M^\gamma)$, so that the set

$$\{(P_1^\gamma,\ D_1^\gamma), \ldots, (P_{A_\gamma}^\gamma,\ D_{A_\gamma}^\gamma), (M_1^\gamma,\ E_1^\gamma), \ldots, (M_{B_\gamma}^\gamma,\ E_{B_\gamma}^\gamma)\} \tag{4.14}$$

formed by the decompositions $P^\gamma = P_1^\gamma + \cdots + P_{A_\gamma}^\gamma$, $M^\gamma = M_1^\gamma + \cdots + M_{B_\gamma}^\gamma$ with $P_i^\gamma \in \mathcal{P}(V)$, $M_j^\gamma \in \mathcal{B}(V)$ and $D^\gamma = D_1^\gamma + \cdots + D_{A_\gamma}^\gamma$, $E^\gamma = E_1^\gamma + \cdots + E_{B_\gamma}^\gamma$ with $D_i^\gamma \in \mathcal{D}(P_i^\gamma)$, $E_j^\gamma \in \mathcal{E}(M_j^\gamma)$ is $w$-connected. Below when we write $(M,\ E) \in \gamma$ and $(P,\ D) \in \gamma$ we are intrinsically be assuming that $M \in \mathcal{B}(\mathbb{Z}^{d+1})$ with $E \in \mathcal{E}(M)$ and that $P \in \mathcal{P}(\mathbb{Z}^{d+1})$ with $D \in \mathcal{D}(P)$.

For a polymer $\gamma = ((P^\gamma,\ D^\gamma),\ (M^\gamma,\ E^\gamma))$ we call the pair $\gamma_g := (P^\gamma,\ M^\gamma)$ the geometrical part of $\gamma$ and the pair $\gamma_c := (D^\gamma,\ E^\gamma)$ is the "colouring" of $\gamma$. Each pair $(D,\ E)$, $D \in \mathcal{D}(P)$, $E \in \mathcal{E}(M)$ with $P \in \mathcal{P}(\mathbb{Z}^{d+1})$, $M \in \mathcal{B}(\mathbb{Z}^{d+1})$ is a colour for $(P,\ M)$. Another important definition is the "size" of a polymer. For reasons which will be clear in Appendix A we define the size of $\gamma$ by $|\gamma| = |\gamma_g| := |P^\gamma| + |M^\gamma|$, where $|P^\gamma|$ (respectively $|M^\gamma|$) is the number of plaquettes (respectively bonds) making up $P^\gamma$ (respectively $M^\gamma$).

A remark which will be of some relevance for the proof of convergence of the polymer expansion we are going to define is the fact that the sets $\mathcal{D}(P)$ and $\mathcal{E}(M)$ above have at most $(N-1)^{|P|}$, respectively $(N-1)^{|M|}$ elements. This estimate comes from the simple fact that the forms $D$ and $E$ can assume at each plaquette, respectively, at each bond, at most $(N-1)$ different values. Hence $\gamma_g$ can have at most $(N-1)^{|\gamma_g|}$ different colourings, i.e., there are at most $(N-1)^{|\gamma_g|}$ different polymers with the same geometrical part $\gamma_g$.

The activity $\mu(\gamma) \in \mathbb{C}$ of a polymer $\gamma$ is defined to be

$$\mu(\gamma) := [D^\gamma : E^\gamma] \left\{ \prod_{i=1}^{A_\gamma} \left[ \prod_{p \in P_i^\gamma} g(D_i^\gamma(p)) \right] \right\} \left\{ \prod_{j=1}^{B_\gamma} \left[ \prod_{b \in M_j^\gamma} h(E_j^\gamma(b)) \right] \right\}, \tag{4.15}$$

with $\mu(\emptyset) = 1$.

We need a notion of "compatibility" for pairs of polymers. Two polymers $\gamma$ and $\gamma'$ are said to be incompatible, $\gamma \not\sim \gamma'$ if at least one of the following conditions hold:

1. There exist $M_a^\gamma \in \gamma_g$ and $M_b^{\gamma'} \in \gamma_g'$, so that $M_a^\gamma$ and $M_b^{\gamma'}$ are connected (i.e., there exists at least one lattice point $x$ so that $x \in \partial b$ and $x \in \partial b'$ for some bonds $b \in M_a^\gamma$ and $b' \in M_b^{\gamma'}$);

2. There exists $P_a^\gamma \in \gamma_g$ and $P_b^{\gamma'} \in \gamma_g'$, so that $P_a^\gamma$ and $P_b^{\gamma'}$ are co-connected (i.e., there exists at least one 3-cube $c$ in the lattice so that $p \in \partial c$ and $p' \in \partial c$ for some plaquettes $p \in P_a^\gamma$ and $p' \in P_b^{\gamma'}$);

3. There exists $(M_a^\gamma, E_a^\gamma) \in \gamma$ and $(P_b^{\gamma'}, D_b^{\gamma'}) \in \gamma'$, so that $(M_a^\gamma, E_a^\gamma)$ and $(P_b^{\gamma'}, D_b^{\gamma'})$ are $w$-connected. Or the same with $\gamma$ and $\gamma'$ interchanged.

They are said to be compatible, $\gamma \sim \gamma'$, otherwise.

We denote by $\mathcal{G}(V)$ the set of all polymers in $V$ and by $\mathcal{G}_{com}(V)$ the set of all finite sets of compatible polymers. Having these definitions at hand we can write (4.10) as:

$$Z_V^1 = \sum_{\Gamma \in \mathcal{G}_{com}} \mu^\Gamma, \tag{4.16}$$

in multi-index notation. We will often identify the elements of $\mathcal{G}_{com}$ with their characteristic functions.

We want to express the expectation of classical observables (3.27) in terms of our polymer expansion. We consider the following

**Definition 4.1** *Let $\alpha$ and $\beta$ be a 1-form, respectively a 2-form with finite support. Define*

$$B(\alpha, \beta) := \exp\left[-\frac{2\pi i}{N}\langle \alpha, u\rangle\right] \prod_p \frac{g((du+\beta)(p))}{g(du(p))} \qquad \square \tag{4.17}$$

Since any classical observable can be written as a linear combination of such functions we consider

$$\langle B(\alpha, \beta)\rangle_V = \frac{1}{Z_V^1} \sum_{\substack{D \in V^2 \\ d(D-\beta)=0}} \sum_{\substack{E \in V^1 \\ d^*(E-\alpha)=0}} [D-\beta : E-\alpha] \left[\prod_{p \in \text{supp} D} g(D(p))\right] \left[\prod_{b \in \text{supp} E} h(E(b))\right]. \tag{4.18}$$

One has the following positivity properties:

$$\langle B(0, \beta)\rangle_V \geq 0, \qquad \langle B(\alpha, 0)\rangle_V \geq 0. \tag{4.19}$$

The first one is obvious from (4.17) and the second follows from the first using the duality results proven in Proposition B.1, Appendix B. Both follow also from Griffiths inequalities. Expectations like $\langle B(\alpha, \beta)\rangle_V$ are generally complex numbers but one can easily check that the following relations hold:

$$\langle B(\alpha, \beta)\rangle_V = \overline{\langle B(-\alpha, \beta)\rangle_V} = \overline{\langle B(\alpha, -\beta)\rangle_V} = \langle B(-\alpha, -\beta)\rangle_V. \tag{4.20}$$

The forms $D$ above can uniquely be decomposed in such a way that $D = D_0 + D_1$ with $d(D_0 - \beta) = 0$ and $dD_1 = 0$ and so that supp $D_0$ is co-connected and co-disconnected from supp $D_1$. If $d\beta = 0$ we choose $D_0 = 0$. The forms $E$ above, in turn, can be decomposed uniquely in such a way that $E = E_0 + E_1$ with $d^*(E_0 - \alpha) = 0$ and $d^* E_1 = 0$ and so that supp $E_0$ is connected and disconnected from supp $E_1$. If $d^*\alpha = 0$ we choose $E_0 = 0$.

We denote by $\mathcal{C}_1(\alpha)$ the set of the supports of all such $E_0$'s, for a given $\alpha$ and by $\mathcal{C}_2(\beta)$ the set of the supports of all such $D_0$'s, for a given $\beta$. For $d^*\alpha = 0$ we have $\mathcal{C}_1(\alpha) = \emptyset$ and for $d\beta = 0$ we have $\mathcal{C}_2(\beta) = \emptyset$. We define the sets of pairs

$$\text{Conn}_1(\alpha) := \Big\{(M, E), \text{ so that } M \in \mathcal{C}_1(\alpha) \text{ and } E \in V^1, \text{ with } \operatorname{supp} E = M \text{ and } d^*E = d^*\alpha\Big\}, \tag{4.21}$$

$$\text{Conn}_2(\beta) := \Big\{(P, D), \text{ so that } P \in \mathcal{C}_2(\beta) \text{ and } D \in V^2, \text{ with } \operatorname{supp} D = P \text{ and } dD = d\beta\Big\}. \tag{4.22}$$

We then write

$$
\begin{aligned}
\langle B(\alpha,\,\beta)\rangle_V \;\;=\;\; & \sum_{\substack{(M,\,E)\in \mathrm{Conn}_1(\alpha)\\ (P,\,D)\in \mathrm{Conn}_2(\beta)}} [D-\beta:\,E-\alpha] \\
& \times \left[\prod_{p\in P} g(D(p))\right] \left[\prod_{b\in M} h(E(b))\right] \frac{\sum_{\Gamma\in\mathcal{G}_{com}} a^{\Gamma}_{(M,E),\,\alpha} b^{\Gamma}_{(P,D),\,\beta}\, \mu^{\Gamma}}{\sum_{\Gamma\in\mathcal{G}_{com}} \mu^{\Gamma}}, \qquad (4.23)
\end{aligned}
$$

for

$$
a_{(M,E),\,\alpha}(\gamma) := \begin{cases} 0, & \text{if } M^{\gamma} \text{ is connected with } M, \\ [D^{\gamma}:\,E-\alpha]\,, & \text{otherwise} \end{cases}, \qquad (4.24)
$$

and

$$
b_{(P,D),\,\beta}(\gamma) := \begin{cases} 0, & \text{if } P^{\gamma} \text{ is co-connected with } P \\ [D-\beta:\,E^{\gamma}]\,, & \text{otherwise} \end{cases}. \qquad (4.25)
$$

It is for many purposes useful to write (4.16) in the form

$$
Z^1_V = \exp\left\{ \sum_{\Gamma\in\mathcal{G}_{clus}(V)} c_{\Gamma}\mu^{\Gamma} \right\}. \qquad (4.26)
$$

Let us explain the symbols used above. Our notation is close to that of [1]. $\mathcal{G}_{clus}(V)$ is the set of all finite clusters of polymers in $V$, i.e., an element $\Gamma \in \mathcal{G}_{clus}$ is a finite set of (not necessarily distinct) polymers building a connected "incompatibility graph". An incompatibility graph is a graph which has polymers as vertices and where two vertices are connected by a line if the corresponding polymers are incompatible. We will often identify an elements $\Gamma \in \mathcal{G}_{clus}$ with a function $\Gamma$: $\mathcal{G} \to \mathbb{N}$, where $\Gamma(\gamma)$ is the multiplicity of $\gamma$ in $\Gamma \in \mathcal{G}_{clus}$. The coefficients $c_{\Gamma}$ are the "Ursell functions" and are of purely combinatorial nature. They are defined (see [1] and [12]) by

$$
c_{\Gamma} := \sum_{n=1}^{\infty} \frac{(-1)^{n+1}}{n} \mathcal{N}_n(\Gamma), \qquad (4.27)
$$

where $\mathcal{N}_n(\Gamma)$ is the number of ways of writing $\Gamma$ in the form $\Gamma = \Gamma_1 + \cdots + \Gamma_n$ where $0 \neq \Gamma_i \in \mathcal{G}_{com}$, $i = 1, \ldots, n$.

Relation (4.26) makes sense provided the sum over clusters is convergent. As discussed in [1] and Appendix A, a sufficient condition for this is $\|\mu\| \leq \|\mu\|_c$, where $\|\mu\| := \sup_{\gamma\in\mathcal{G}} |\mu(\gamma)|^{1/|\gamma|}$, and $\|\mu\|_c$ is a constant defined in [1] (see also Appendix A below). By (4.15),

$$
|\mu(\gamma)| \leq [\max\{g(1), \ldots, g(N-1), h(1), \ldots, h(N-1)\}]^{|\gamma|} \qquad (4.28)
$$

and by the condition $g(0) = 1$ and $h(0) = 1$ we have $\mathcal{F}[\beta_g](0) = 0$ and $\mathcal{F}[\gamma_h](0) = 0$, from which it follows that, for $n \in \{1, \ldots, N-1\}$,

$$
\mathcal{F}[\beta_g](n) = \sum_{m=0}^{N-1} \beta_g(m) \left(\cos\left(\frac{2\pi i n m}{N}\right) - 1\right) / \sqrt{N} < 0, \qquad (4.29)
$$

and

$$
\mathcal{F}[\gamma_h](n) = \sum_{m=0}^{N-1} \gamma_h(m) \left(\cos\left(\frac{2\pi i n m}{N}\right) - 1\right) / \sqrt{N} < 0, \qquad (4.30)
$$

if $\beta_g(m) > 0$ and $\gamma_h(m) > 0$ for all $m = 1, \ldots, N-1$. Therefore one has $\|\mu\| \leq e^{-b}$, where this $b$ can be chosen to be arbitrarily large, for each $N$ fixed, by choosing $\min\{\beta_g(1), \ldots, \beta_g(N-1), \gamma_h(1), \ldots, \gamma_h(N-1)\}$ to be large enough. All results concerning charged states presented below hold inside of the convergence region above.

We also write

$$
\begin{aligned}
\langle B(\alpha,\,\beta)\rangle_V \;=\; & \sum_{\substack{(M,\,E)\in \mathrm{Conn}_1(\alpha)\\ (P,\,D)\in \mathrm{Conn}_2(\beta)}} [D-\beta:\,E-\alpha]\left[\prod_{p\in P} g(D(p))\right]\left[\prod_{b\in M} h(E(b))\right] \\
& \times \exp\left(\sum_{\Gamma\in\mathcal{G}_{clus}(V)} c_\Gamma\left(a^\Gamma_{(M,E),\alpha}\, b^\Gamma_{(P,D),\beta}-1\right)\mu^\Gamma\right). \qquad (4.31)
\end{aligned}
$$

# 5 Duality Transformations. Algebraic Aspects.

In this section we will review and extend some results of Gaebler [8] on duality transformations and apply them in the definition and study of algebraic properties of global transfer matrices in the vacuum sector.

Let us start defining $S_{\underline{V}} = \exp(B_{\underline{V}}/2)\exp(A_{\underline{V}}/2)$, $A_{\underline{V}}$ and $B_{\underline{V}}$ defined in (2.14)-(2.15). Defining the automorphism of $\mathfrak{F}$

$$\beta_{\underline{V}}(A) := S_{\underline{V}} A S_{\underline{V}}^{-1}, \qquad A\in\mathfrak{F}, \qquad (5.1)$$

one has $\alpha_i(\cdot)_{\underline{V}} = \beta_{\underline{V}}^{*\,-1}\circ\beta_{\underline{V}}(\cdot)$. The adjoint $\gamma^*$ of an automorphism $\gamma$ was previously defined through $\gamma^*(A) := (\gamma(A^*))^*$. The limit $\underline{V}\uparrow\mathbb{Z}^d$ of $\beta_{\underline{V}}$ exists on $\mathfrak{F}_0$ for the same reason why it exists for $\alpha_{i,\underline{V}}$ and defines a non-$*$ automorphism we will denote by $\beta$. One has $\alpha_i = \beta^{*\,-1}\circ\beta$. Since $\beta$ keeps the ideal $J$ invariant one naturally defines the action of $\beta$ on $\mathfrak{B}_0$, which we again denote by the same symbol $\beta$, by $\beta(A+J) = \beta(A)+J$, $A\in\mathfrak{A}_0$, as a non-$*$ automorphism.

Below we will also be considering the dynamics defined by the automorphism $\alpha_i^T := \beta\circ\beta^{*\,-1}$, which is obtained by interchanging $T_{\underline{V}}$ by $T_{\underline{V}}^T$ in the definition of $\alpha_i$.

The automorphism $\beta$ has been introduced in [8] and plays an important role in the study of duality transformations. At algebraic level duality transformations are introduced by a $*$-endomorphism $\Delta$ of the observable algebra which, in the model we are considering, is defined on the generators of $\mathfrak{A}_0$ by

$$
\begin{aligned}
\Delta(\mathbb{1}) &:= \mathbb{1}, & (5.2)\\
\Delta(P_G(\underline{b})) &:= [\delta Q_H]\,(-*\underline{b})Q_G(-*\underline{b})^*, & (5.3)\\
\Delta([\delta Q_H]\,(\underline{b})Q_G(\underline{b})^*) &:= P_G(*\underline{b}), & (5.4)\\
\Delta(P_H(\underline{x})) &:= \delta Q_G(*\underline{x}), & (5.5)
\end{aligned}
$$

where the geometric duality transformations on $\mathbb{Z}^2$ and its cells are presented in Figure 1.

Above $\underline{x}$, $*\underline{p}\in\underline{l}_0$; $\underline{b}$, $*\underline{b}\in\underline{l}_1$ and $\underline{p}$, $*\underline{x}\in\underline{l}_2$. With these definitions one has on the $a$-cells $** = (-1)^a f_{(1,1)}$, where $f_{(x,\,y)}$ is a shift of the cells by $(x,\,y)$ in $\mathbb{Z}^2$.

**Definition 5.1** *Consider a 1-form $\gamma$. Define $(*\gamma)(\underline{b}) := \gamma(-f_{-(1,1)}(*\underline{b}))$ and the translation on forms $(g_{(a,b)}\gamma)(\underline{b}) := \gamma(f_{-(a,b)}(\underline{b}))$. The operation $*$ is analogous to the Hodge-$*$ operation. One also has $(**\gamma)(\underline{b}) = -(g_{(1,1)}\gamma)(\underline{b})$* □

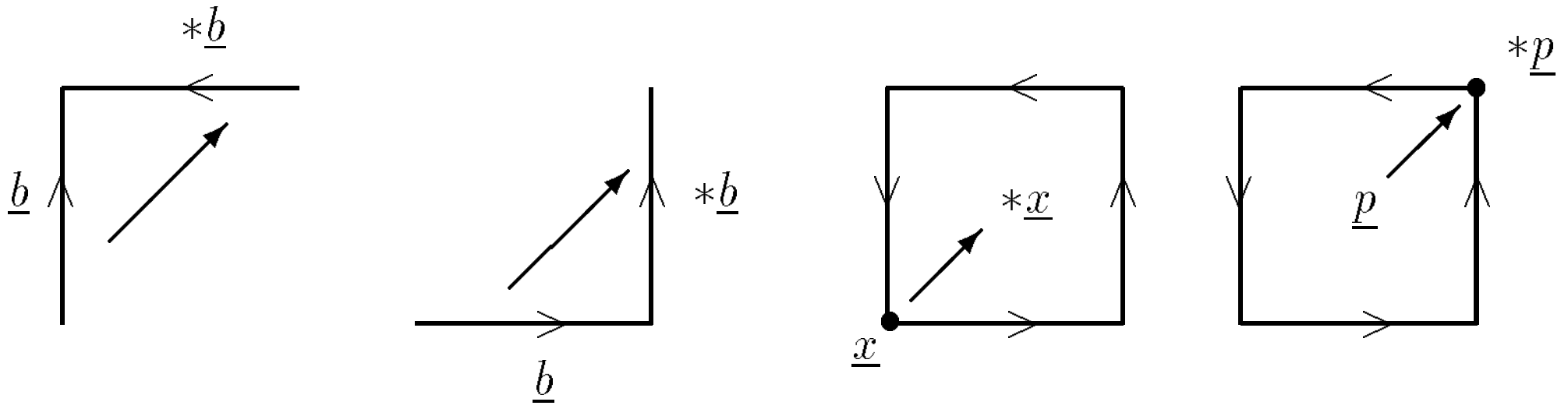

Figure 1: *The transformations* $\underline{b} \to *\underline{b}$, $\underline{x} \to *\underline{x}$ *and* $\underline{p} \to *\underline{p}$.

Since $\Delta(\mathcal{Q}(\underline{x})) = \mathbb{1}$, $\Delta$ annihilates the ideal $J$, and therefore the action of $\Delta$ on $\mathfrak{B}$ is a $*$-automorphism. We denote this action by the same symbol $\Delta$. On the generators of $\mathfrak{B}_0$ it acts like

$$\begin{aligned} \Delta(\mathbb{1} + \mathrm{J}) &:= \mathbb{1} + \mathrm{J}, & (5.6)\\ \Delta(U_1(\gamma)) &:= U_3(-(*\gamma)), & (5.7)\\ \Delta(U_3(\gamma)) &:= U_1(*\gamma). & (5.8) \end{aligned}$$

for a 1-form $\gamma$. One can check that on $\mathfrak{B}$ one has $\Delta^2 = \tau_{(1,1)}$.

In order to analyze the interplay between $\Delta$ and $\beta$ let us define the *duality transformations of the couplings*. Let $\mathcal{Z}$ be the set of all functions $\{0, \ldots, N-1\} \to \mathbb{C}$ let $D$ and $D^{-1}$ be the transformations $\mathcal{Z} \to \mathcal{Z}$ defined by

$$\begin{aligned} D[a] &:= \mathcal{F}^{-1}\left[\ln\left(\mathcal{F}\left[\exp(\mathcal{F}[a])\right]\right)\right], & (5.9)\\ D^{-1}[a] &:= \mathcal{F}^{-1}\left[\ln\left(\mathcal{F}^{-1}\left[\exp(\mathcal{F}[a])\right]\right)\right] & \\ &= \mathcal{F}^{-1}\left[\ln\left(\mathcal{F}\left[\exp(\mathcal{F}^{-1}[a])\right]\right)\right]. & (5.10) \end{aligned}$$

One easily checks that really $D \circ D^{-1} = D^{-1} \circ D = \mathrm{id}$. Let $(a, b)$ be a pair of functions in $\mathcal{Z}$. We call the map

$$(a, b) \to (a, b)' := (D^{-1}[b], D[a]) \qquad (5.11)$$

a duality transformation. Note that, in general, $(a, b)'' = (a, b)$ and that, according to (3.11)-(3.13), for the coupling functions $\beta_g$, $\beta_h$ the duality transformations are simply

$$(\beta_g, \beta_h) \to (\beta_g, \beta_h)' = (\gamma_h, \gamma_g). \qquad (5.12)$$

We extend this notation to arbitrary functions of the couplings with values in $\mathbb{C}$. For a map $f : \mathcal{Z} \times \mathcal{Z} \to \mathbb{C}$ we define $f' : \mathcal{Z} \times \mathcal{Z} \to \mathbb{C}$ by $f'(a, b) := f(D^{-1}[b], D[a])$. In particular $f'(\beta_g, \beta_h)$ denotes $f(\gamma_h, \gamma_g)$. We also generalize this notion to operators, states and automorphisms in $\mathfrak{F}_0$. For this, denote by $E_i$ a generator of $\mathfrak{F}_0$, i.e., $E_i$ is a finite product involving $\mathbb{1}$, $Q_H(\underline{x})$, $P_H(\underline{x})$, $Q_G(\underline{b})$ and $P_G(\underline{b})$ for $\underline{x}$, $\underline{b} \in \mathbb{Z}^d$. Writing for $A \in \mathfrak{F}_0$, $A = \sum_i c_i E_i$ as a finite sum, where the $c_i$'s eventually depend on the couplings, we define $A' := \sum_i c_i' E_i$ as a mapping $\mathfrak{F}_0 \to \mathfrak{F}_0$. One easily sees that this definition is independent of the basis of generators chosen, if the elements of the base do not depend on the couplings. For states we define $\omega'(A) := (\omega(A'))'$, i.e., $\omega'(A) = \sum_i c_i(\omega(E_i))'$, with $\omega(E_i)$ taken as a function of the couplings. Finally we define for an automorphism $\gamma$: $\gamma'(A) := (\gamma(A'))'$, i.e. $\gamma'(A) = \sum_{i,j} c_i d'_{i,j} E_j$, where $\gamma(E_i) = \sum_j d_{i,j} E_j$, the $d_{i,j}$'s being eventually functions of the couplings. Since this duality map keeps the ideal $J$ invariant these definitions extend to $\mathfrak{B}_0$ as well. Finally note that for all objects $a$ above one has $a'' = a$.

With these definitions at hand and assuming $\beta_g$, $\beta_h$, $\gamma_g$ and $\gamma_h$ to be <u>real</u> couplings one can easily verify the following relations (c.p. [8]):

$$\begin{array}{ll} \Delta(A_{\underline{V}} + J) = B'_{*\underline{V}} + J, & \Delta(B_{\underline{V}} + J) = A'_{*\underline{V}} + J, \\ \Delta(S_{\underline{V}} + J) = (S'_{*\underline{V}})^* + J, & \Delta(T_{\underline{V}} + J) = (T^{T\,\prime}_{*\underline{V}})^* + J. \end{array} \tag{5.13}$$

which justify calling $\Delta$ a duality transformation. They imply the following relation between $\Delta$ and $\beta$, as automorphisms on $\mathfrak{B}_0$:

$$\Delta \circ \beta \;=\; (\beta^{*\,-1})' \circ \Delta, \tag{5.14}$$

from which we derive the following useful relations on $\mathfrak{B}_0$:

$$\Delta \circ \beta^{*\,-1} \;=\; \beta' \circ \Delta, \tag{5.15}$$

$$\Delta \circ \alpha_i \;=\; \alpha_i^{T\,\prime} \circ \Delta, \tag{5.16}$$

$$((\beta^{*\,-1})' \circ \Delta) \circ \alpha_i \;=\; \alpha_i' \circ ((\beta^{*\,-1})' \circ \Delta), \tag{5.17}$$

$$((\beta^{-1})' \circ \Delta) \circ \alpha_i \;=\; \alpha_i' \circ ((\beta^{-1})' \circ \Delta). \tag{5.18}$$

We also remark that, generally

$$\beta \circ \alpha_i \;=\; \alpha_i^T \circ \beta, \tag{5.19}$$

$$\beta^* \circ \alpha_i \;=\; \alpha_i^T \circ \beta^*. \tag{5.20}$$

In order to establish some properties of duality transformations we need a general abstract result. Let $\{\omega_a\}$ be a finite set of states on a unital $*$-algebra $\mathfrak{C}$ and let us assume the existence of automorphisms $\gamma_{a,b}$ of $\mathfrak{C}$ such that, for all pairs $(a,\, b)$,

$$\omega_a = \omega_b \circ \gamma_{b,\,a} = \omega_b \circ \gamma^*_{b,\,a}, \tag{5.21}$$

$$\gamma_{a,\,b} = \gamma^{-1}_{b,\,a} \qquad \text{and} \qquad \gamma_{a,\,a} = \text{id}. \tag{5.22}$$

Define

$$\alpha_{a,\,b} \;:=\; \gamma_{a,\,b} \circ \gamma^*_{b,\,a} \tag{5.23}$$

$$\alpha^{-1}_{a,\,b} \;:=\; \gamma^*_{a,\,b} \circ \gamma_{b,\,a} = \alpha^*_{a,\,b}. \tag{5.24}$$

Clearly one has $\omega_a \circ \alpha_{a,\,b} = \omega_a \circ \gamma_{a,\,b} \circ \gamma^*_{b,\,a} = \omega_b \circ \gamma^*_{b,\,a} = \omega_a$. The following Theorem generalizes a result of [8].

**Theorem 5.1** *For a fixed pair $(a,\, b)$ the statements*

$$0 \le \omega_a(A^* \alpha_{a,b}(A)) \le K \omega_a(A^* A), \quad \forall A \in \mathfrak{C}, \tag{5.25}$$

*and*

$$0 \le \omega_b(A^* \alpha^{-1}_{b,\,a}(A)) \le K \omega_b(A^* A), \quad \forall A \in \mathfrak{C}, \tag{5.26}$$

*for some $K > 0$, are equivalent* □

**Proof:** (Taken in adapted form from [8]).

*i)* (5.25) $\Longrightarrow$ (5.26). First note that

$$0 \leq \omega_a((\gamma_{a,b}(A))^* \gamma_{a,b}(A)) = \omega_b \circ \gamma^*_{b,a}((\gamma_{a,b}(A))^* \gamma_{a,b}(A)) = \omega_b(A \alpha^{-1}_{b,a}(A)), \tag{5.27}$$

the first inequality to be proven. Beyond this one has

$$\omega_b(A^*A) = \omega_a \circ \gamma_{a,b}(A^*A) = \omega_a(\gamma^*_{a,b}(A)^* \alpha_{a,b} \circ \gamma^*_{a,b}(A)). \tag{5.28}$$

By the hypothesis this shows that $\omega_b(A^*A) \geq 0$ and that $\omega_b(A^*A) \leq K\omega_a(\gamma^*_{a,b}(A)^* \gamma^*_{a,b}(A))$. The right hand side equals $\omega_a \circ \gamma_{a,b}(A^* \gamma_{b,a} \gamma^*_{a,b}(A)) = \omega_b(A^* \alpha_{b,a}(A))$ and we arrive at

$$\omega_b(A^*A) \leq K\omega_b(A^* \alpha_{b,a}(A)). \tag{5.29}$$

Assuming without loss that $\omega_b(A^* \alpha^{-1}_{b,a}(A))) \neq 0$ we get by the Cauchy-Schwarz inequality that

$$\begin{aligned} \omega_b(A^* \alpha^{-1}_{b,a}(A))^2 &\leq \omega_b(A^*A)\omega_b(\alpha^{-1}_{b,a}(A)^* \alpha^{-1}_{b,a}(A)) && (5.30)\\ &\underset{by\,(5.29)}{\leq} K\omega_b(A^*A)\omega_b(\alpha^{-1}_{b,a}(A)^*A) = K\omega_b(A^*A)\omega_b(A^* \alpha^{-1}_{b,a}(A)), && (5.31) \end{aligned}$$

that means:

$$\omega_b(A^* \alpha^{-1}_{b,a}(A)) \leq K\omega_b(A^*A), \tag{5.32}$$

completing the proof of the statement.

*ii)* (5.26) $\Longrightarrow$ (5.25). This proof is analogous to the previous one. We can get to it through the replacement $a \longleftrightarrow b$ and by interchanging $\gamma \longleftrightarrow \gamma^*$, because these automorphisms play a symmetric role by (5.21) ■

Using the automorphism $\beta$ we can consider four possible dynamics:

$$\begin{array}{ll} \alpha^0_i = \alpha_i := \beta^{*\,-1} \circ \beta & ; \;\; \alpha^1_i = \alpha^T_i := \beta \circ \beta^{*\,-1}; \\ \alpha^2_i = \alpha_i{}' := (\beta^{*\,-1})' \circ \beta' & ; \;\; \alpha^3_i = \alpha^{T'}_i := \beta' \circ (\beta^{*\,-1})'. \end{array} \tag{5.33}$$

From Theorem 5.1 we get the following

**Corollary 5.1** *Consider the following four states on* $\mathfrak{B}_0$

$$\omega_0, \quad \omega_1 := \omega_0 \circ \beta^{*\,-1}, \quad \omega_2 := \omega_0 \circ \beta^{*\,-1} \circ \Delta, \quad \omega_3 := \omega_0 \circ \Delta. \tag{5.34}$$

*Then the claims that, for all* $j = 0, \ldots, 3$, $\omega_j$ *is a ground state w.r.t.* $\alpha^j_i$, *are equivalent claims* □

**Remarks.** Actually, since we are assuming that $\omega_0$ is a ground state w.r.t. $\alpha^0_i$, this Corollary says that $\omega_j$ is a ground state w.r.t. $\alpha^j_i$ for $j = 1$, 2 and 3 as well [8]. Note also that $\omega_1$ is really a state, i.e., a positive linear functional, since $0 \leq \omega_0(\beta^{-1}(A)^* \alpha_i(\beta^{-1}(A))) = \omega_0 \circ \beta^{*\,-1}(A^*A)$. For $\omega_2$ and $\omega_3$ the proof is analogous, since $\Delta$ is a $*$-automorphism. Finally remark that $\omega_0$ is a ground state w.r.t. $\alpha^0_i$ for the whole algebra $\mathfrak{F}_0$ and therefore the same holds for $\omega_1$ □

**Proof.** We need a family of automorphisms $\gamma_{a,b}$ satisfying (5.21) and (5.22) for all these $\omega_i$'s. A possible choice can be represented in matrix notation as

$$\begin{pmatrix} \gamma_{0,0} & \gamma_{0,1} & \gamma_{0,2} & \gamma_{0,3} \\ \gamma_{1,0} & \gamma_{1,1} & \gamma_{1,2} & \gamma_{1,3} \\ \gamma_{2,0} & \gamma_{2,1} & \gamma_{2,2} & \gamma_{2,3} \\ \gamma_{3,0} & \gamma_{3,1} & \gamma_{3,2} & \gamma_{3,3} \end{pmatrix} = \begin{pmatrix} \mathrm{id} & \beta^{*\,-1} & \beta^{*\,-1} \circ \Delta & \Delta \\ \beta^* & \mathrm{id} & \Delta & \beta^* \circ \Delta \\ \Delta^{-1} \circ \beta^* & \Delta^{-1} & \mathrm{id} & \Delta^{-1} \circ \beta^* \circ \Delta \\ \Delta^{-1} & \Delta^{-1} \circ \beta^{*\,-1} & \Delta^{-1} \circ \beta^{*\,-1} \circ \Delta & \mathrm{id} \end{pmatrix}. \tag{5.35}$$

The reader is invited to check that (5.21) and (5.22) are satisfied by this choice. From this, using (5.14) and (5.15) we get for the $\alpha$-automorphisms

$$\begin{pmatrix} \alpha_{0,0} & \alpha_{0,1} & \alpha_{0,2} & \alpha_{0,3} \\ \alpha_{1,0} & \alpha_{1,1} & \alpha_{1,2} & \alpha_{1,3} \\ \alpha_{2,0} & \alpha_{2,1} & \alpha_{2,2} & \alpha_{2,3} \\ \alpha_{3,0} & \alpha_{3,1} & \alpha_{3,2} & \alpha_{3,3} \end{pmatrix} = \begin{pmatrix} \mathrm{id} & \beta^{*\,-1}\circ\beta & \beta^{*\,-1}\circ\beta & \mathrm{id} \\ \beta^{*}\circ\beta^{-1} & \mathrm{id} & \mathrm{id} & \beta^{*}\circ\beta^{-1} \\ \beta'^{-1}\circ\beta'^{*} & \mathrm{id} & \mathrm{id} & \beta'^{-1}\circ\beta'^{*} \\ \mathrm{id} & \beta'\circ\beta'^{*\,-1} & \beta'\circ\beta'^{*\,-1} & \mathrm{id} \end{pmatrix}$$

$$= \begin{pmatrix} \mathrm{id} & \alpha_i^0 & \alpha_i^0 & \mathrm{id} \\ (\alpha_i^1)^{-1} & \mathrm{id} & \mathrm{id} & (\alpha_i^1)^{-1} \\ (\alpha_i^2)^{-1} & \mathrm{id} & \mathrm{id} & (\alpha_i^2)^{-1} \\ \mathrm{id} & \alpha_i^3 & \alpha_i^3 & \mathrm{id} \end{pmatrix} = \begin{pmatrix} \mathrm{id} & \alpha_i & \alpha_i & \mathrm{id} \\ \alpha_i^{T\,-1} & \mathrm{id} & \mathrm{id} & \alpha_i^{T\,-1} \\ \alpha_i^{-1\prime} & \mathrm{id} & \mathrm{id} & \alpha_i^{-1\prime} \\ \mathrm{id} & \alpha_i^{T\prime} & \alpha_i^{T\prime} & \mathrm{id} \end{pmatrix}. \tag{5.36}$$

Now applying Theorem 5.1 and looking at the matrix of $\alpha$-automorphisms above, we get the following chain:

$\omega_0$ is a ground state w.r.t. $\alpha_i^0 = \alpha_{0,1} \Longleftrightarrow \omega_1$ is a ground state w.r.t. $\alpha_{1,0}^{-1} = \alpha_i^1$,
$\omega_0$ is a ground state w.r.t. $\alpha_i^0 = \alpha_{0,2} \Longleftrightarrow \omega_2$ is a ground state w.r.t. $\alpha_{2,0}^{-1} = \alpha_i^2$,
$\omega_1$ is a ground state w.r.t. $\alpha_i^1 = \alpha_{1,3}^{-1} \Longleftrightarrow \omega_3$ is a ground state w.r.t. $\alpha_{3,1} = \alpha_i^3$,
$\omega_2$ is a ground state w.r.t. $\alpha_i^2 = \alpha_{2,3}^{-1} \Longleftrightarrow \omega_3$ is a ground state w.r.t. $\alpha_{2,3} = \alpha_i^3$ ■

Let us now investigate the properties of the transfer matrix associated to the states $\omega_i$. Call $(\pi_{\omega_i}, \mathcal{H}_{\omega_i}, \Omega_{\omega_i})$ the GNS-triple associated to the state $\omega_i$ and the algebra of observables $\mathfrak{B}_0$.

For a fixed pair $(a, b)$ we will assume that $\omega_a$ is a ground state w.r.t. $\alpha_{a,b}$. Define $U_{a\to b}$: $\mathcal{H}_{\omega_a} \to \mathcal{H}_{\omega_b}$ by

$$U_{a\to b}\pi_{\omega_a}(A)\Omega_{\omega_a} := \pi_{\omega_b}\circ\gamma^*_{b,a}(A)\Omega_{\omega_b}. \tag{5.37}$$

$U_{a\to b}$ is densely defined and is actually well defined since, in case $\pi_{\omega_a}(A)\Omega_{\omega_a} = 0$, one has

$$0 = \omega_a(A^*\gamma_{a,b}\circ\gamma^*_{b,a}(A)) = \omega_b(\gamma^*_{b,a}(A)^*\gamma^*_{b,a}(A)) = \|\pi_{\omega_b}\circ\gamma^*_{b,a}(A)\Omega_{\omega_b}\|^2. \tag{5.38}$$

Analogously, if $\pi_{\omega_b}\circ\gamma^*_{b,a}(A)\Omega_{\omega_b} = 0$ then

$$0 = \omega_b(\gamma^*_{b,a}(A)^*\gamma_{b,a}(A)) = \omega_b(\gamma_{b,a}(A^*A)) = \omega_a(A^*A) = \|\pi_{\omega_a}(A)\Omega_{\omega_a}\|^2 \tag{5.39}$$

and so $\mathrm{Ker}\, U_{a\to b} = \{0\}$. One also has $\overline{\mathrm{Ran}\, U_{a\to b}} = \mathcal{H}_{\omega_b}$, since $\gamma^*_{b,a}$ is invertible. One easily sees that $U_{a\to b}$ is bounded since, by the hypothesis,

$$\|U_{a\to b}\pi_{\omega_a}(A)\Omega_{\omega_a}\|^2_{\mathcal{H}_{\omega_b}} = \omega_b(\gamma_{b,a}(A^*)\gamma^*_{b,a}(A)) = \omega_a(A^*\alpha_{a,b}(A)) \le \omega_a(A^*A) = \|\pi_{\omega_a}(A)\Omega_{\omega_a}\|^2_{\mathcal{H}_{\omega_a}}. \tag{5.40}$$

Since $\gamma^*_{b,a}$ is invertible $U_{a\to b}$ is also invertible and the inverse is $U_{b\to a}$, which is densely defined. Note that this inverse is not, in general, bounded since $\omega_b$ is not a ground state w.r.t. $\alpha_{b,a}$ but w.r.t. $\alpha^{-1}_{b,a}$. One can also easily check that

$$(U_{a\to b})^*\pi_{\omega_b}(B)\Omega_{\omega_b} = \pi_{\omega_a}\circ\gamma_{a,b}(B)\Omega_{\omega_a}. \tag{5.41}$$

We can now define two transfer matrices $T_{\omega_a,\alpha_{a,b}}$ and $T_{\omega_b,\alpha^{-1}_{b,a}}$, acting on the spaces $\mathcal{H}_{\omega_a}$ and $\mathcal{H}_{\omega_b}$, respectively, by

$$T_{\omega_a,\alpha_{a,b}} := (U_{a\to b})^*U_{a\to b} \quad \text{and} \quad T_{\omega_b,\alpha^{-1}_{b,a}} := U_{a\to b}(U_{a\to b})^*. \tag{5.42}$$

and one easily sees that the natural definitions for such transfer matrices (as proposed in [1]), namely: $T_{\omega_a,\alpha_{a,b}}\pi_{\omega_a}(A)\Omega_{\omega_a} := \pi_{\omega_a}\circ\alpha_{a,b}(A)\Omega_{\omega_a}$ and $T_{\omega_b,\alpha^{-1}_{b,a}}\pi_{\omega_b}(A)\Omega_{\omega_b} := \pi_{\omega_b}\circ\alpha^{-1}_{b,a}(A)\Omega_{\omega_b}$, hold. These two transfer matrices are positive and bounded (since $U_{a\to b}$ is bounded) and have densely defined inverses. We now establish the following Proposition:

**Proposition 5.1** *The transfer matrices $T_{\omega_a,\alpha_{a,b}}$ and $T_{\omega_b,\alpha^{-1}_{b,a}}$ defined above are unitarily equivalent* □

**Proof.** By the polar decomposition one has $U_{a\to b} = \mathcal{U}_{a\to b}T^{1/2}_{\omega_a,\alpha_{a,b}}$ where $\mathcal{U}_{a\to b}$: $\mathcal{H}_{\omega_a}\to\mathcal{H}_{\omega_b}$ is a unitary map. By (5.42) one has $U_{a\to b}T_{\omega_a,\alpha_{a,b}} = T_{\omega_b,\alpha^{-1}_{b,a}}U_{a\to b}$ and using the polar decomposition above one easily verifies that $\mathcal{U}_{a\to b}T_{\omega_a,\alpha_{a,b}} = T_{\omega_b,\alpha^{-1}_{b,a}}\mathcal{U}_{a\to b}$ ■

From this it immediately follows the

**Corollary 5.2** *The transfer matrices $T_{\omega_0,\alpha^0_i}$, $T_{\omega_1,\alpha^1_i}$, $T_{\omega_2,\alpha^2_i}$ and $T_{\omega_3,\alpha^3_i}$ acting on the spaces $\mathcal{H}_{\omega_j}$, $j = 0,\ldots,3$, respectively, associated to the states $\omega_j$, and defined by $T_{\omega_a,\alpha^a_i}\pi_{\omega_a}(A)\Omega_{\omega_a} := \pi_{\omega_a}(\alpha^a_i(A))\Omega_{\omega_a}$, $A\in\mathfrak{B}_0$, are unitarily equivalent* □

We will simplify the notation and call $T_{\omega_a} := T_{\omega_a,\alpha^a_i}$. If $\tau_{\underline{x}}$ is the $*$-automorphism group generating translations by $\underline{x}\in\mathbb{Z}^d$ we can define the unitary operators

$$U_{\omega_a}(\underline{x})\pi_{\omega_a}(A)\Omega_{\omega_a} := \pi_{\omega_a}(\tau_{\underline{x}}(A))\Omega_{\omega_a} \tag{5.43}$$

implementing the translations on $\mathcal{H}_{\omega_a}$. Since $\tau_{\underline{x}}$ commutes with all $\gamma_{a,b}$ and with all $\alpha^j_i$ one easily verifies that $U_{a\to b}U_{\omega_a}(\underline{x}) = U_{\omega_b}(\underline{x})U_{a\to b}$ and that $U_{\omega_a}(\underline{x})T_{\omega_a} = T_{\omega_a}U_{\omega_a}(\underline{x})$. Using the polar decomposition above for $U_{a\to b}$ we get $\mathcal{U}_{a\to b}U_{\omega_a}(\underline{x}) = U_{\omega_b}(\underline{x})\mathcal{U}_{a\to b}$. Define the momentum operators by $U_{\omega_a}(\underline{x}) = e^{i\mathbb{P}_{\omega_a}\underline{x}}$, $\mathrm{sp}\,\mathbb{P}_{\omega_a} = (-\pi,\,\pi]^d$. Since the unitary operators intertwining the operators $U(\underline{x})$ are the same which intertwine the transfer matrices we have established the following

**Corollary 5.3** *The joint spectrum of the transfer matrix and the momentum operator, $sp\,(T_{\omega_a},\,\mathbb{P}_{\omega_a})$, is the same for all $a$* □

It is interesting to see that the operators $U_{0\to 3}$ and $U_{1\to 2}$ are related to the algebraic duality transformations in a simple way. One has namely $U_{0\to 3}\pi_{\omega_0}(A)\Omega_{\omega_0} = \mathcal{U}_{0\to 3}\pi_{\omega_0}(A)\Omega_{\omega_0} = \pi_{\omega_3}(\Delta^{-1}(A))\Omega_{\omega_3}$ and $U_{1\to 2}\pi_{\omega_1}(A)\Omega_{\omega_1} = \mathcal{U}_{1\to 2}\pi_{\omega_1}(A)\Omega_{\omega_1} = \pi_{\omega_2}(\Delta^{-1}(A))\Omega_{\omega_2}$.

At this point of our analysis an important question rises. Since $\omega_0'$ and $\omega_2$ are ground states w.r.t. the same dynamics, namely $\alpha_i'$, one could suppose that the identification $\omega_0' = \omega_2$ holds. However, this is by no means a trivial statement since the ground state w.r.t. a given dynamics must not be unique. In spite of this, for the $\mathbb{Z}_2$ case, this identification has been proven to be true in the region of convergence of the polymer expansions [8]. We have the following

**Theorem 5.2** *For the $\mathbb{Z}_N$ model considered here one has the following relations for states on $\mathfrak{B}_0$, valid at least in the region of convergence of the polymer expansions already described:*

$$i)\ \omega_0' = \omega_2 \qquad and \qquad ii)\ \omega_3' = \omega_1 \qquad \square \tag{5.44}$$

Note that *i)* and *ii)* are equivalent. We are going to present a proof of *i)* for the $\mathbb{Z}_N$ case in Appendix B. We remark that this Theorem holds as far as one can establish the existence of a unique thermodynamic limit for the classical expectations.

Theorem 5.2 shows that the states $\omega_2$ and $\omega_3$, originally defined on $\mathfrak{B}_0$, have natural extensions to $\mathfrak{F}_0$, namely $\omega_0'$ and $\omega_1'$, respectively.

**Corollary 5.4** *Under the assumptions of Theorem 5.2 the operators* $T_{\omega_a'}$ *and* $T_{\omega_b}$ *are, for all pairs* $(a, b)$*, unitarily equivalent. The same holds for the operators* $U_{\omega_a'}(\underline{x})$ *and* $U_{\omega_b}(\underline{x})$ *and so the joint spectra* $sp(T_{\omega_a'}, \mathbb{P}_{\omega_a'})$ *and* $sp(T_{\omega_b}, \mathbb{P}_{\omega_b})$ *are identical for all pairs* $(a, b)$ $\square$

Actually, under Theorem 5.2, we can identify $T_{\omega_0'} = T_{\omega_2}$, $T_{\omega_3'} = T_{\omega_1}$, etc. This last corollary describes in which sense duality transformations are a symmetry of the quantum spin system. The corollary says in particular that the particle content of the sectors described by all $\omega_a$ and $\omega_a'$ is the same. A generalization of this corollary to the charged sectors constructed below will also be found.

# 6 The Construction of Electrically and of Magnetically Charged States.

We start defining some operators which will naturally emerge in the discussion presented below and show some useful relations among them. Defining $X^{(n)}(\underline{y})^{1/2} := \beta^*(Q_H(\underline{y})^n)Q_H(\underline{y})^{*n} \in \mathfrak{A}_0$, one has

$$X^{(n)}(\underline{y}) = \exp\left\{\frac{1}{\sqrt{N}}\sum_{j=0}^{N-1}\gamma_h(j)(e^{2\pi i\frac{jn}{N}}-1)P_H(\underline{y})^j\right\} = Y^{(n)}(\underline{y}) + J, \tag{6.1}$$

for

$$Y^{(n)}(\underline{y}) := \exp\left\{\frac{1}{\sqrt{N}}\sum_{j=0}^{N-1}\gamma_h(j)(e^{\frac{2\pi ijn}{N}}-1)(\delta^*U_1(\underline{y}))^j\right\} \qquad \in \mathfrak{B}_0. \tag{6.2}$$

From this we notice that $X^{(n)}(\underline{y})$ and $Y^{(n)}(\underline{y})$ are self-adjoint and therefore one has

$$X^{(n)}(\underline{x})^{1/2} = Q_H(\underline{x})^n\beta(Q_H(\underline{x})^{n\,*}). \tag{6.3}$$

**Definition 6.1** *Let* $\underline{s}$ *be an a-cell on* $\mathbb{Z}^2$*. Then* $\tilde{\underline{s}}$ *denotes* $(-1)^a f_{(-1,-1)}(*\underline{s})$*, where* $f_{(x,y)}$ *denotes translation by* $(x, y) \in \mathbb{Z}^2$ $\square$

Define also, for $\underline{y} \in \mathbb{Z}^2$,

$$Z^{(n)}(\tilde{\underline{y}})' := \Delta^{-1}(Y^{(n)}(\underline{y})) \qquad \in \mathfrak{B}_0, \tag{6.4}$$

that means

$$Z^{(n)}(\tilde{\underline{y}}) = \exp\left\{\frac{1}{\sqrt{N}}\sum_{j=0}^{N-1}\beta_g(j)(e^{\frac{2\pi ijn}{N}}-1)\left(\delta U_3(\tilde{\underline{y}})^*\right)^j\right\}, \tag{6.5}$$

which is also self-adjoint. In (6.4), the symbol $'$ refers to the duality transformation among the couplings, as defined before, mapping $\beta_g \to \gamma_h$, and has been used there for further purposes.

The following operators from $\mathfrak{A}_0$ are also important:

$$
\begin{aligned}
f_0^{(n)}(\underline{x}) &:= \alpha_i^0(Q_H(\underline{x})^n)Q_H(\underline{x})^{*\,n}, &(6.6)\\
f_1^{(n)}(\underline{x}) &:= \alpha_i^1(Q_H(\underline{x})^n)Q_H(\underline{x})^{*\,n}. &(6.7)
\end{aligned}
$$

Define also in $\mathfrak{B}_0$ the operators,

$$
f_2^{(n)}(\tilde{\underline{x}}) := \Delta^{-1}(f_1^{(n)}(\underline{x}) + J) \quad \text{and} \quad f_3^{(n)}(\tilde{\underline{x}}) := \Delta^{-1}(f_0^{(n)}(\underline{x}) + J), \tag{6.8}
$$

for $\underline{x} \in \mathbb{Z}^2$. Clearly all $f_a^{(n)}$ are invertible. We will frequently assume $f_0$ and $f_2$ as elements of $\mathfrak{B}_0$. As such one can establish that:

$$
\begin{aligned}
f_0^{(n)}(\underline{y}) &= \beta^{*\,-1}(Y^{(n)}(\underline{y})^{-1}), &(6.9)\\
f_2^{(n)}(\tilde{\underline{y}})' &= \alpha_i\left(Z^{(n)}(\tilde{\underline{y}})^{-1/2}\right) Z^{(n)}(\tilde{\underline{y}})^{-1/2}, &(6.10)
\end{aligned}
$$

as well as the identities

$$
\begin{aligned}
\left(f_0^{(n)}(\underline{x})\right)^* &= \alpha_{-i}(f_0^{(n)}(\underline{x})),\\
\left(f_2^{(n)}(\tilde{\underline{x}})\right)^* &= \alpha_{-i}(f_2^{(n)}(\tilde{\underline{x}})). &(6.11)
\end{aligned}
$$

For all $\underline{x}$, $\underline{y}$, $n$ and $m$, one can see that the operator $Z^{(n)}(\tilde{\underline{x}})$ commutes with $Y^{(m)}(\underline{y})$, with $\beta^{*\,-1}(Y^{(m)}(\underline{y}))$ and with $\beta^{-1}(Y^{(m)}(\underline{y}))$. This in particular implies that $f_0^{(n)}(\underline{y})$ and $f_2^{(m)}(\tilde{\underline{x}})'$ also commute.

It is interesting to compute the classical functions of some of these operators. One has

$$
\begin{aligned}
\left[Z^{(m)}(\tilde{\underline{0}})\right]^{cl} &= \frac{g(du(p_0) - m)}{g(du(p_0))}, &(6.12)\\
\left[f_0^{(n)}(\underline{0})^{-1}\right]^{cl} &= \exp\left(-\frac{2\pi i n u(b_0)}{N}\right), &(6.13)
\end{aligned}
$$

in the unitary gauge, where $p_0$ is the plaquette $\tilde{\underline{0}}$ placed at the Euclidean time hyperplane zero and $b_0$ is the bond spanned by $(\underline{0},\, 0)$ and $(\underline{0},\, 1)$ oriented in this direction. One can say that $Z^{(m)}(\tilde{\underline{0}})$ creates a $m$-frustrated plaquette at $p_0$, or a magnetic vortex with charge $m$ and that $f_0^{(n)}(\underline{0})^{-1}$ creates a frustrated bond at the vertical bond $b$ with charge $n$. All the operators above appear naturally in our construction.

Now we go over to the construction of charged states. Following [1] an electrically charged state on $\mathfrak{F}_0$ with a $\mathbb{Z}_N$-charge $n$ can be produced as the limit of the following sequence of "dipole" states:

$$
\omega_r^{E,\,(n)}(A) := \frac{\omega_0\left(F_0^{(n)}(r)^* A F_0^{(n)}(r)\right)}{\omega_0\left(F_0^{(n)}(r)^* F_0^{(n)}(r)\right)}, \qquad A \in \mathfrak{F}_0 \tag{6.14}
$$

for $r \in \mathbb{N}$, with $F_0^{(n)}(r) := F_0^{(n)}(r,\, r)$, where $F_0^{(n)}(a,\, b)$, $a,\, b \in \mathbb{N}$, is defined as

$$
F_0^{(n)}(a,\, b) := Q_H(\underline{0})^n Q_H(\underline{x}_a)^{n\,*} \alpha_{ib}(Q_G(\underline{L}_a)^n) \in \mathfrak{A}_0, \tag{6.15}
$$

where $\underline{x}_a$ has coordinates $(2a, 0, \ldots, 0)$, $\underline{L}_a$ is a finite set of bonds with $\partial \underline{L}_a = \{\underline{0},\, \underline{x}_a\}$ and $Q_G(\underline{L}_a) = \prod_{\underline{b} \in \underline{L}_a} Q_G(\underline{b})$. The number $a$ measures the distance between the charges of the dipole and $b$ the Euclidean time evolution applied to the Mandelstam string connecting both charges. As an element of $\mathfrak{B}_0$ we write

$$F_0^{(n)}(r) = \left\{ \prod_{a=0}^{r-1} \alpha_{ai} \left( f_0^{(n)}(\underline{0})^{-1} f_0^{(N-n)}(\underline{x}_r)^{-1} \right) \right\} \alpha_{ri}(U_3(\underline{L}_r)^{*n}), \tag{6.16}$$

for $U_3(\underline{L}_r) := \prod_{\underline{b}\in\underline{L}_r} U_3(\underline{b})$.

For all elements of $\mathfrak{F}_0$ the states $\omega_r^{E,(n)}$ converge in the considered region of the phase diagram to a state which we denote by $\omega^{E\,(n)}$. We omit the proof since it is analogous to the case of [1]. The interpretation of $\omega^{E\,(n)}$ as a charged state is confirmed by the following. Let $\underline{V} \in \mathbb{Z}^d$ be a finite set of lattice sites, say, a cube centered at $\underline{0}$ and let $\Phi^E(\underline{V}) := \prod_{\underline{x}\in\underline{V}}[\delta^* P_G](\underline{x}) = \prod_{\underline{b}\in\partial^*\underline{V}} P_G(\underline{b})$ be the operator measuring the $\mathbb{Z}_N$-electric flux through $\partial\underline{V}$, where $\partial^*\underline{V}$ is the set of all oriented bonds $\underline{b}$ so that $\partial\underline{b}\cap\underline{V}$ consists of only one element. Then

$$\lim_{\underline{V}\uparrow\mathbb{Z}^d} \lim_{r\to\infty} \frac{\omega_r^{E\,(n)}(\Phi^E(\underline{V}))}{\omega_0(\Phi^E(\underline{V}))} = e^{-\frac{2\pi i}{N}n}. \tag{6.17}$$

Using our polymer expansion the proof is again essentially analogous to the corresponding one in [1]. We present this proof together with the proof of Theorem 6.2 (below) in Appendix C.

Since the $F_0^{(n)}(r)$ are gauge-invariant one has $\omega^{E\,(n)}(J) = 0$ and hence $\omega^{E\,(n)}(A+J) := \omega^{E\,(n)}(A)$, $A \in \mathfrak{A}_0$, defines $\omega^{E\,(n)}$ on $\mathfrak{B}_0$.

Important for the physical interpretation of these dipole states is the fact that their energy remains bounded for increasing values of $r$. Precisely one has that for all $m \in \mathbb{N}$,

$$\omega_0 \left( F_0^{(n)}(r)^* \alpha_{-im} \left( F_0^{(n)}(r) \right) \right) / \omega_0 \left( F_0^{(n)}(r)^* F_0^{(n)}(r) \right) < c_m, \tag{6.18}$$

where $c_n$ is a positive constant independent of $r$. The proof is found in [1]. The "perimeter law" of the Wilson loops, needed for that proof, also holds in our model (see [13]).

Before we introduce the magnetically charged states we need some results on states with *external* electric charge which can be constructed from $\omega^{E,(n)}$. Defining the $*$-automorphism $\rho_{\underline{0}}(A) := Q_H(\underline{0})AQ_H(\underline{0})^*$, $A \in \mathfrak{F}$, we can define a state on $\mathfrak{F}$ with an external electric charge $n$ located at $\underline{0}$ through $\omega_{ext}^{E\,(n)} := \omega^{E\,(n)} \circ \rho_{\underline{0}}^n$. Note that, for any $\underline{x}$, $\underline{y}$, $\rho_{\underline{x}}^n \circ \mathrm{ad}_{\mathcal{Q}(\underline{y})^n} = \mathrm{ad}_{\mathcal{Q}(\underline{y})^n} \circ \rho_{\underline{x}}^n$ and for this reason $\omega_{ext}^{E\,(n)}$ is gauge invariant.

Let us now define the following states on $\mathfrak{F}_0$ (to keep the notation as simple as possible we will often omit the reference to the charge $n$ and to the point $\underline{0}$):

$$\lambda_0 := \omega_{ext}^{E\,(n)}, \tag{6.19}$$
$$\lambda_1 := \lambda_0 \circ \beta^{*\,-1}. \tag{6.20}$$

Note that the $\lambda_1$ is indeed a state for the reasons explained in the remarks after the statement of Corollary 5.1. We have the following important Theorem.

**Theorem 6.1** *For the states $\lambda_0$ and $\lambda_1$ above the two following statements hold:*

*I) For $a = 0$ and $a = 1$, $\lambda_a$ is a ground state with respect to the dynamics $\alpha_i^a$ and the observable algebra $\mathfrak{A}_0$, i.e.:*

$$0 \le \lambda_0(A^*\alpha_i(A)) \le \lambda_0(A^*A), \qquad \forall A \in \mathfrak{A}_0, \tag{6.21}$$
$$0 \le \lambda_1(A^*\alpha_i^1(A)) \le \lambda_1(A^*A), \qquad \forall A \in \mathfrak{A}_0. \tag{6.22}$$

*II) For the field algebra $\mathfrak{F}_0$ we have the following generalization of the inequalities above: there exists a finite constant $K_e \geq 1$ so that*

$$0 \leq \lambda_0(F^*\alpha_i(F)) \leq K_e \lambda_0(F^*F), \qquad \forall F \in \mathfrak{F}_0, \tag{6.23}$$
$$0 \leq \lambda_1(F^*\alpha_i^1(F)) \leq K_e \lambda_1(F^*F), \qquad \forall F \in \mathfrak{F}_0 \qquad \Box \tag{6.24}$$

**Proof.** We will prove the statements only for $\lambda_0$. Then for $\lambda_1$ they will follow from the general Theorem 5.1.

*Proof of I.* By the definition one has, for all $A \in \mathfrak{A}_0$,

$$\begin{aligned}\lambda_0(A^*\alpha_i(A)) &= \lim_{r\to\infty} \frac{\omega_0\left((\alpha_{ir}(Q_G(\underline{L}_r)^n))^* A^* \alpha_i(A)(\alpha_{ir}(Q_G(\underline{L}_r)^n))\right)}{\omega_0\left((\alpha_{ir}(Q_G(\underline{L}_r)^n))^* \alpha_{ir}(Q_G(\underline{L}_r)^n)\right)} = \\ &= \lim_{r\to\infty} \frac{\omega_0\left((\alpha_{ir}(Q_G(\underline{L}_r)^n))^* A^* \alpha_i\left(A(\alpha_{ir}(Q_G(\underline{L}_r)^n))\right)\right)}{\omega_0\left((\alpha_{ir}(Q_G(\underline{L}_r)^n))^* \alpha_{i(r+1)}(Q_G(\underline{L}_r)^n)\right)} \geq 0,\end{aligned} \tag{6.25}$$

from the fact that $\omega_0$ is a ground state for $\mathfrak{F}_0$. The second equality is crucial and follows from the representation of the states in terms of classical expectations and from the cluster expansions using the fact that, for a local observable $A$, the classical function $[A]^{cl}$ is also local in the unitary gauge, i.e., has finite support. This last fact is not true in general for elements of $\mathfrak{F}_0$ and for this reason we have in that case only bounds like II (see below).

In order to complete the proof we need to show, according to Lemma 2.1, that $\lambda_0$ has the cluster property for $\alpha_i$ and $\mathfrak{A}_0$. This again follows, using (3.24), from the representation of the state $\lambda_0$ as

$$\lambda_0(A\alpha_{ik}(B)) = \lim_{r\to\infty} \frac{\langle B(\alpha_r^n, 0)[A]^{cl}[B]^{cl}(k)\rangle}{\langle B(\alpha_r^n, 0)\rangle} \tag{6.26}$$

($B(\alpha_r^n, 0)$ is defined after (6.47)-(6.48) and $[B]^{cl}(k)$ is the function $[B]^{cl}$ translated by $k \in \mathbb{Z}$ in time direction), and from the cluster expansions, using again the fact that, for a local observable $A$, the classical function $[A]^{cl}$ is also local in the unitary gauge, i.e., has finite support.

*Proof of II.* Call $G_{\underline{V}}$ the group of all gauge-transformations inside of a finite volume $\underline{V} \in \mathbb{Z}^2$; for $\lambda \in G_{\underline{V}}$, call $g_\lambda := \mathrm{ad}_{\mathcal{Q}(\lambda)}$ and $E_G := \lim_{\underline{V}\uparrow\mathbb{Z}^2} |G_{\underline{V}}|^{-1} \sum_{\lambda\in G_{\underline{V}}} g_\lambda$, the projector of $\mathfrak{F}_0$ into $\mathfrak{A}_0$. Since $\lambda_0$ is gauge invariant one has $\lambda_0 \circ E_G(F) = \lambda_0(F)$, $\forall F \in \mathfrak{F}_0$. Hence, using the same trick as in (6.25) we have, for $F \in \mathfrak{F}_0$,

$$\lambda_0(F^*\alpha_i(F)) = \lim_{r\to\infty} \frac{\omega_0\left((\alpha_{ir}(Q_G(\underline{L}_r)^n))^*\, A_F\, \alpha_{i(r+1)}(Q_G(\underline{L}_r)^n)\right)}{W(r,\, r+1)} \tag{6.27}$$

where $A_F := E_G(F^*\alpha_i(F)) \in \mathfrak{A}_0$ and, to simplify the notation, we used, for $a, b \in \mathbb{N}$,

$$W(a,\, b) := \omega_0\left((\alpha_{ia}(Q_G(\underline{L}_r)^n))^* (\alpha_{ib}(Q_G(\underline{L}_r)^n))\right). \tag{6.28}$$

Now we write this, using the gauge invariance of $\omega_0$, as

$$\lim_{\underline{V}\uparrow\mathbb{Z}^2} \frac{1}{|G_{\underline{V}}|} \sum_{\lambda\in G_{\underline{V}}} \lim_{r\to\infty} \frac{\omega_0\left(\left(\alpha_{ir}\left(g_\lambda(Q_G(\underline{L}_r)^n)\right)\right)^* F^*\alpha_i(F) \left(\alpha_{i(r+1)}\left(g_\lambda(Q_G(\underline{L}_r)^n)\right)\right)\right)}{W(r,\, r+1)} =$$

$$\lim_{r\to\infty} \frac{\omega_0\left( (\alpha_{ir}(Q_G(\underline{L}_r)^n))^* F^* \alpha_i(F) \left(\alpha_{i(r+1)}(Q_G(\underline{L}_r)^n)\right)\right)}{W(r,\, r+1)}, \tag{6.29}$$

because, in general, $g_\lambda(Q_G(\underline{L}_r)^n) = \theta Q_G(\underline{L}_r)^n$, for some $\theta \in \mathbb{C}$ with $|\theta| = 1$. From, this and from the fact that $\omega_0$ is a ground state for $\alpha_i$ and $\mathfrak{F}_0$ it follows that

$$0 \leq \lambda_0(F^*\alpha_i(F)) \leq K_{\underline{L}}\lambda_0(F^*F), \tag{6.30}$$

with

$$K_{\underline{L}} := \lim_{r\to\infty} \frac{W(r,\, r)}{W(r,\, r+1)} = \lim_{r\to\infty} \frac{\|\pi_{\omega_0}(\alpha_{ir}(Q_G(\underline{L}_r)^n))\|}{\|T_{\omega_0}^{1/2}\pi_{\omega_0}(\alpha_{ir}(Q_G(\underline{L}_r)^n))\|}. \tag{6.31}$$

Note that by the last equality $K_{\underline{J}} \geq 1$. This constant $K_{\underline{L}}$ already appears in [1]. The existence of the limit in the definition of $K_{\underline{L}}$ can be seen with the cluster expansions. We do not enter into details. ($K_{\underline{L}}$ also depends on the path $\underline{L}_r$ connecting $\underline{0}$ and $\underline{x}_r$). The constant $K_e$ of the Theorem is the infimum over all constants satisfying $\lambda_0(A^*\alpha_i(A)) \leq K\omega(A^*A)$ for all $A \in \mathfrak{F}_0$. Clearly $K_e \geq 1$ (take the case $A = \mathbb{1}$) and in particular we have seen that $K_e$ is finite ■

Now we want to define the magnetically charged states using the electrically charged ones and duality transformations. Consider now the following four states:

$$\mu_0 := \lambda_0 \circ \rho_{\underline{0}}^{-n} = \omega^{E\,(n)}, \tag{6.32}$$

$$\mu_1 := \lambda_1 \circ \rho_{\underline{0}}^{-n} = \omega^{E\,(n)} \circ \beta^{*\,-1} \circ \mathrm{ad}_{X^{(n)}(\underline{0})^{1/2}}\,, \tag{6.33}$$

$$\mu_2 := \mu_1 \circ \Delta, \tag{6.34}$$

$$\mu_3 := \mu_0 \circ \Delta, \tag{6.35}$$

where $\mu_0$ and $\mu_1$ are states on $\mathfrak{F}_0$ and $\mu_2$ and $\mu_3$ are states on $\mathfrak{B}_0$, which are well defined, since $\mu_0(J) = \mu_1(J) = 0$. Above we used the identity $\rho_{\underline{y}}^n \circ \beta^{*\,-1} \circ \rho_{\underline{y}}^{-n} = \beta^{*\,-1} \circ \mathrm{ad}_{X^{(n)}(\underline{y})^{1/2}}$. The automorphism $\beta^{*\,-1} \circ \mathrm{ad}_{X^{(n)}(\underline{y})^{1/2}}$ is naturally defined on $\mathfrak{B}_0$ since $X^{(n)}(\underline{y})$ is gauge-invariant.

The states $\mu_0$ and $\mu_1$ are electrically charged, $\mu_2$ and $\mu_3$ are magnetically charged. The precise meaning of this claim is explained in the next Theorem. Notice that by the magnetic states are the duals of electric ones, what makes the definitions (6.34) and (6.35) very natural. The definition (6.33) is also natural since $\lambda_1$ is, as already discussed, a state with an external electric charge.

In order to understand in which sense the states above are charged we need some definitions. For $\underline{V} \subset \mathbb{Z}^2$, finite, define the charge measuring operators

$$\Phi^E(\underline{V}) := \prod_{\underline{x}\in\underline{V}} \delta^* P_G(\underline{x}) \qquad \text{and} \qquad \Phi^M(\tilde{\underline{V}}) := \prod_{\underline{p}\in\tilde{\underline{V}}} \delta U_3(\underline{p}), \tag{6.36}$$

both being unitary and related through $\Phi^M(\tilde{\underline{V}}) := \Delta^{-1}(\Phi^E(\underline{V}))$. They are interpreted as operators measuring the $\mathbb{Z}_N$-electric flux through $\delta^*\underline{V}$, respectively the $\mathbb{Z}_N$-magnetic flux through $\tilde{\underline{V}}$. We use these operators in the next Theorem to justify why the states above deserve the interpretation of being (electrically or magnetically) charged states.

**Theorem 6.2** *If $\underline{V} \subset \mathbb{Z}^2$ is e.g. a square centered at the origin, one has:*

$$\lim_{\underline{V}\uparrow\mathbb{Z}^2} \frac{\mu_0(\Phi^E(\underline{V}))}{\omega_0(\Phi^E(\underline{V}))} = \lim_{\underline{V}\uparrow\mathbb{Z}^2} \frac{\mu_3(\Phi^M(\tilde{\underline{V}}))}{\omega_3(\Phi^M(\tilde{\underline{V}}))} = e^{\frac{-2\pi i n}{N}}, \tag{6.37}$$

$$\lim_{\underline{V}\uparrow\mathbb{Z}^2} \frac{\mu_2(\Phi^M(\tilde{\underline{V}}))}{\omega_2(\Phi^M(\tilde{\underline{V}}))} = \lim_{\underline{V}\uparrow\mathbb{Z}^2} \frac{\mu_1(\Phi^E(\underline{V}))}{\omega_1(\Phi^E(\underline{V}))} = e^{\frac{-2\pi i n}{N}} \tag{6.38}$$

*and*

$$\lim_{\underline{V}\uparrow\mathbb{Z}^2}\frac{\mu_0(\Phi^M(\tilde{\underline{V}}))}{\omega_0(\Phi^M(\tilde{\underline{V}}))} = \lim_{\underline{V}\uparrow\mathbb{Z}^2}\frac{\mu_3(\Phi^E(\underline{V}))}{\omega_3(\Phi^E(\underline{V}))} = 1, \tag{6.39}$$

$$\lim_{\underline{V}\uparrow\mathbb{Z}^2}\frac{\mu_2(\Phi^E(\underline{V}))}{\omega_2(\Phi^E(\underline{V}))} = \lim_{\underline{V}\uparrow\mathbb{Z}^2}\frac{\mu_1(\Phi^M(\tilde{\underline{V}}))}{\omega_1(\Phi^M(\tilde{\underline{V}}))} = 1 \qquad \Box \tag{6.40}$$

The proof of this Theorem is found in Appendix C. An interesting and important point is that it shows that the different states are charged with respect to different ground states of different dynamics. In particular one sees that there is a special interest on the states $\mu_0$ and $\mu_2'$, since they are, respectively, electrically and magnetically charged states with respect to the same state: $\omega_0$.

For this reason we turn back until the end of this section to the previous notation and call $\omega^{E\,(n)} := \mu_0$ and $\omega^{M\,(n)} := \mu_2'$.

**Theorem 6.3** *In analogy with (6.14) and (6.16), one has:*

$$\omega_r^{M,\,(n)}(A) := \frac{\omega_0\left(F_2^{(n)}(r)^* A F_2^{(n)}(r)\right)}{\omega_0\left(F_2^{(n)}(r)^* F_2^{(n)}(r)\right)}, \qquad A \in \mathfrak{F}_0 \tag{6.41}$$

*with* $F_2^{(n)}(r) := F_2^{(n)}(r,\, r)$*, where, for* $a,\, b \in \mathbb{N}$*,*

$$F_2^{(n)}(a,\, b) := \left\{\prod_{j=0}^{b-1}\alpha_{ji}\left((f_2^{(n)}(\tilde{\underline{0}})')^{-1}(f_2^{(N-n)}(\tilde{\underline{x}}_a)')^{-1}\right)\right\}\alpha_{bi}\left(Z^{(n)}(\tilde{\underline{0}})^{1/2}Z^{(N-n)}(\tilde{\underline{x}}_a)^{1/2}\beta^{*\,-1}(U_1(\tilde{\underline{L}}_a)^n)\right). \tag{6.42}$$

The operator $F_2^{(n)}(a,\, b)$ creates a "dipole" of magnetic vortices separated by $a$ and connected by a magnetic vortex string translated by $b$ in Euclidean time direction.

**Proof.** To simplify the notation we frequently drop the reference to the ideal $J$. Let us look more closely at state $\mu_2$. By definition one has, for $A \in \mathfrak{B}_0$,

$$\mu_2(A) = \lim_{r\to\infty}\frac{\omega_0\circ\beta^{*\,-1}\circ\Delta\left[\Delta^{-1}\circ\beta^*(F_0^{(n)}(r)^*)\,\mathrm{ad}_{Z^{(n)}(\tilde{\underline{0}})^{1/2}{}'}(A)\,\Delta^{-1}\circ\beta^*(F_0^{(n)}(r))\right]}{\omega_0\circ\beta^{*\,-1}\circ\Delta\left(\Delta^{-1}\circ\beta^*(F_0^{(n)}(r)^*)\,\Delta^{-1}\circ\beta^*(F_0^{(n)}(r))\right)}. \tag{6.43}$$

Using the representation (6.16), (6.9), (5.14) and the fact that $\Delta^{-1}\circ\beta^*\circ\alpha_{ai} = \alpha'_{(a-1)i}\circ(\beta^{*\,-1})'\circ\Delta^{-1}$ we get $\Delta^{-1}\circ\beta^*(F_0^{(n)}(r)) = H'_{r,\,r-1}$, where, for general $p,\, q \in \mathbb{N}$,

$$H_{p,\,q} := Z^{(n)}(\tilde{\underline{0}})^{1/2}Z^{(N-n)}(\tilde{\underline{x}}_p)^{1/2}F_2^{(n)}(p,\, q). \tag{6.44}$$

Since

$$\Delta^{-1}\circ\beta^*(F_0^{(n)}(r)^*) = \left(\alpha_i'\circ\Delta^{-1}\circ\beta^*(F_0^{(n)}(r))\right)^* = (\alpha_i'(H'_{r,\,r-1}))^* = \left(Z^{(n)}(\tilde{\underline{0}})^{-1}Z^{(N-n)}(\tilde{\underline{x}}_r)^{-1}H_{r,\,r}\right)^{\prime\,*} \tag{6.45}$$

it is possible write the state $\mu_2'$ in the form

$$\mu_2'(A) = \lim_{r\to\infty}\frac{\omega_0\left(F_2^{(n)}(r)^*\,A\,F_2^{(n)}(r,\, r-1)\right)}{\omega_0\left(F_2^{(n)}(r)^*\,F_2^{(n)}(r,\, r-1)\right)} = \lim_{r\to\infty}\frac{\omega_0\left(F_2^{(n)}(r)^*\,A\,F_2^{(n)}(r)\right)}{\omega_0\left(F_2^{(n)}(r)^*\,F_2^{(n)}(r)\right)}. \tag{6.46}$$

The last equality in (6.46) comes from the polymer and cluster expansions ■

Considering $F_2^{(n)}(r)$ as an element of $\mathfrak{A}_0$ the last expression above also defines an extension of $\mu_2$ on $\mathfrak{F}_0$, provided the limit exists, what can also be proven using the polymer expansion.

One can express $\omega^{E\,(n)}(A)$ and $\omega^{M\,(n)}(A)$ in terms of classical expectation values. Using Definition 4.1 one has

$$\omega^{E\,(n)}(A) = \lim_{r\to\infty} \frac{\langle B(\alpha_r^n,\,0)[A]^E\rangle}{\langle B(\alpha_r^n,\,0)\rangle}, \tag{6.47}$$

$$\omega^{M\,(n)}(A) = \lim_{r\to\infty} \frac{\langle B(0,\,-\beta_r^n)[A]^M\rangle}{\langle B(0,\,-\beta_r^n)\rangle}, \tag{6.48}$$

where $\alpha_r^n$ and $\beta_r^n$ is a 1-form, respectively a 2-form, with $d^*\alpha_r^n = 0$ and $d\beta_r^n = 0$, as indicated in the Figure 2. In this figure we indicate the support of the forms. The value that the form $\beta_r^n$ (respectively $\alpha_r^n$) assumes at a plaquette $p$ of its support (respectively, at a bond $b$ of its support) is $n$ for $p$ (respectively, $b$) oriented in the sense indicated by the curly arrow.

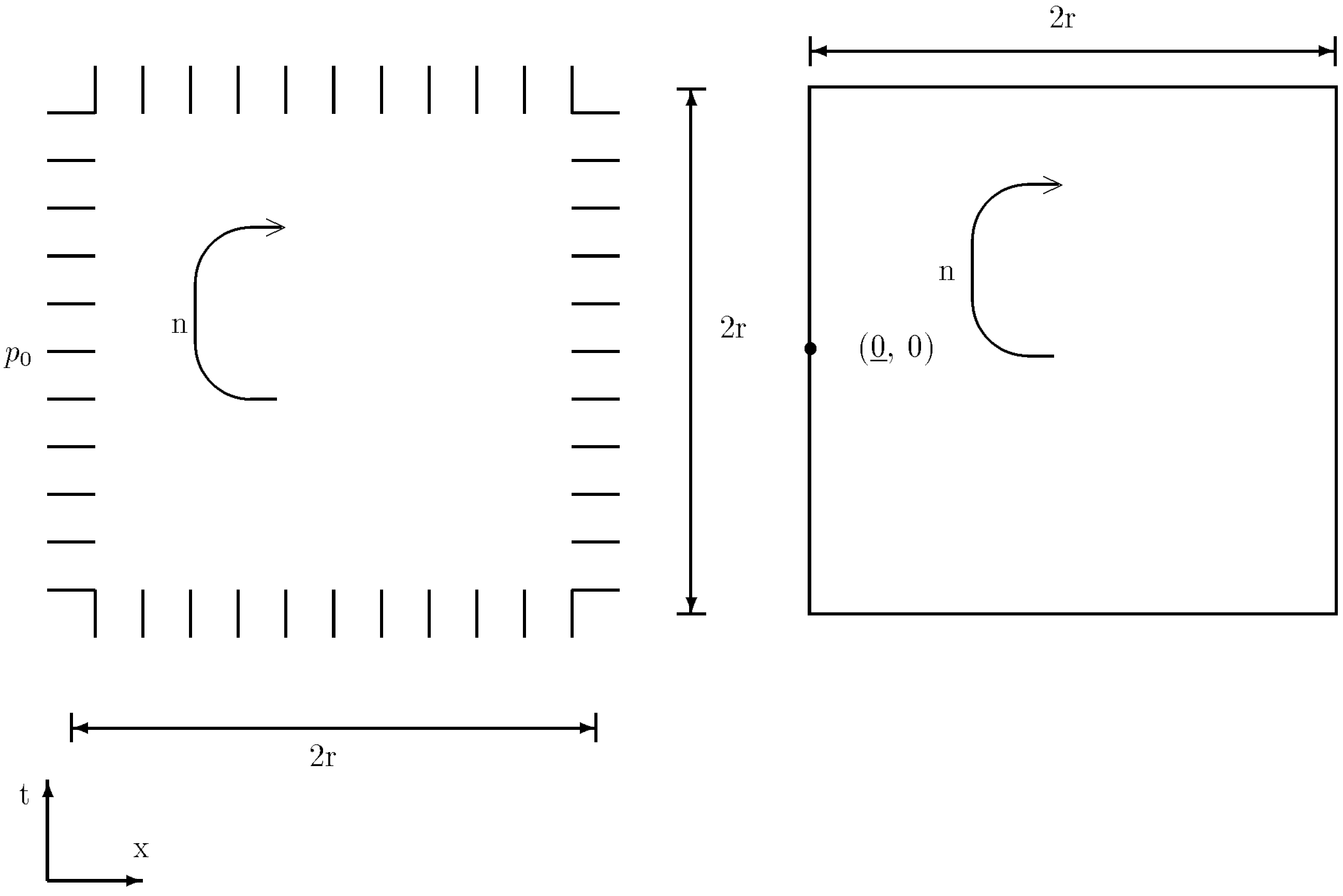

Figure 2: *The supports of the forms $\beta_r^n$ (left) and $\alpha_r^n$ (right). $p_0$ is the plaquette $\tilde{\underline{0}}$ at Euclidean time 0. The horizontal axis indicates the x-direction and the vertical the Euclidean time-direction.*

The functions $[A]^E$ and $[A]^M$ are given (in the unitary gauge) for $r$ large enough by

$$[A]^E := \exp\left(\frac{2\pi i n(u(b_1)+u(b_2))}{N}\right)\left[\left(f_0^{(n)}(\underline{0})^{-1}\right)^* A f_0^{(n)}(\underline{0})^{-1}\right]^{cl}, \tag{6.49}$$

$$[A]^M := \left(\frac{g(du(p_0)+n)}{g(du(p_0))}\right)^{-1}\left[Z^{(n)}(\tilde{\underline{0}})^{1/2} A Z^{(n)}(\tilde{\underline{0}})^{1/2}\right]^{cl}, \tag{6.50}$$

where $b_1$ and $b_2$ are the bonds spanned by $(\underline{0},\,-1)$ and $(\underline{0},\,0)$ and by $(\underline{0},\,0)$ and by $(\underline{0},\,1)$, respectively, $p_0$ is the plaquette $\tilde{\underline{0}}$ at Euclidean time 0. Actually a straightforward computation shows that, for $A \in \mathfrak{A}_0$, $[A]^E = [Q_H^{-n}(\underline{0}) A Q_H^n(\underline{0})]^{cl}$ and $[A]^M = [A]^{cl}$.

**Remark.** We warn the reader that, due to the appearance of other factors in the expectation values in the right hand side of (6.47) and (6.48), relation (3.24) cannot be used for the functions $[A]^E$ and $[A]^M$ □

Let us now study some properties of the states $\omega^E$ and $\omega^M$. We start with a simple Lemma:

**Lemma 6.1** *i) For any $p \in \mathbb{N}$ and for both $a = 0$ and $a = 2$ we have*

$$\lim_{r\to\infty} \frac{\omega_0(F_a(r)^* A F_a(r))}{\omega_0(F_a(r)^* F_a(r))} = \lim_{r\to\infty} \frac{\omega_0(F_a(r)^* A F_a(r,\, r+p))}{\omega_0(F_a(r)^* F_a(r,\, r+p))}, \qquad A \in \mathfrak{F}_0. \tag{6.51}$$

*ii) One has $\omega_0(F_a(r)^* F_a(r,\, r+2)) = \omega_0(F_a(r,\, r+1)^* F_a(r,\, r+1))$ and $\omega_0(F_a(r)^* F_a(r,\, r+p)) \geq 0$ for any $p \in \mathbb{N}$* □

**Proof.** The equality in part ii) is evident from the representation as classical expectations and translation invariance. The other claim follows from (4.19) or from Griffiths inequalities. Part i) can be proven with the polymer and cluster expansions, as one can see from the proof of the existence of the limit states. We do not repeat the details ■

Now we establish some important properties of the electrically and magnetically charged states. We will denote by $\omega_{\underline{x}}^{E\,(n)} := \omega^{E\,(n)} \circ \tau_{-\underline{x}}$ and $\omega_{\underline{\tilde{y}}}^{M\,(n)} := \omega^{M\,(n)} \circ \tau_{-\underline{y}}$, the electrically and magnetically charged states with charges centered at $\underline{x}$ and $\underline{\tilde{y}}$.

**Theorem 6.4** *For the states $\omega_{\underline{x}}^{E\,(n)}$ and $\omega_{\underline{\tilde{y}}}^{M\,(n)}$ defined above the following invariance properties and inequalities can be established:*

1. *For the charged states one has $\omega_{\underline{x}}^{E\,(n)} \circ \delta_{\underline{x}}^{E\,(n)} = \omega_{\underline{x}}^{E\,(n)}$ and $\omega_{\underline{\tilde{y}}}^{M\,(n)} \circ \delta_{\underline{\tilde{y}}}^{M\,(n)} = \omega_{\underline{\tilde{y}}}^{M\,(n)}$ where the automorphism $\delta_{\underline{x}}^{E\,(n)}$ and $\delta_{\underline{\tilde{y}}}^{M\,(n)}$ are defined by*

$$\delta_{\underline{x}}^{E\,(n)} := ad_{(f_0^{(n)}(\underline{x}))^{-1}} \circ \alpha_i = \alpha_i \circ ad_{(f_0^{(n)}(\underline{x}))^{*\,-1}}\,, \tag{6.52}$$

$$\delta_{\underline{\tilde{y}}}^{M\,(n)} := ad_{(f_2^{(m)}(\underline{\tilde{y}})')^{-1}} \circ \alpha_i = \alpha_i \circ ad_{(f_2^{(m)}(\underline{\tilde{y}})')^{*\,-1}}\,. \tag{6.53}$$

*Note that $\left(\delta_{\underline{x}}^{E\,(n)}\right)^* = \left(\delta_{\underline{x}}^{E\,(n)}\right)^{-1}$ and $\left(\delta_{\underline{\tilde{y}}}^{M\,(n)}\right)^* = \left(\delta_{\underline{\tilde{y}}}^{M\,(n)}\right)^{-1}$.*

2. *For all $A,\, B \in \mathfrak{B}_0$ the following cluster properties hold in the region of convergence of the polymer and cluster expansions:*

$$\lim_{a\to\infty} \omega_{\underline{x}}^{E\,(n)}\left(A \left(\delta_{\underline{x}}^{E\,(n)}\right)^a (B)\right) = \omega_{\underline{x}}^{E\,(n)}(A)\,\omega_{\underline{x}}^{E\,(n)}(B). \tag{6.54}$$

$$\lim_{a\to\infty} \omega_{\underline{\tilde{y}}}^{M\,(n)}\left(A \left(\delta_{\underline{\tilde{y}}}^{M\,(n)}\right)^a (B)\right) = \omega_{\underline{\tilde{y}}}^{M\,(n)}(A)\,\omega_{\underline{\tilde{y}}}^{M\,(n)}(B). \tag{6.55}$$

3. *For all $A \in \mathfrak{B}_0$ the following inequalities hold:*

$$0 \leq \omega_{\underline{x}}^{E\,(n)}(A^* \delta_{\underline{x}}^{E\,(n)}(A)) \leq \omega_{\underline{x}}^{E\,(n)}(A^* A), \tag{6.56}$$

$$0 \leq \omega_{\underline{\tilde{y}}}^{M\,(n)}(A^* \delta_{\underline{\tilde{y}}}^{M\,(n)}(A)) \leq \omega_{\underline{\tilde{y}}}^{M\,(n)}(A^* A), \tag{6.57}$$

*that means, $\omega_{\underline{x}}^{E\,(n)}$ is a ground state with respect to $\delta_{\underline{x}}^{E\,(n)}$ and $\omega_{\underline{\tilde{y}}}^{M\,(n)}$ is a ground state with respect to $\delta_{\underline{\tilde{y}}}^{M\,(n)}$.*

*4. For all $A \in \mathfrak{B}_0$ we have the following inequalities*

$$0 \leq \omega_{\underline{x}}^{E\,(n)} \left( A^* \alpha_i(A) f_0^{(n)}(\underline{x}) \right), \tag{6.58}$$

$$0 \leq \omega_{\underline{\tilde{y}}}^{M\,(n)} \left( A^* \alpha_i(A) f_2^{(n)}(\underline{\tilde{y}})' \right) \qquad \square \tag{6.59}$$

**Proof of Theorem 6.4.** We present the proof for the magnetic states $\omega^{M\,(n)}$. The generalization to $\omega_{\underline{\tilde{y}}}^{M\,(n)}$ is trivial. The case of the electric states is similar.

**Part 1.** Invariance of the states already follows from the first inequality in (6.57). We give another more direct proof, which follows from the fact that, by (6.51),

$$\omega^{M\,(n)}(A) = \lim_{r \to \infty} \frac{\omega_0(F_2(r)^* A F_2(r,\, r+2))}{\omega_0(F_2(r)^* F_2(r,\, r+2))}, \qquad A \in \mathfrak{B}_0, \tag{6.60}$$

from the fact that

$$F_2(r,\, r+1) = \left( f_2^{(n)}(\underline{\tilde{0}})' f_2^{(M-n)}(\underline{\tilde{x}}_r)' \right)^* \alpha_{-i}(F_2(r,\, r+2)), \tag{6.61}$$

and from the fact that $\omega_0(F_2(r,\, r+1)^* F_2(r,\, r+1)) = \omega_0(F_2(r)^* F_2(r,\, r+2))$.

**Part 2.** Writing $\omega^{M\,(n)}$ as the limit (6.41), using

$$\left( \delta_{\underline{\tilde{0}}}^{M\,(n)} \right)^a = \mathrm{ad}_{(f_2^{(m)}(\underline{\tilde{0}})^{-1})'} \circ \cdots \circ \mathrm{ad}_{\alpha_{i(a-1)}((f_2^{(m)}(\underline{\tilde{0}})^{-1})')} \circ \alpha_{ai}, \tag{6.62}$$

and using (3.24) we get, after some simple manipulations,

$$\omega^{M\,(n)} \left( A \left( \delta_{\underline{\tilde{0}}}^{M\,(n)} \right)^a (B) \right) = \lim_{r \to \infty} \frac{\langle B(0,\, -\beta_r^n)[A]^M(0)[B]^M(a) \rangle}{\langle B(0,\, -\beta_r^n) \rangle}, \tag{6.63}$$

where $[B]^M(a)$, is the classical function $[B]^M$ translated to the $a$-th Euclidean time-slice. The desired cluster property is obtained writing the expectation values above in terms of the polymer and cluster expansions and using the standard techniques. From the same expansions one sees that the clustering is exponential.

**Part 3.** We consider $\omega^{M\,(n)} \left( A^* \delta_{\underline{\tilde{0}}}^{M\,(n)}(A) \right)$ and write the state as in (6.60). A computation then shows that

$$\omega^{M\,(n)} \left( A^* \delta_{\underline{\tilde{0}}}^{M\,(n)}(A) \right) = \lim_{r \to \infty} \frac{\omega_0(L^* \alpha_i(L))}{\omega_0(F_2(r)^* F_2(r,\, r+1))} \geq 0 \tag{6.64}$$

where $L := Z^{(n)}(\underline{\tilde{0}}) A Z^{(N-n)}(\underline{\tilde{x}}_r) F_2(r)$. The second inequality follows from the cluster property and Lemma 2.1.

**Part 4.** Again by (6.51) we can write

$$\omega^{M\,(n)}(B) = \lim_{r \to \infty} \frac{\omega_0(F_2(r)^* B F_2(r,\, r+1))}{\omega_0(F_2(r)^* F_2(r,\, r+1))}, \qquad \forall B \in \mathfrak{B}_0. \tag{6.65}$$

Using (6.42) and taking $B := A^* \alpha_i(A) f_2^{(n)}(\underline{\tilde{0}})'$ the operator $F_2(r)^* A F_2(r,\, r+1)$ can be written as $F^* \alpha_i(F)$, for $F := A Z^{(N-m)}(\underline{\tilde{x}}_r) F_2(r)$ and the result follows from the ground state property of $\omega_0$ ■

We finish this sections with some observations concerning the automorphisms $\delta_{\underline{0}}^{E\,(n)}$ and $\delta_{\tilde{\underline{0}}}^{M\,(m)}$. As automorphisms acting on $\mathfrak{A}_0$ we can write then as the limit $\underline{V} \uparrow \mathbb{Z}^2$ of ad $T_{\underline{V}(n,0)}$ and ad $T_{\underline{V}(0,m)}$ respectively, where we introduced the finite-volume *modified* transfer matrices:

$$T_{\underline{V}}(n,\, 0) \;:=\; Q_H(\underline{0})^n T_{\underline{V}} Q_H(\underline{0})^{-n}, \tag{6.66}$$

$$T_{\underline{V}}(0,\, m) \;:=\; Z^{(m)}(\tilde{\underline{0}})^{1/2} T_{\underline{V}} Z^{(m)}(\tilde{\underline{0}})^{1/2}. \tag{6.67}$$

Recalling relations (2.13)-(2.15) one easily sees that $T_{\underline{V}}(n,\, 0)$ differs from $T_{\underline{V}}$ by the replacement $P_H(\underline{0}) \to e^{\frac{2\pi i n}{N}} P_H(\underline{0})$ and $T_{\underline{V}}(0,\, m)$ by the replacement $[\delta Q_G](\tilde{\underline{0}}) \to e^{\frac{2\pi i m}{N}}[\delta Q_G](\tilde{\underline{0}})$. This means that these modified transfer matrices differ from the usual one by the introduction of a "shift" in a vertical bond starting at $\underline{0}$, respectively, in a horizontal plaquette located at $\tilde{\underline{0}}$. A generalization of this idea will be used in the construction of dyonic states with multiple electric and magnetic charges located at different points.

# 7 The Global Transfer Matrices on the Charged Sectors.

In this section we are interested in defining transfer matrices on the sectors defined by the electrically and magnetically states constructed before and in studying the relations among them. The question of the translation invariance of the global transfer matrices will be discussed in Section 8 below.

Let us first consider the electrically charged states $\lambda_0$, $\lambda_1$, $\mu_0$ and $\mu_1$. Call $(\pi_{\lambda_a},\, \mathcal{H}_{\lambda_a},\, \phi_{\lambda_a})$ the GNS-triple associated to the states $\lambda_a$ and the algebra $\mathfrak{F}_0$. Based on our experience with the vacuum sector we define the operator $K_{0\to 1}$: $\mathcal{H}_{\lambda_0} \to \mathcal{H}_{\lambda_1}$ by

$$K_{0\to 1}\pi_{\lambda_0}(B)\phi_{\lambda_0} := \pi_{\lambda_1}(\beta(B))\phi_{\lambda_1}, \qquad B \in \mathfrak{F}_0, \tag{7.1}$$

with

$$K^*_{0\to 1}\pi_{\lambda_1}(B)\phi_{\lambda_1} := \pi_{\lambda_0}({\beta^*}^{-1}(B))\phi_{\lambda_0}, \qquad B \in \mathfrak{F}_0. \tag{7.2}$$

This operator $K_{0\to 1}$ is analogous to the operator $U_{0\to 1}$ previously defined. Using Theorem 6.1, part II, one easily checks that $\|K_{0\to 1}\| = K_e^{1/2}$.

We then define transfer matrices on $\mathcal{H}_{\lambda_0}$ and $\mathcal{H}_{\lambda_1}$, respectively, by

$$T_{\lambda_0} \;:=\; K^*_{0\to 1}K_{0\to 1}, \tag{7.3}$$

$$T_{\lambda_1} \;:=\; K_{0\to 1}K^*_{0\to 1}, \tag{7.4}$$

with the result that, as expected, one has

$$T_{\lambda_0}\pi_{\lambda_0}(B)\phi_{\lambda_0} \;:=\; \pi_{\lambda_0}(\alpha_i^0(B))\phi_{\lambda_0}, \qquad B \in \mathfrak{F}_0, \tag{7.5}$$

$$T_{\lambda_1}\pi_{\lambda_1}(B)\phi_{\lambda_1} \;:=\; \pi_{\lambda_1}(\alpha_i^1(B))\phi_{\lambda_1}, \qquad B \in \mathfrak{F}_0. \tag{7.6}$$

Analogously to the previously treated case $T_{\lambda_0}$ and $T_{\lambda_1}$ are unitarily equivalent but $\|T_{\lambda_a}\| = K_e$, $a = 0,\, 1$. Next we want to define transfer matrices on the gauge invariant sectors generated by the states $\mu_0$ and $\mu_1$ and the relevant algebra of observables $\mathfrak{B}_0$. Let us first consider the GNS triple $(\pi_{\mu_b},\, \mathcal{H}_{\mu_b},\, \phi_{\mu_b})$ associated to the states $\mu_b$ $(b = 0,\, 1)$ and the larger algebra $\mathfrak{F}_0$.

Define the unitary operators $L_a$: $\mathcal{H}_{\lambda_a} \to \mathcal{H}_{\lambda_a}$, $R_a$: $\mathcal{H}_{\mu_a} \to \mathcal{H}_{\lambda_a}$ and $S_a$: $\mathcal{H}_{\mu_a} \to \mathcal{H}_{\lambda_a}$, $a = 0,\, 1$, by

$$L_a\pi_{\lambda_a}(A)\phi_{\lambda_a} \;:=\; \pi_{\lambda_a}(Q_H(\underline{0})^{-n})\pi_{\lambda_a}(A)\phi_{\lambda_a}, \qquad A \in \mathfrak{F}_0, \tag{7.7}$$

$$R_a\pi_{\mu_a}(A)\phi_{\mu_a} \;:=\; \pi_{\lambda_a}(AQ_H(\underline{0})^n)\phi_{\lambda_a}, \qquad A \in \mathfrak{F}_0, \tag{7.8}$$

$$S_a\pi_{\mu_a}(A)\phi_{\mu_a} \;:=\; \pi_{\lambda_a}(\rho_{\underline{0}}^{-n}(A))\phi_{\lambda_a}, \qquad A \in \mathfrak{F}_0. \tag{7.9}$$

with $S_a = L_a R_a$. Define also, $W_{0\to 1}$: $\mathcal{H}_{\mu_0} \to \mathcal{H}_{\mu_1}$ by

$$
\begin{aligned}
W_{0\to 1} &:= R_1^* K_{0\to 1} R_0, \quad \text{with} && (7.10)\\
W_{0\to 1}^* &:= R_0^* K_{0\to 1}^* R_1. && (7.11)
\end{aligned}
$$

We then define the transfer matrices associated to $\mu_{0,1}$ and $\mathfrak{F}_0$ by

$$
\begin{aligned}
T_{\mu_0} &:= W_{0\to 1}^* W_{0\to 1} = R_0^* T_{\lambda_0} R_0, \quad \text{and} && (7.12)\\
T_{\mu_1} &:= W_{0\to 1} W_{0\to 1}^* = R_1^* T_{\lambda_1} R_1. && (7.13)
\end{aligned}
$$

The operator $R_a$ defines a canonical map between the GNS Hilbert spaces $\mathcal{H}_{\mu_a}$ and $\mathcal{H}_{\lambda_a}$, what makes the definitions above particularly natural. A simple computation shows that

$$
\begin{aligned}
T_{\mu_0}\pi_{\mu_0}(A)\phi_{\mu_0} &= \pi_{\mu_0}\left(\alpha_i^0(A) f_0^{(n)}(\underline{0})\right)\phi_{\mu_0}, \qquad A \in \mathfrak{F}_0, && (7.14)\\
T_{\mu_1}\pi_{\mu_1}(A)\phi_{\mu_1} &= \pi_{\mu_1}\left(\alpha_i^1(A) f_1^{(n)}(\underline{0})\right)\phi_{\mu_1}, \qquad A \in \mathfrak{F}_0. && (7.15)
\end{aligned}
$$

The inverses of the transfer matrices are given by

$$
T_{\mu_a}^{-1}\pi_{\mu_a}(A)\phi_{\mu_a} = \pi_{\mu_a}\left(\alpha_{-i}^a\left(A f_a^{(n)}(\underline{0})^{-1}\right)\right)\phi_{\mu_a}, \qquad A \in \mathfrak{F}_0, \tag{7.16}
$$

$a = 0, 1$, and are densely defined.

Inspired in the previous construction consider also $X_{0\to 1}$: $\mathcal{H}_{\mu_0} \to \mathcal{H}_{\mu_1}$ by

$$
\begin{aligned}
X_{0\to 1} &:= S_1^* K_{0\to 1} S_0, \quad \text{with} && (7.17)\\
X_{0\to 1}^* &:= S_0^* K_{0\to 1}^* S_1. && (7.18)
\end{aligned}
$$

We then define the global <u>*modified*</u> transfer matrices associated to $\mu_{0,1}$ and $\mathfrak{F}_0$ by

$$
\begin{aligned}
V_{\mu_0} &:= X_{0\to 1}^* X_{0\to 1} = S_0^* T_{\lambda_0} S_0, \quad \text{and} && (7.19)\\
V_{\mu_1} &:= X_{0\to 1} X_{0\to 1}^* = S_1^* T_{\lambda_1} S_1. && (7.20)
\end{aligned}
$$

As one easily checks, one has for $a = 0, 1$,

$$
V_{\mu_a}\pi_{\mu_a}(A)\phi_{\mu_a} := \pi_{\mu_a}\left(\delta_{\underline{0}}^{E(n)}(A)\right)\phi_{\mu_a}, \qquad A \in \mathfrak{F}_0. \tag{7.21}
$$

Clearly

$$
V_{\mu_a} = R_a^* T_{\mu_a} R_a, \tag{7.22}
$$

and

$$
V_{\mu_a} = \pi_{\mu_a}\left(f_0^{(n)}(\underline{0})^{-1}\right) T_{\mu_a}. \tag{7.23}
$$

Observe that $V_{\mu_0} \sim T_{\mu_0} \sim T_{\lambda_0} \sim T_{\lambda_1} \sim T_{\mu_1} \sim V_{\mu_1}$.

As in the vacuum sector case the definitions (7.12)-(7.13) imply $\mathcal{W}_{0\to 1} T_{\mu_0} = T_{\mu_1}\mathcal{W}_{0\to 1}$, where $\mathcal{W}_{0\to 1}$ is the unitary operator defined through the polar decomposition of $W_{0\to 1}$: $W_{0\to 1} = \mathcal{W}_{0\to 1} T_{\mu_0}^{1/2}$. Note that by (6.3)

$$
W_{0\to 1}\pi_{\mu_0}(A)\phi_{\mu_0} = \pi_{\mu_1}(\beta(A) X^{(n)}(\underline{0})^{-1/2})\phi_{\mu_1}, \qquad A \in \mathfrak{F}_0. \tag{7.24}
$$

Hence $W_{0\to 1}\mathcal{H}^g_{\mu_0} \subset \mathcal{H}^g_{\mu_1}$, where $\mathcal{H}^g_{\mu_a} := \overline{\{\pi_{\mu_a}(A)\phi_{\mu_a},\ A \in \mathfrak{A}_0\}}$ is the subspace without external charges. This holds also for $\mathcal{W}_{0\to 1}$, since $T_{\mu_a}$ keeps $\mathcal{H}^g_{\mu_a}$ invariant for both $a = 0, 1$. Defining $T^g_{\mu_a} := T_{\mu_a} \restriction \mathcal{H}^g_{\mu_a}$ we conclude that

$$\mathcal{W}_{0\to 1}T^g_{\mu_0} = T^g_{\mu_1}\mathcal{W}_{0\to 1}. \tag{7.25}$$

Since $\mu_0(J) = \mu_1(J) = 0$ there is, for both $b = 0, 1$, a canonical identification between the GNS-triple associated to $\mathfrak{A}_0$ and the GNS-triple associated to $\mathfrak{B}_0$. For this last we have

$$W_{0\to 1}\pi_{\mu_0}(A)\phi_{\mu_0} = \pi_{\mu_1}\left(\beta(A)Y^{(n)}(\underline{0})^{-1/2}\right)\phi_{\mu_1}, \qquad A \in \mathfrak{B}_0, \tag{7.26}$$

$$T^g_{\mu_0}\pi_{\mu_0}(A)\phi_{\mu_0} = \pi_{\mu_0}\left(\alpha^0_i(A)f^{(n)}_0(\underline{0})\right)\phi_{\mu_0}, \qquad A \in \mathfrak{B}_0, \tag{7.27}$$

$$T^g_{\mu_1}\pi_{\mu_1}(A)\phi_{\mu_1} = \pi_{\mu_1}\left(\alpha^1_i(A)f^{(n)}_1(\underline{0})\right)\phi_{\mu_1}, \qquad A \in \mathfrak{B}_0. \tag{7.28}$$

Now we treat the magnetic states. Call $(\pi_{\mu_b}, \mathcal{H}^g_{\mu_b}, \phi_{\mu_b})$ the GNS-triple associated to the states $\mu_b$, $b = 2, 3$ and the algebra $\mathfrak{B}_0$. There are natural <u>unitary</u> maps $\mathcal{H}^g_{\mu_0} \to \mathcal{H}^g_{\mu_3}$ and $\mathcal{H}^g_{\mu_1} \to \mathcal{H}^g_{\mu_2}$ given by

$$W_{0\to 3}\pi_{\mu_0}(A)\phi_{\mu_0} := \pi_{\mu_3}(\Delta^{-1}(A))\phi_{\mu_3}, \quad A \in \mathfrak{B}_0 \tag{7.29}$$

$$W_{1\to 2}\pi_{\mu_1}(A)\phi_{\mu_1} := \pi_{\mu_2}(\Delta^{-1}(A))\phi_{\mu_2}, \quad A \in \mathfrak{B}_0. \tag{7.30}$$

This naturally invites to the following definitions for the transfer matrices on the magnetic sectors associated to $\mu_2$ and $\mu_3$:

$$T^g_{\mu_2} := W_{1\to 2}T^g_{\mu_1}W^{-1}_{1\to 2} \quad \text{and} \quad T^g_{\mu_3} := W_{0\to 3}T^g_{\mu_0}W^{-1}_{0\to 3}, \tag{7.31}$$

which leads to

$$T^g_{\mu_2}\pi_{\mu_2}(A)\phi_{\mu_2} = \pi_{\mu_2}\left(\alpha^2_i(A)f^{(n)}_2(\underline{\tilde{0}})\right)\phi_{\mu_2}, \qquad A \in \mathfrak{B}_0, \tag{7.32}$$

$$T^g_{\mu_3}\pi_{\mu_3}(A)\phi_{\mu_3} = \pi_{\mu_3}\left(\alpha^3_i(A)f^{(n)}_3(\underline{\tilde{0}})\right)\phi_{\mu_3}, \qquad A \in \mathfrak{B}_0. \tag{7.33}$$

The inverses are given by

$$\left(T^g_{\mu_a}\right)^{-1}\pi_{\mu_a}(A)\phi_{\mu_a} = \pi_{\mu_a}\left(\alpha^a_{-i}(Af^{(n)}_a(\underline{\tilde{0}})^{-1})\right)\phi_{\mu_a}, \qquad A \in \mathfrak{B}_0, \tag{7.34}$$

$a = 2, 3$, and are densely defined.

From (7.25) and (7.31) we conclude that $T^g_{\mu_0} \sim T^g_{\mu_1} \sim T^g_{\mu_2} \sim T^g_{\mu_3}$, where $\sim$ means unitary equivalence. After a simple computation one can see that

$$\left(\pi_{\mu_a}(A)\phi_{\mu_a},\ T^g_{\mu_a}\pi_{\mu_a}(B)\left(T^g_{\mu_a}\right)^{-1}\pi_{\mu_a}(C)\phi_{\mu_a}\right) = \mu_a(A^*\alpha^a_i(B)C) \tag{7.35}$$

for all $a \in \{0, \ldots, 3\}$, and for all $A$, $B$, $C \in \mathfrak{B}_0$.

For completeness and further uses we also write down the explicit definitions of the operators $W_{0\to 2}$, $W_{1\to 3}$ and $W_{2\to 3}$:

$$W_{0\to 2}\pi_{\mu_0}(A)\phi_{\mu_0} = \pi_{\mu_2}\left(\Delta^{-1}\circ\beta(A)\left(Z^{(n)}(\underline{\tilde{0}})^{-1/2}\right)'\right)\phi_{\mu_2}, \qquad A \in \mathfrak{B}_0, \tag{7.36}$$

$$W_{1\to 3}\pi_{\mu_1}(A)\phi_{\mu_1} = \pi_{\mu_3}\left(\Delta^{-1}\circ\beta^{-1}(A)\left(\beta^*\left(Z^{(n)}(\underline{0})^{1/2}\right)\right)'\right)\phi_{\mu_3}, \qquad A \in \mathfrak{B}_0, \tag{7.37}$$

$$W_{2\to 3}\pi_{\mu_2}(A)\phi_{\mu_2} = \pi_{\mu_3}\left(\beta^{*\prime}\left(A\left(Z^{(n)}(\underline{\tilde{0}})^{1/2}\right)'\right)\right)\phi_{\mu_3}, \qquad A \in \mathfrak{B}_0. \tag{7.38}$$

We close this section mentioning without proof an elementary Proposition we will use.

**Proposition 7.1** *Using the definitions given above one has:*

$$\left(T^g_{\mu_a}\right)^B \phi_{\mu_a} = \pi_{\mu_a}\left(\alpha^a_{(B-1)i}\left(f^{(n)}_a(\underline{0})\right)\cdots\alpha^a_i\left(f^{(n)}_a(\underline{0})\right)f^{(n)}_a(\underline{0})\right)\phi_{\mu_a} \tag{7.39}$$

*for all $a$ , and for all $B \in \mathbb{N}$, $B \geq 1$, and finally*

$$\left(T^g_{\mu'_2}\right)^B \psi_{\mu'_2} = \pi_{\mu'_2}\left(\alpha_{(B-1)i}\left(Z^{(n)}(\tilde{\underline{0}})^{-1}\right)\cdots\alpha_i\left(Z^{(n)}(\tilde{\underline{0}})^{-1}\right)Z^{(n)}(\tilde{\underline{0}})^{-1}\right)\psi_{\mu'_2}, \tag{7.40}$$

*for all $B \in \mathbb{N}$, $B \geq 1$, where $\psi_{\mu'_2} := \pi_{\mu'_2}(Z^{(n)}(\tilde{\underline{0}})^{1/2})\phi_{\mu'_2}$* □

# 8 The Translation Operators on the Charged Sectors.

As far, we have not discussed the question of the translation invariance of the global transfer matrices defined above. Let us first analyze this question for the GNS triple associated to the electric state $\mu_0$ and $\mathfrak{B}_0$.

Our task is to exhibit a unitary operator acting in $\mathcal{H}^g_{\mu_0}$ implementing the translations and to show the translation invariance of the global transfer matrix $T^g_{\mu_0}$. We will be following the steps of [1] with some adaptations. We remember that the state $\mu_0$ and all GNS objects associated to it have been constructed with a charge "located" at $\underline{0}$. For an arbitrary point $\underline{x} \in \mathbb{Z}^2$, we have to show the existence of vectors in $\mathcal{H}^g_{\mu_0}$ implementing the states $\mu_0 \circ \tau_{-\underline{x}}$, the electrically charged state with charge "sitting" at $\underline{x}$.

Let us start defining the following operator on $\mathcal{H}^g_{\mu_0}$:

$$V_{\underline{0}}\,\pi_{\mu_0}(A)\phi_{\mu_0} = \pi_{\mu_0}\left(\delta_{\underline{0}}^{E\,(n)}(A)\right)\phi_{\mu_0}, \qquad A \in \mathfrak{B}_0. \tag{8.1}$$

This operator is well-defined, bounded with $\|V_{\underline{0}}\| = 1$, self-adjoint and, by (6.56), positive. Clearly $V_{\underline{0}}$ coincides with $V_{\mu_0} \restriction \mathcal{H}^g_{\mu_0}$, defined in the previous section. Boundedness follows from the Cauchy-Schwarz inequality and from the cluster property of $\mu_0$ with respect to $\delta_{\underline{0}}^{E\,(n)}$. The same cluster property also implies the uniqueness up to a phase of the eigenvector of $V_{\underline{0}}$ with eigenvalue one. Note that this operator has been defined here only for for the point $\underline{0}$. Later we will extend the definition to arbitrary $\underline{x}$.

Let $\underline{L}_{\underline{a}\to\underline{b}}$ denote a finite connected set of bonds in $\mathbb{Z}^2$ having $\underline{a}$ and $\underline{b}$ as end points, oriented from $\underline{a}$ to $\underline{b}$. We will call these sets of bonds transporter bonds. For such transporter bonds and for $p \in \mathbb{N}$ define the operator

$$A_p(\underline{L}_{\underline{a}\to\underline{b}}) := \left[\prod_{a=0}^{p-1}\alpha_{ai}(f_0^{(n)}(\underline{b})^{-1})\right]\alpha_{pi}\left(U_3(\underline{L}_{\underline{a}\to\underline{b}})^{n\,*}\right)\left[\prod_{b=0}^{p-1}\alpha_{bi}(f_0^{(n)}(\underline{a})^{-1})\right]^{-1}. \tag{8.2}$$

The idea behind this definition is that the operator $A_p(\underline{L}_{\underline{x}\to\underline{x}'})\tau_{\underline{x}}(F_0(r))$, $p < r$, is associated to a classical function described in Figure 3.

**Proposition 8.1** *For fixed $\underline{L}_{\underline{0}\to\underline{x}}$ the sequence defined by*

$$\psi_p(\underline{L}_{\underline{0}\to\underline{x}}) := \pi_{\mu_0}\left(A_p(\underline{L}_{\underline{0}\to\underline{x}})\right)\phi_{\mu_0}/\|\pi_{\mu_0}\left(A_p(\underline{L}_{\underline{0}\to\underline{x}})\right)\phi_{\mu_0}\|, \qquad p \in \mathbb{N}, \tag{8.3}$$

*is a Cauchy sequence in $\mathcal{H}_{\mu_0}$. Call $\psi_{\underline{x}}(\underline{L}_{\underline{0}\to\underline{x}})$ the limit vector of the Cauchy sequence above. Then $\psi_{\underline{x}}(\underline{L}_{\underline{0}\to\underline{x}})$ is independent of the transporter bonds $\underline{L}_{\underline{0}\to\underline{x}}$* □

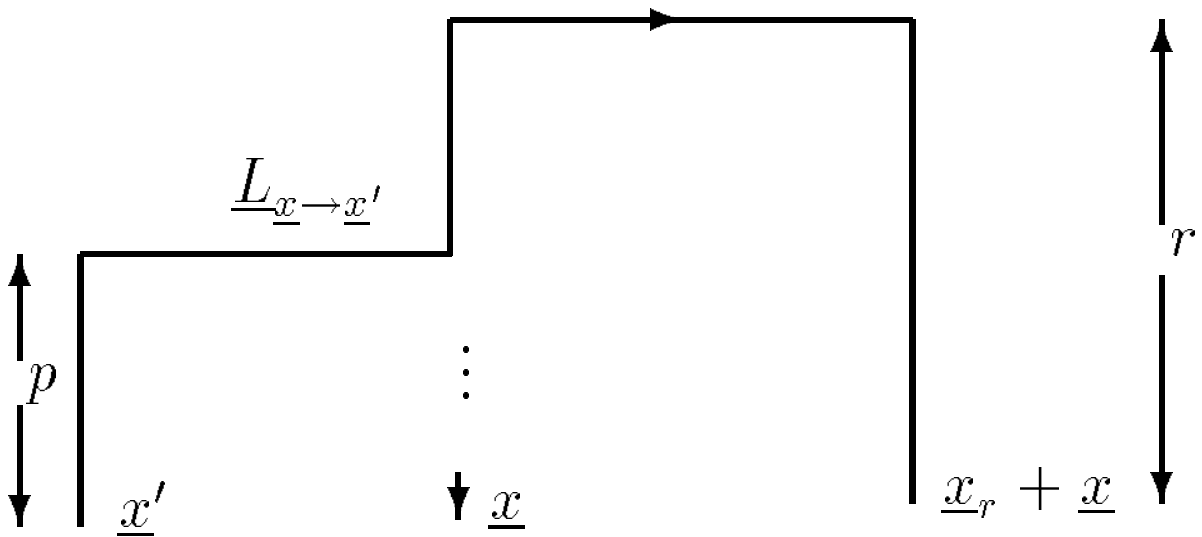

Figure 3: *Semi-loop of bounds associated to the classical function of $A_p(\underline{L}_{\underline{x}\to\underline{x}'})\tau_{\underline{x}}(F_0(r))$. The arrow indicates the orientation sense of the corresponding 1-form.*

**Proof:** Call $\psi_p := \psi_p(\underline{L}_{\underline{0}\to\underline{x}})$. By (4.19), $(\psi_n,\, \psi_m) \geq 0$ and can be written as $[(\psi_n,\, \psi_m)(\psi_m,\, \psi_n)]^{1/2}$. Expressing this in terms of cluster expansions one sees, for $n$ and $m$ large, that the only clusters which contribute are of size larger that $2\min(n,\, m)$ and that for $|\ln[(\psi_n,\, \psi_m)]|$ we have the bound $(const.)(|\underline{L}_{\underline{0}\to\underline{x}}|)e^{-(const.)\min(n,\,m)}$. So $(\psi_n,\, \psi_m) \to 1$, for $n,\, m \to \infty$ and this implies convergence. Now we are going to prove the independence of the limit vectors on the transporter bonds. The argument is analogous to the previous one. Consider

$$\left(\psi_{\underline{x}}(\underline{R}_{\underline{0}\to\underline{x}}), \psi_{\underline{x}}(\underline{L}_{\underline{0}\to\underline{x}})\right) = \lim_{p\to\infty} \left[\left(\psi_p(\underline{R}_{\underline{0}\to\underline{x}}), \psi_p(\underline{L}_{\underline{0}\to\underline{x}})\right)\left(\psi_p(\underline{L}_{\underline{0}\to\underline{x}}), \psi_p(\underline{R}_{\underline{0}\to\underline{x}})\right)\right]^{1/2}. \tag{8.4}$$

Expressing this in terms of cluster expansions one sees that, for $p$ large the only clusters which contribute are of size larger than $2p$ and that for $\left|\ln\left(\psi_p(\underline{R}_{\underline{0}\to\underline{x}}), \psi_p(\underline{L}_{\underline{0}\to\underline{x}})\right)\right|$ we have the upper bound $(const.)(|\underline{L}_{\underline{0}\to\underline{x}}| + |\underline{R}_{\underline{0}\to\underline{x}}|)e^{-(const.)p} \to 0$ for $p \to \infty$ ■

**Definition 8.1** *Call $\phi_{\mu_0}(\underline{x}) := \psi_{\underline{x}}(\underline{L}_{\underline{0}\to\underline{x}})$ for any $\underline{L}_{\underline{0}\to\underline{x}}$, with $\phi_{\mu_0}(\underline{0}) := \phi_{\mu_0}$.*

**Proposition 8.2** *i) One has for all $A \in \mathfrak{B}_0$*

$$(\phi_{\mu_0}(\underline{x}),\, \pi_{\mu_0}(A)\phi_{\mu_0}(\underline{x})) = \mu_0(\tau_{-\underline{x}}(A)). \tag{8.5}$$

*ii) $\phi_{\mu_0}(\underline{x})$ is a cyclic vector with respect to $\pi_{\mu_0}(\mathfrak{B}_0)$* □

**Proof:** The proof of part *i)* is easily obtained using the definitions and the cluster expansions. We omit the details. Part *ii)* holds if there exists a sequence

$$\psi_n := \pi_{\mu_0}(B_n)\phi_{\mu_0}(\underline{x})/\|\pi_{\mu_0}(B_n)\phi_{\mu_0}(\underline{x})\|, \tag{8.6}$$

$B_n \in \mathfrak{B}_0$, $n \in \mathbb{N}$, converging to $e^{if}\phi_{\mu_0}$, $f \in \mathbb{R}$ which is cyclic, by construction. Write $\phi_{\mu_0}(\underline{x}) = \psi_{\underline{x}}(\underline{L}_{\underline{0}\to\underline{x}})$. We will show that one has such a sequence for a choice like $B_q = A_q(\underline{R}_{\underline{x}\to\underline{0}})$, in particular with $\underline{R}_{\underline{x}\to\underline{0}} = -\underline{L}_{\underline{0}\to\underline{x}}$. The fact that, for this choice, $\psi_n$ is a Cauchy sequence can be proven similarly to the previous case using part i) of this Proposition. In order to show the convergence to $e^{if}\phi_{\mu_0}$ we observe that, for arbitrary $\underline{x}$, $\underline{L}$ and $\underline{R}$, one has

$$\delta_{\underline{0}}^{E(n)}\left(A_q(\underline{R}_{\underline{x}\to\underline{0}})A_p(\underline{L}_{\underline{0}\to\underline{x}})\right) = A_{q+1}(\underline{R}_{\underline{x}\to\underline{0}})A_{p+1}(\underline{L}_{\underline{0}\to\underline{x}}), \tag{8.7}$$

since, in general

$$\delta_{\underline{a}}^{E(n)}\left(A_p(\underline{L}_{\underline{a}\to\underline{b}})\right) = f_0^{(n)}(\underline{a})^{-1} f_0^{(n)}(\underline{b})A_{p+1}(\underline{L}_{\underline{a}\to\underline{b}}), \tag{8.8}$$

and

$$\delta_{\underline{b}}^{E\,(n)}\left(A_p(\underline{L}_{\underline{a}\to\underline{b}})\right) = A_{p+1}(\underline{L}_{\underline{a}\to\underline{b}})f_0^{(n)}(\underline{a})^{-1}f_0^{(n)}(\underline{b}). \tag{8.9}$$

This implies that

$$V_{\underline{0}}\psi = \lim_{q\to\infty}\lim_{p\to\infty}\left[\frac{\left\|\pi_{\mu_0}\left(A_{q+1}(-\underline{L}_{\underline{0}\to\underline{x}})A_{p+1}(\underline{L}_{\underline{0}\to\underline{x}})\right)\phi_{\mu_0}\right\|}{\left\|\pi_{\mu_0}\left(A_{q}(-\underline{L}_{\underline{0}\to\underline{x}})A_{p}(\underline{L}_{\underline{0}\to\underline{x}})\right)\phi_{\mu_0}\right\|}\right]\psi, \tag{8.10}$$

where $\psi$ is the limit of the sequence $\psi_n$. In Appendix D we will prove that the factor inside of the brackets converges to one in the limits above. Therefore, $\psi$ is an eigenvector of $V_{\underline{0}}$ with eigenvalue one and has to be equal to $e^{if}\phi_{\mu_0}$ ■

Now we are able to generalize the definition of the operator $V_{\underline{0}}$ on $\mathcal{H}^g_{\mu_0}$. Define

$$V_{\underline{x}}\,\pi_{\mu_0}(A)\phi_{\mu_0}(\underline{x}) = \pi_{\mu_0}(\delta_{\underline{x}}^{E\,(n)}(A))\phi_{\mu_0}(\underline{x}), \qquad A \in \mathfrak{B}_0. \tag{8.11}$$

This operator is well-defined, densely defined, bounded with $\|V_{\underline{x}}\| = 1$, self-adjoint and, by (6.56), positive. Boundedness follows from the Cauchy-Schwarz inequality and from the cluster property of $\mu_0$ with respect to $\delta_{\underline{x}}^{E\,(n)}$. The same cluster property also implies the uniqueness up to a phase of the eigenvector of $V_{\underline{x}}$ with eigenvalue one.

Let us now introduce the generators of the translations. Following [1], define

$$U_{\mu_0}(\underline{z})\pi_{\mu_0}(A)\phi_{\mu_0} := \pi_{\mu_0}(\tau_{\underline{z}}(A))\phi_{\mu_0}(\underline{z}), \qquad A \in \mathfrak{B}_0 \tag{8.12}$$

for all $\underline{z} \in \mathbb{Z}^2$ and $A \in \mathfrak{B}_0$. This is a densely defined isometry with dense range and so it defines a unitary operator. The family $\{U_{\mu_0}(\underline{z}),\ \underline{z} \in \mathbb{Z}^2\}$ defines a unitary representation of $\mathbb{Z}^2$ in $\mathcal{H}^g_{\mu_0}$. This follows from the same arguments as in [1] and we will not repeat the details. As we will discuss in [9], the corresponding situation in the dyonic case is more complicated, since the positivity argument used in [1] does not hold for expectations involving loops of bonds and of plaquettes.

With this definition we easily check that $U_{\mu_0}(\underline{z})V_{\underline{x}} = V_{\underline{x}+\underline{z}}U_{\mu_0}(\underline{z})$ and $U_{\mu_0}(\underline{z})T^g_{\mu_0} = T_{\underline{z}}U_{\mu_0}(\underline{z})$ where, for $\underline{x} \in \mathbb{Z}^2$, we define the operator

$$T_{\underline{x}} := \pi_{\mu_0}(f_0^{(n)}(\underline{x}))V_{\underline{x}}. \tag{8.13}$$

As one sees, this is a self-adjoint operator and, by (6.58), one has $0 \leq T_{\underline{x}} \leq \|f_0^{(n)}(\underline{x})\|$. Clearly $T_{\underline{0}} = T^g_{\mu_0}$ and we want to show that $T_{\underline{x}} = T^g_{\mu_0}$ for all $\underline{x}$. For, observe that

$$T_{\underline{x}} = \left(\lim_{q\to\infty}\frac{\|\pi_{\mu_0}(A_{q+1}(\underline{L}_{\underline{x}\to\underline{0}}))\phi_{\mu_0}(\underline{x})\|}{\|\pi_{\mu_0}(A_{q}(\underline{L}_{\underline{x}\to\underline{0}}))\phi_{\mu_0}(\underline{x})\|}\right)T_{\mu_0}, \tag{8.14}$$

what can be seen applying the definitions on the dense set $\{\pi_{\mu_0}(A)\phi_{\mu_0},\quad A \in \mathfrak{B}_0\}$ and using (8.8) and (8.9). Using arguments analogous to that used in Appendix D we can see that the factor between parenthesis in (8.14) is equal to one. This can be also more directly seen from the fact that $T_{\underline{x}}$ and $T_{\mu_0}$ have the same norm since $U_{\mu_0}(\underline{x})$ is unitary and intertwines both. From this it follows that $U_{\mu_0}(\underline{z})$ and $T^g_{\mu_0}$ commute for all $\underline{z} \in \mathbb{Z}^2$.

Now we present the definitions of the translation operators in the GNS-sectors associated to the states $\mu_1$, $\mu_2$ and $\mu_3$. We simply define, for all $\underline{z} \in \mathbb{Z}^2$ and $a = 1,\,2,\,3$: $U_{\mu_a}(\underline{z}) := W_{0\to a}U_{\mu_0}(\underline{z})W_{0\to a}^{-1}$. Let us look at each case more closely. Using the previously discussed polar decomposition of the operator $W_{0\to 1}$ one has

$$U_{\mu_1}(\underline{z}) := W_{0\to 1}U_{\mu_0}(\underline{z})W_{0\to 1}^{-1} = \mathcal{W}_{0\to 1}(T^g_{\mu_0})^{1/2}U_{\mu_0}(\underline{z})(T^g_{\mu_0})^{-1/2}\mathcal{W}_{0\to 1}^{-1}, \tag{8.15}$$

and since $U_{\mu_0}$ commutes with $T^g_{\mu_0}$ we conclude that, for each $\underline{z}$, $U_{\mu_1}(\underline{z})$ and $U_{\mu_0}(\underline{z})$ are unitarily equivalent. Analogously we conclude that $U_{\mu_2}(\underline{z})$ and $U_{\mu_0}(\underline{z})$ are unitarily equivalent since $W_{0\to 2} = W_{1\to 2}W_{0\to 1}$ and $W_{1\to 2}$ is unitary. Finally $U_{\mu_3}(\underline{z})$ and $U_{\mu_0}(\underline{z})$ are unitarily equivalent because $W_{0\to 3}$ is unitary. Note that the unitary operators intertwining $U_{\mu_a}(\underline{z})$ and $U_{\mu_b}(\underline{z})$ are the same intertwining $T^g_{\mu_a}$ and $T^g_{\mu_b}$, as found in the previous section. In this way we have found an equivalent to Corollary 5.3 for the charged sectors:

**Corollary 8.1** *The joint spectrum of the transfer matrix and the momentum operator, $sp\,(T^g_{\mu_a},\, \mathbb{P}_{\mu_a})$, is the same for all $a$* □

This in particular says that, if there exists an electrically charged particle in the sector associated to $\mu_0$ there must be a magnetically charged particle in the sector associated to $\mu_2$ with the same mass and dispersion relation.

It is interesting to study in more detail how $U_{\mu_2}(\underline{z})$ acts. We will in particular derive a result which will be useful in the proof of the existence of magnetic particles.

**Definition 8.2** *Let $L_{\underline{0}\to\underline{z}}$, $\underline{z} \in \mathbb{Z}^2$ be transporter bounds. Define for $p \in \mathbb{N}$,*

$$B_p(L_{\underline{0}\to\underline{z}}) :=$$

$$\left(Z^{(n)}(\tilde{\underline{z}})^{1/2}\right)' \left[\prod_{a=1}^{p-1} \alpha'_{ai}\left(Z^{(n)}(\tilde{\underline{z}})'\right)\right] \alpha'_{pi}\left(\beta^{*\,-1\,\prime}(U_1(\tilde{\underline{L}}_{\underline{0}\to\underline{z}})^n)\right) \left[\prod_{b=1}^{p-1} \alpha'_{bi}\left(Z^{(n)}(\tilde{\underline{0}})'\right)\right]^{-1} \left(Z^{(n)}(\tilde{\underline{0}})^{-1/2}\right)'. \tag{8.16}$$

*Call*

$$\psi_{\mu_2}(\tilde{\underline{z}}) := \lim_{p\to\infty} \frac{\pi_{\mu_2}\left(\left(Z^{(n)}(\tilde{\underline{z}})^{1/2}\right)' B_p(L_{\underline{0}\to\underline{z}})\right)\phi_{\mu_2}}{\left\|(T^g_{\mu_2})^{1/2}\,\pi_{\mu_2}\left(\left(Z^{(n)}(\tilde{\underline{0}})^{1/2}\right)' B_p(L_{\underline{0}\to\underline{z}})\right)\phi_{\mu_2}\right\|}, \tag{8.17}$$

*with $\psi_{\mu_2}(\tilde{\underline{0}}) := \pi_{\mu_2}\left(\left(Z^{(n)}(\tilde{\underline{0}})^{1/2}\right)'\right)\phi_{\mu_2}$ as in Proposition 7.1* □

In (8.17) we used the fact that

$$\left\|\left(T^g_{\mu_2}\right)^{1/2}\pi_{\mu_2}\left(\left(Z^{(n)}(\tilde{\underline{0}})^{1/2}\right)'\right)\phi_{\mu_2}\right\| = \left(\pi_{\mu_2}\left(\left(Z^{(n)}(\tilde{\underline{0}})^{1/2}\right)'\right)\phi_{\mu_2},\, T^g_{\mu_2}\pi_{\mu_2}\left(\left(Z^{(n)}(\tilde{\underline{0}})^{1/2}\right)'\right)\phi_{\mu_2}\right) = 1. \tag{8.18}$$

The existence of the limit in (8.17) can be established with the same methods used in the electric case. Analogously to Proposition 7.1 one can show that

$$T^g_{\mu_2}\psi_{\mu_2}(\underline{z}) = \ \pi_{\mu_2}\left(\left(Z^{(n)}(\tilde{\underline{z}})^{-1}\right)'\right)\psi_{\mu_2}(\underline{z}). \tag{8.19}$$

**Theorem 8.1** *With the definitions above*

$$U_{\mu_2}(\underline{z})\pi_{\mu_2}(A)\psi_{\mu_2}(\tilde{\underline{0}}) = \pi_{\mu_2}(\tau_{\underline{z}}(A))\psi_{\mu_2}(\tilde{\underline{z}}) \tag{8.20}$$

*for all $A \in \mathfrak{B}_0$, $\underline{z} \in \mathbb{Z}^2$* □

**Proof.** Using the definition of $U_{\mu_2}(\underline{z}) := W_{0\to 2} U_{\mu_0}(\underline{z}) W_{0\to 2}^{-1}$ one can show that

$$U_{\mu_2}(\underline{z})\pi_{\mu_2}(A)\phi_{\mu_2} = \lim_{p\to\infty} \frac{\pi_{\mu_2}\left(\tau_{\underline{z}}(A) K_p(\underline{L}_{\underline{0}\to\underline{z}})\right)\phi_{\mu_2}}{\left\|\pi_{\mu_0}(A_p(\underline{L}_{\underline{0}\to\underline{z}}))\phi_{\mu_0}\right\|}, \tag{8.21}$$

with

$$K_p(\underline{L}_{\underline{0}\to\underline{z}}) := \left(Z^{(n)}(\tilde{\underline{z}})^{1/2}\right)' \Delta^{-1} \circ \beta(A_p(\underline{L}_{\underline{0}\to\underline{z}})) \left(Z^{(n)}(\tilde{\underline{0}})^{-1/2}\right)'. \tag{8.22}$$

A lengthy but straightforward computation shows that

$$K_p(\underline{L}_{\underline{0}\to\underline{z}}) = \left(Z^{(n)}(\underline{z})^{1/2}\right)' \alpha_i' \left(\left(Z^{(n)}(\tilde{\underline{z}})^{1/2}\right)' B_{p-1}(L_{\underline{0}\to\underline{z}})\right) f_2^{(n)}(\tilde{\underline{0}}). \tag{8.23}$$

Hence, the numerator in (8.21) can be written as

$$\pi_{\mu_2}\left(\tau_{\underline{z}}(A)\left(Z^{(n)}(\underline{z})^{1/2}\right)'\right) T^g_{\mu_2}\pi_{\mu_2}\left(\left(Z^{(n)}(\tilde{\underline{z}})^{1/2}\right)' B_{p-1}(\underline{L}_{\underline{0}\to\underline{z}})\right)\phi_{\mu_2}. \tag{8.24}$$

Concerning the denominator in (8.21) one can show that it equals

$$\left\|\left(T^g_{\mu_2}\right)^{1/2}\pi_{\mu_2}\left(\left(Z^{(n)}(\tilde{\underline{z}})^{1/2}\right)' B_{p-1}(\underline{L}_{\underline{0}\to\underline{z}})\right)\phi_{\mu_2}\right\|. \tag{8.25}$$

Therefore, after the limit is taken we get

$$U_{\mu_2}(\underline{z})\pi_{\mu_2}(A)\phi_{\mu_2} = \pi_{\mu_2}\left(\tau_{\underline{z}}\left(A\left(Z^{(n)}(\tilde{\underline{0}})^{1/2}\right)'\right)\right) T^g_{\mu_2}\psi_{\mu_2}(\underline{z}) = \pi_{\mu_2}\left(\tau_{\underline{z}}\left(A\left(Z^{(n)}(\tilde{\underline{0}})^{-1/2}\right)'\right)\right)\psi_{\mu_2}(\underline{z}), \tag{8.26}$$

where here we used (8.19). To finish the proof, replace $A \to A\left(Z^{(n)}(\tilde{\underline{0}})^{1/2}\right)'$ ■

# 9 The Existence of Electrically and of Magnetically Charged Particles.

The existence of an electrically charged particle in the $\mathbb{Z}_2$ case in $d \geq 2$ was established in [3] using methods previously developed by Schor and collaborators in the vacuum sector (see references in [3]). Here we will use the same techniques to show the existence of $N-1$ electrically and magnetically charged particles in our 3-dimensional $\mathbb{Z}_N$ model. We will restrict ourself to present only the basic results concerning the existence of electric charged particles in the $\mathbb{Z}_N$ case. Further details for the proof of existence of the one-particle states can be inferred from the basic discussion found in [3]. By duality, or more precisely, by Corollary 8.1, we conclude the existence of $N-1$ magnetically charged particles as well. Nonetheless a direct proof of the existence of magnetic particles can be found repeating the steps of the electric case. As in the previous sections our results are restricted to the region of couplings with $\max\{g(1), \ldots, g(N-1), h(1), \ldots, h(N-1)\}$ small enough.

Due to some special problems already observed in [3] we have to represent $\langle B(\alpha,\, \beta)\rangle$ in a slightly different way from (4.31). Here we write

$$\begin{aligned}\langle B(\alpha,\, \beta)\rangle \;=\; & \sum_{\substack{(M,\,E)\in \mathrm{Conn}_1(\alpha;\, X)\\ (P,\,D)\in \mathrm{Conn}_2(\beta;\, C)}} [D-\beta:\, E-\alpha]\left[\prod_{p\in P} g(D(p))\right]\left[\prod_{b\in M} h(E(b))\right] \\ & \times \exp\left(\sum_{\Gamma\in \mathcal{G}_{clus}(V)} c_\Gamma\left(a^\Gamma_{(M,E),\,\alpha;\,X}\, b^\Gamma_{(P,D),\,\beta;\,C} - 1\right)\mu^\Gamma\right).\end{aligned} \tag{9.1}$$

We used the following definitions:

- $\mathrm{Conn}_1(\alpha;\,X)$ equals $\mathrm{Conn}_1(\alpha)$ if $d\alpha \neq 0$ and otherwise equals the set $\mathcal{G}_X$ of all polymers whose geometric part is formed a simple connected set of bonds which are connected to at least one point of the finite set $X \subset l_0 = \mathbb{Z}^3$. One also has $\emptyset \in \mathcal{G}_X$.

- $\mathrm{Conn}_2(\beta;\,C)$ equals $\mathrm{Conn}_2(\beta)$ if $d^*\beta \neq 0$ and otherwise equals the set $\mathcal{G}_C$ of all polymers whose geometric part is formed a simple co-connected set of plaquettes which are co-connected to at least one cube of the finite set $C \subset l_3$. One also has $\emptyset \in \mathcal{G}_C$.

- $a_{(M,E),\alpha;X}(\gamma)$ equals $a_{(M,E),\alpha}(\gamma)$ but is zero in the case $d\alpha = 0$ if there are bonds composing $\gamma$ connected with at least one point of $X$.

- $b_{(P,D),\beta;C}(\gamma)$ equals $b_{(P,D),\beta}(\gamma)$ but is zero in the case $d^*\beta = 0$ if there are plaquettes composing $\gamma$ co-connected with at least one cube of $C$.

The convergence of this representations can be proven by the same methods. We note that (9.1) does not depend on $X$ and $C$, which can be chosen arbitrarily. We indicate the choice of $X$ and $C$ by writing $\langle \cdots \rangle_{X;C}$. This representation only differs from (4.31) if $d\alpha = 0$ or $d^*\beta = 0$.

Starting from the states $\mu_0$ and $\mu_2'$ on the algebra $\mathfrak{B}_0$ we associate via the GNS construction the objects

$$\mu_0 \;\to\; \phi_E,\ \pi_E,\ \mathcal{H}_E,\ T_E,\ U_E(\underline{x}); \tag{9.2}$$
$$\mu_2' \;\to\; \phi_M,\ \pi_M,\ \mathcal{H}_M,\ T_M,\ U_M(\underline{x}); \tag{9.3}$$

where we introduced the new notations $\phi_E = \phi_{\mu_0}$, etc, and $\phi_M = \phi_{\mu_2'}$, etc.

We will always be assuming that these states have (electric or magnetic) charge $n$. We will analyze the following two-point functions of vector states with charge $(n)$:

$$G^{E,(n)}(x_0,\,\underline{x}) \;:=\; \left(\phi_E,\ U_E(\underline{x}) T_E^{|x_0|} \phi_E\right), \tag{9.4}$$
$$G^{M,(n)}(x_0,\,\tilde{\underline{x}}) \;:=\; \left(T_M^{1/2}\psi_M, U_M(\underline{x}) T_M^{|x_0|} T_M^{1/2}\psi_M\right), \tag{9.5}$$

for $(\underline{x},\,x_0) \in \mathbb{Z}^3$, with

$$\psi_M := \pi_M(Z^{(n)}(\tilde{\underline{0}})^{1/2})\phi_M. \tag{9.6}$$

The factor $T_M^{1/2}\pi_M(Z^{(n)}(\tilde{\underline{0}})^{1/2})$ was introduced for convenience. The results we are going to establish say that the vectors $\phi_E$ and $T_M^{1/2}\psi_M$ have non-vanishing components on one-particle subspaces. The same holds for the vector $\psi_M$ since $T_M$ commutes with the momentum operator. Note that

$$G^{M,(n)}(x_0,\,\tilde{\underline{x}}) = \left(\pi_M(Z^{(n)}(\tilde{\underline{x}})^{1/2})\phi_M,\ \left(U_M(\underline{x}) T_M^{|x_0|+1}\right) \pi_M(Z^{(n)}(\tilde{\underline{x}})^{1/2})\phi_M\right). \tag{9.7}$$

We now introduce a convention frequently used below.

**Notation 9.1** *If $\gamma$ is a form in $\mathbb{Z}^3$ with finite support we will denote by $\gamma[\underline{x}]$ the form $\gamma$ translated by $\underline{x} \in \mathbb{Z}^2$. Clearly $\gamma[\underline{0}] = \gamma$* □

We represent these two two-point functions as the square root of the Green functions of two infinitely separated charges. Having (7.39) in mind we write, for two fixed points $x = (x_0, \underline{x})$ and $y = (y_0, \underline{y}) \in \mathbb{Z}^3$,

$$G^{E,(n)}(x, y) = \lim_{r\to\infty} \left( \frac{\langle B\,(\gamma_r^n[\underline{x}],\, 0)\,\rangle}{\langle B\,(\alpha_r^n[\underline{x}],\, 0)\,\rangle} \right)^{1/2}, \tag{9.8}$$

$$G^{M,(n)}(x, y) = \lim_{r\to\infty} \left( \frac{\langle B\,(0,\, -\delta_r^n[\underline{x}])\,\rangle}{\langle B\,(0,\, -\beta_r^n[\underline{x}])\,\rangle} \right)^{1/2}, \tag{9.9}$$

where the forms $\gamma_r^n$ and $\alpha_r^n$ are defined in Figure 4 (see also Notation 9.1). The forms $\delta_r^n$ and $\beta_r^n$ are the analogous of the forms $\gamma_r^n$ and $\alpha_r^n$, respectively, on the dual lattice.

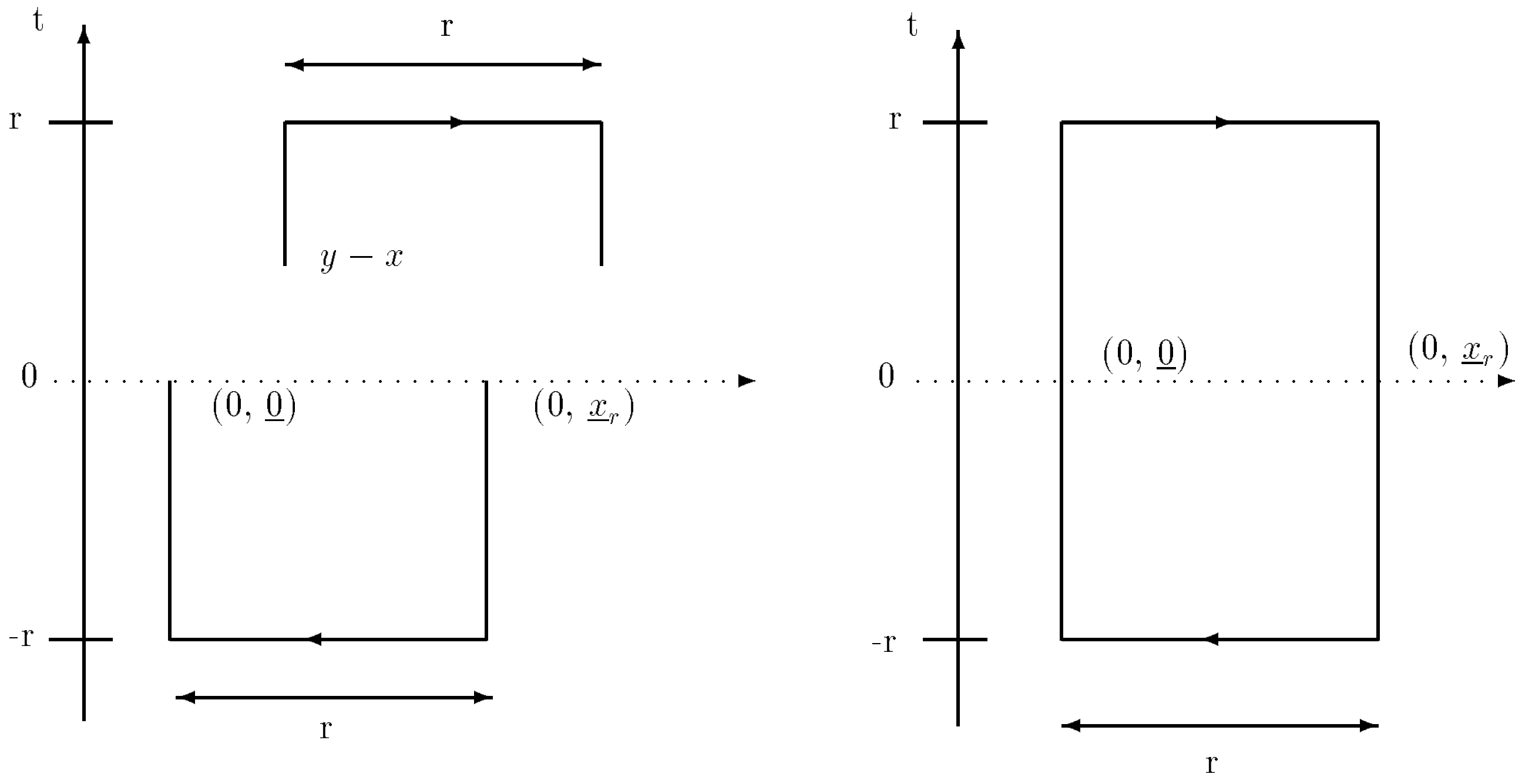

Figure 4: *The forms* $\gamma_r^n = \gamma_r^n[\underline{0}]$ *(left) and* $\alpha_r^n = \alpha_r^n[\underline{0}]$ *(right).*

In the method we follow we consider configurations of complex space-time dependent coupling constants $k := \{h(1)(b), \ldots, h(N)(b), g(1)(p), \ldots, g(N)(p), \quad b \in l_1,\, p \in l_2\}$. The configurations we will consider are invariant under space translations. We denote by $\langle\cdot\rangle(k)$ the classical expectations associated to the configuration of couplings $k$.

We define the Green functions for space-time dependent couplings for $x_0 \le y_0$ by

$$G^{E,(n)}(x, y)_k = \lim_{r\to\infty} \left( \frac{\langle B(\gamma_r^n[\underline{x}],\, 0)\rangle_{X;\emptyset}(k)}{\langle B(\alpha_r^n[\underline{x}],\, 0)\rangle(k)} \right)^{1/2}, \tag{9.10}$$

$$G^{M,(n)}(*x, *y)_k = \lim_{r\to\infty} \left( \frac{\langle B(0,\, -\delta_r^n[\underline{x}])\rangle_{\emptyset;C}(k)}{\langle B(0,\, -\beta_r^n[\underline{x}])\rangle(k)} \right)^{1/2}, \tag{9.11}$$

where we have chosen $X = \{y, y_r\}$ and $C = \{c_y, c_{y_r}\}$, where $c_y$ the cube spanned by the plaquettes $(\underline{\tilde{y}}, y_0)$ and $(\underline{\tilde{y}}, y_0 + 1)$. In [3] another representation of $G^E$ for variable couplings has been used. Both lead to the same results.

We identify bonds, plaquettes, and cubes with their geometric central points, (which are points in $\mathbb{Z}^3/2$). For $a < b$ we call by $\mathcal{G}_{a,b}$ the subset of $\mathcal{G}_{clus}$ composed by clusters whose bonds and plaquettes are contained in the time-slice $a < x_0 < b$. We also define $\mathcal{G}^*_{a,b} := \mathcal{G}_{clus} \setminus (\mathcal{G}_{-\infty,a} \cup \mathcal{G}_{b,\infty})$.

To handle with (9.10) and (9.11) we need some abbreviations and call

$$a^r_{(M,E);x,y}(\gamma) \quad := \quad a_{(M,E),\gamma^n_r[\underline{x}];y}(\gamma), \tag{9.12}$$

$$b^r_{(P,D);x,y}(\gamma) \quad := \quad b_{(P,D),-\delta^n_r[\underline{x}];c_y}(\gamma), \tag{9.13}$$

$$a^r_{\underline{z}}(\gamma) \quad := \quad a_{\emptyset,\alpha^n_r[\underline{z}];\emptyset}(\gamma), \tag{9.14}$$

$$b^r_{\underline{z}}(\gamma) \quad := \quad b_{\emptyset,-\beta^n_r[\underline{z}];\emptyset}(\gamma). \tag{9.15}$$

From now one we concentrate on the analysis of $G^{E,(n)}$. The treatment of $G^{M,(n)}$ is analogous. Before the limit $r \to \infty$, the right hand side of (9.10) can be written as

$$\sum_{\substack{(M,E)\in \\ \mathrm{Conn}_1(\gamma^n_r[\underline{x}];X)}} \left[\prod_{b\in M} h(E(b))_k\right] \exp\left\{ \sum_{\Gamma\in\mathcal{G}_{y_0,\infty}} c_\Gamma \mu^\Gamma_k \left( (a^r_{(M,E);x,y})^\Gamma - (a^r_{\underline{x}})^\Gamma \right) \right.$$

$$\left. + \sum_{\Gamma\in\mathcal{G}_{-\infty,x_0}} c_\Gamma \mu^\Gamma_k \left( (a^r_{(M,E);x,y})^\Gamma - (a^r_{\underline{x}})^\Gamma \right) + \sum_{\Gamma\in\mathcal{G}^*_{x_0,y_0}} c_\Gamma \mu^\Gamma_k \left( (a^r_{(M,E);x,y})^\Gamma - (a^r_{\underline{x}})^\Gamma \right) \right\}, \tag{9.16}$$

where $\mu(\gamma)_k$, etc, means that the activity is defined on the configuration of couplings $k$. We have

$$\sum_{\Gamma\in\mathcal{G}_{y_0,\infty}} c_\Gamma \mu^\Gamma_k \left( (a^r_{\underline{x}})^\Gamma - 1 \right) = \sum_{\Gamma\in\mathcal{G}_{y_0,\infty}} c_\Gamma \mu^\Gamma_k \left( (a^r_{\underline{y}})^\Gamma - 1 \right) \tag{9.17}$$

because the sum is convergent for finite $r$ and the result is translation invariant. An analogous argument implies that

$$\sum_{\Gamma\in\mathcal{G}^*_{x_0,y_0}} c_\Gamma \mu^\Gamma_k \left( (a^r_{\underline{x}})^\Gamma - 1 \right) = \sum_{\Gamma\in\mathcal{G}^*_{x_0,y_0}} c_\Gamma \mu^\Gamma_k \left( (a^r_{\underline{x}'})^\Gamma - 1 \right) \tag{9.18}$$

for an arbitrary $\underline{x}' \in \mathbb{Z}^2$. The limit $r \to \infty$ of this expression is also convergent.

Using now these facts and applying the polymer and cluster expansion machinery we can control the limit above and get, for $x, y \in \mathbb{Z}^3$,

$$G^{E,(n)}(x,y)_k = \sum_{\substack{(M,E)\in \\ \mathrm{Conn}_1(\gamma^n[\underline{x}];x,y)}} \left[\prod_{b\in M} h(E(b))_k\right] \varrho((M,E);x,y), \tag{9.19}$$

where $\gamma^n[\underline{x}] = \lim_{r\to\infty} \gamma^n_r[\underline{x}]$ and

$$\varrho((M,E);x,y) = \exp\left(D((M,E);x,y)\right), \tag{9.20}$$

with

$$D((M,E);x,y) = A_+((M,E);x,y) + A_-((M,E);x,y) + B((M,E);x,y), \tag{9.21}$$

where, finally,

$$A_+((M,\,E);\,x,\,y)\;:=\;\sum_{\Gamma\in\mathcal{G}_{y_0,\,\infty}} c_\Gamma\mu_k^\Gamma\left(\left(a_{(M,E);x,y}\right)^\Gamma-\left(a_{\underline{y}}\right)^\Gamma\right),\tag{9.22}$$

$$A_-((M,\,E);\,x,\,y)\;:=\;\sum_{\Gamma\in\mathcal{G}_{-\infty,\,x_0}} c_\Gamma\mu_k^\Gamma\left(\left(a_{(M,E);x,y}\right)^\Gamma-\left(a_{\underline{x}}\right)^\Gamma\right),\tag{9.23}$$

$$B((M,\,E);\,x,\,y)\;:=\;\sum_{\Gamma\in\mathcal{G}^*_{x_0,\,y_0}} c_\Gamma\mu_k^\Gamma\left(\left(a_{(M,E);x,y}\right)^\Gamma-\left(a_{\underline{x}'}\right)^\Gamma\right),\tag{9.24}$$

for arbitrary $\underline{x}'\in\mathbb{Z}^2$, as described above and with the definition $a_{\underline{z}}(\gamma):=\lim_{r\to\infty}a^r_{\underline{z}}(\gamma)$ and, for $(M,\,E)\in\mathrm{Conn}_1(\gamma^n[\underline{x}];\,x,\,y)$, with $a_{(M,E);x,y}:=\lim_{r\to\infty}a^r_{(M',E');x,y}$ for some $(M',\,E')\in\mathrm{Conn}_1(\gamma^n_r[\underline{x}];\,x,\,y)$ containing $(M,\,E)$.

Using the polymer expansion techniques we can prove that

$$|D((M,\,E);\,x,\,y)|\leq k_1|M|+k_2,\tag{9.25}$$

for positive constants $k_1,\,k_2$.

Now we start to collect the results concerning these Green functions which lead to the proof of the existence of one particle states. We omit the proofs since they can be established in the same way as in [3]. First for the Green function with constant couplings one can show that

$$G^{E,\,(n)}(x_0=1,\,\underline{x}=\underline{0})\geq kh(n),\tag{9.26}$$

for some positive constant $k$ and $h$ small enough.

For the Green functions with variable couplings we have to use a special sort of configurations $k$. We will namely consider $g(a)(p)=g_t(a)$, $a=1,\ldots,N-1$ for all plaquettes lying on the plane $x_0=t$, $h(a)(b)=h_t(a)$ for all time-like bonds connecting the planes $x_0=t$ and $x_0=t+1$. For all time-like plaquettes and space-like bonds we fix $g(a)(p)=g(a)$, $h(a)(b)=h(a)$.

The next step is to study the analytical dependence of the Green function on the variable couplings, which lead to analytical properties of the original Green function in momentum-space. We performed this analysis using the techniques of [3] and the results are captured in the following

**Theorem 9.1** *For the electric Green function with charge $n$ and variable couplings as defined above one has the following facts:*

1. *At $h_t(1)=\cdots=h_t(N-1)=0$ and for $x_0\leq t<y_0$ one has*

$$G^{E,\,(n)}((\underline{x},\,x_0);\,((\underline{y},\,y_0))_k=0.\tag{9.27}$$

2. *At $g_t(1)=h_t(1)=\cdots=g_t(N-1)=h_t(N-1)=0$, for $x_0\leq t<y_0$ one has:*

$$\partial_{h_t(n)}G^{E,\,(n)}(x;\,y)_k=\exp(f_{t+1})\sum_{a\in\Lambda_t}G^{E,\,(n)}(x;\,a)_kG^{E,\,(n)}(a+e_0;\,y)_k,\tag{9.28}$$

*where $\Lambda_t$ is the plane $x_0=t$, $e_0$ is the unit vector in positive time direction and $f$ is a holomorphic function of the couplings.*

3. *At $h_t(1)=\cdots=h_t(N-1)=0$, for $x_0\leq t<y_0$ one has:*

$$\partial^{b_1}_{h_t(1)}\cdots\partial^{b_{N-1}}_{h_t(N-1)}G^{E,\,(n)}(x;\,y)_k=0,\tag{9.29}$$

*in the case $\sum_{a=1}^{N-1}ab_a\neq n$ mod $N$.*

*4. At $g_t(1) = \cdots = g_t(N-1) = 0$, for all $t$ one has:*

$$\partial^{c_1}_{g_t(1)} \cdots \partial^{c_{N-1}}_{g_t(N-1)} G^{E,(n)}(x;\, y)_k = 0, \tag{9.30}$$

*in the case $\sum_{a=1}^{N-1} ac_a \neq 0 \ mod\ N$ □*

An analogous result can be proven for $G^{M,(n)}$ by interchanging $g_t \leftrightarrow h_t$ above. According to the methods explained in [3] the results above lead to the following

**Theorem 9.2** *For a given $n$, for $\max\{g(1), \ldots, g(N-1), h(1), \ldots, h(N-1)\}$ small enough and under the condition*

$$h(n) > \max\left\{\prod_{a\neq n} h(a)^{b_a}, \quad \forall b_a,\ 0 \leq b_a \in \mathbb{N} \ with \ \sum ab_a = n \ mod\ N\right\}, \tag{9.31}$$

*the Fourier transform of the 2-point function $G^{E,(n)}(x_0, \underline{x})$ can be analytically extended, for each $\underline{p} \in (-\pi, \pi]^2$, to a meromorphic function of $p_0$ in the region $Im\, p_0 < \hat{\nu}^{E,(n)}(\underline{p})$ with an isolated simple pole at $p_0 = i\nu^{E,(n)}(\underline{p})$, where $\nu^{E,(n)}(\underline{p})$, the energy-momentum relation of the particle, is smooth and $\hat{\nu}^{E,(n)}(\underline{p})$ is continuous with $\hat{\nu}^{E,(n)}(\underline{p}) > \nu^{E,(n)}(\underline{p}) \geq m^{E,(n)}$, $m^{E,(n)}$ being the mass gap. The group velocity $grad\, \nu^{E,(n)}(\underline{p})$ is nowhere constant. For $G^{M,(n)}(x_0, \underline{x})$ one has the same results with dispersion relation $\nu^{M,(n)}(\underline{p})$, etc. Concerning the dependence on the couplings one has $\nu^{M,(n)}(\underline{p})(g,\, h) = \nu^{E,(n)}(\underline{p})(g, h)' = \nu^{E,(n)}(\underline{p})(h, g)$, etc, i.e., dual particles have dispersion relation related by dual couplings. One also has $\nu^{E,(n)}(\underline{p})(g,\, h) = \nu^{E,(N-n)}(\underline{p})(g, h)$, etc, i.e., particles and anti-particles have the same dispersion relation* □

**Remark.** The condition (9.31) is for technical reasons necessary in order to guarantee that we have an upper mass gap, i.e. the "mass shell" related to the particle with charge $n$ is isolated from the absolutely continuous spectrum associated to scattering states with *total* charge $n$, since it essentially says that the mass of the particle with charge $n$ has to be smaller than the sum over all masses of particles whose charges sum up to $n$ mod $N$. But the particles may exist without this condition. Note that condition (9.31) can be satisfied simultaneously for all $n$, for instance if all $h(a)$, $a = 1, \ldots, N-1$, are approximately equal □

This Theorem implies the existence of closed subspaces $\mathcal{H}^1_E \subset \mathcal{H}_E$, $\mathcal{H}^1_M \subset \mathcal{H}_M$ (the single particle subspaces) on which the relations $\left(T_{E/M} - e^{-\nu^{E/M,(n)}(\underline{P})}\right) \restriction \mathcal{H}^1_{E/M} = 0$, hold. Here $\underline{P}$ is the momentum operator. $\mathcal{H}^1_E$ and $\mathcal{H}^1_M$ are the closures of the linear spaces $\mathcal{D}^{(1)}_E$ and $\mathcal{D}^{(1)}_M$ where

$$\mathcal{D}^{(1)}_E = \Bigg\{\Psi^E_f, \quad \Psi^E_f = \sum_{\underline{x}} \int dt f(\underline{x}, t) T^{it}_E U_E(\underline{x}) \phi_E,$$

$$\text{supp}\, \tilde{f} \cap \text{sp}(H_E, \underline{P}) \subset \left\{(\nu^E(\underline{p}), \underline{p}),\ \underline{p} \in (-\pi, \pi]^2\right\},\ \ \tilde{f} \in \mathcal{D}(\mathbb{R}^3)\Bigg\}, \tag{9.32}$$

and $\mathcal{D}^{(1)}_M$ is defined analogously replacing $E \to M$ and with $\psi_M$ replacing $\phi_E$. $H_{E/M}$ are the Hamiltonians defined as $H_{E/M} =: -\ln T_{E/M}$.

# Appendices

## A The Convergence of the Cluster Expansion.

In this Appendix we will present a proof of the convergence of the cluster expansion together with some useful estimates. Our proof uses some ideas contained in [1], Appendix A.3, but we organize the material differently. Adaptations to our case have been done in the proof of Lemma A.2 below. We made no attempt to find optimal estimates and so, no concrete numerical predictions for the size of convergence regions for the couplings will be presented. We refer the reader to [1] where this has been performed for the $\mathbb{Z}_2$ case.

Let $\Gamma$ be a cluster of polymers. We say that a polymer $\gamma$ is incompatible with $\Gamma$, i.e., $\gamma \not\sim \Gamma$, if there is at least one $\gamma' \in \Gamma$ with $\gamma \not\sim \gamma'$. For two clusters $\Gamma$, $\Gamma'$ we have $\Gamma \not\sim \Gamma'$ if there is at least one $\gamma \in \Gamma$ with $\gamma \not\sim \Gamma'$.

For the polymer system discussed in this work we are going to prove the following result:

**Theorem A.1** *There is a convex, differentiable, monotonically decreasing function $F_0$: $(a_0, \infty) \to \mathbb{R}_+$, for some $a_0 \geq 0$, with $\lim_{a\to\infty} F_0(a) = 0$ such that, for all sets of polymers $\Gamma$, and for all $a > a_0$,*

$$\sum_{\gamma \not\sim \Gamma} e^{-a|\gamma|} \leq F_0(a)\, \|\Gamma\|, \tag{A.1}$$

*where $\|\Gamma\| = \sum \Gamma(\gamma')|\gamma'|$, $\Gamma(\gamma')$ being the multiplicity of $\gamma'$ in $\Gamma$* □

Once inequality (A.1) has been established, it has been proven in [1], Appendix A.1, that the two following results hold:

$$\sum_{\substack{\Gamma \in \mathcal{G}_{clus} \\ \Gamma \not\sim \Gamma_0}} |c_\Gamma|\, |\mu^\Gamma| \leq F_1(-\ln \|\mu\|)\, \|\Gamma_0\|, \tag{A.2}$$

$$\sum_{\substack{\Gamma \in \mathcal{G}_{clus} \\ \Gamma \not\sim \Gamma_0 \\ \|\Gamma\| \geq n}} |c_\Gamma|\, |\mu^\Gamma| \leq \left(\frac{\|\mu\|}{\|\mu_c\|}\right)^n \|\Gamma_0\| F_0(a_c), \tag{A.3}$$

where $a_c$ and $\|\mu_c\| > 0$ are constants defined in [1], $F_1$: $(a_c + F_0(a_c), \infty) \to \mathbb{R}_+$ is the solution of $F_1(a + F_0(a)) = F_0(a)$ and $\|\mu\| := \sup_\gamma |\mu(\gamma)|^{1/|\gamma|}$. For a proof we refer the reader to [1].

The inequalities (A.2) and (A.3) are of central importance in the theory of cluster expansions and are often used in this work. This makes relevant to prove Theorem A.1.

Since, in general

$$\sum_{\gamma \not\sim \Gamma} e^{-a|\gamma|} \leq \sum_{\gamma' \in \Gamma} \sum_{\gamma \not\sim \gamma'} e^{-a|\gamma|} \leq \sum_{\gamma' \in \Gamma} \Gamma(\gamma') \sum_{\gamma \not\sim \gamma'} e^{-a|\gamma|}, \tag{A.4}$$

it is enough to prove (A.1) for the case in which $\Gamma$ is composed by a unique polymer $\gamma'$, i.e., $\Gamma = \gamma'$. This will be performed in Theorem A.2 below. We first need some definitions:

**Definition A.1** *Let be the sets*

$$C_b(M) := \{M' \in \mathcal{B}(\mathbb{Z}^{d+1}) \text{ so that } M' \text{ is connected with } M\}, \tag{A.5}$$

$M \in \mathcal{B}_{total}(\mathbb{Z}^{d+1})$,

$$C_p(P) := \{P' \in \mathcal{P}(\mathbb{Z}^{d+1}) \text{ so that } P' \text{ is co-connected with } P\}, \tag{A.6}$$

*$P \in \mathcal{P}_{total}(\mathbb{Z}^{d+1})$; and the sets*

$$W_b(P) := \{M' \in \mathcal{B}(\mathbb{Z}^{d+1}) \text{ so that } w(M',\, P) \neq 0\}, \tag{A.7}$$

*$P \in \mathcal{P}_{total}(\mathbb{Z}^{d+1})$,*

$$W_p(M) := \{P' \in \mathcal{P}(\mathbb{Z}^{d+1}) \text{ so that } w(P',\, M) \neq 0\}, \tag{A.8}$$

*$M \in \mathcal{B}_{total}(\mathbb{Z}^{d+1})$,*

$$W_b^{total}(P) := \{M' \in \mathcal{B}_{total}(\mathbb{Z}^{d+1}) \text{ so that } w(M',\, P) \neq 0\}, \tag{A.9}$$

*$P \in \mathcal{P}_{total}(\mathbb{Z}^{d+1})$,*

$$W_p^{total}(M) := \{P' \in \mathcal{P}_{total}(\mathbb{Z}^{d+1}) \text{ so that } w(P',\, M) \neq 0\}, \tag{A.10}$$

*$M \in \mathcal{B}_{total}(\mathbb{Z}^{d+1})$, where, for $M \in \mathcal{B}_{total}(\mathbb{Z}^{d+1})$ and $P \in \mathcal{P}_{total}(\mathbb{Z}^{d+1})$, $w(M,\, P) = w(P,\, M) \in \{0, \ldots, N-1\}$ is the "$\mathbb{Z}_N$-winding number" of $M$ around $P$:*

$$w(M,\, P) = w(P,\, M) := \max_{\substack{D \in \mathcal{D}(P) \\ E \in \mathcal{E}(M)}} w((M,\, E),\, (P,\, D)) = \max_{\substack{D,\, \mathrm{supp}\, D = P \\ E,\, \mathrm{supp}\, E = M}} \Big( \langle u^D,\, E \rangle \bmod N \Big)\,. \tag{A.11}$$

*For $M \in \mathcal{B}_{total}(\mathbb{Z}^{d+1})$ and $P \in \mathcal{P}_{total}(\mathbb{Z}^{d+1})$ we define $|M|$ as the number of bonds contained in $M$ and $|P|$ an the number of plaquettes contained in $P$ $\Box$*

**Proposition A.1** *Let two polymers $(P,\, D)$ with $P \in \mathcal{P}(\mathbb{Z}^{d+1})$ and $(M,\, E)$ with $M \in \mathcal{B}(\mathbb{Z}^{d+1})$ be given. Then there are convex, differentiable, monotonically decreasing functions $F_b,\, F_p$: $(a_0,\, \infty) \to \mathbb{R}_+$, for some $a_0 \geq 0$, with $\lim_{a\to\infty} F_{b,\,p}(a) = 0$ such that, for all $a > a_0$,*

$$\sum_{\gamma \not\sim (P,\, D)} e^{-a|\gamma|} \leq F_p(a)|P| \tag{A.12}$$

*and*

$$\sum_{\gamma \not\sim (M,E)} e^{-a|\gamma|} \leq F_b(a)|M| \qquad \Box \tag{A.13}$$

The proof of Proposition A.1 is given immediately after the proof of Corollary A.1. We now establish the main result of this Appendix:

**Theorem A.2** *There is a convex, differentiable, monotonically decreasing function $F_0$: $(a_0,\, \infty) \to \mathbb{R}_+$, for some $a_0 \geq 0$, with $\lim_{a\to\infty} F_0(a) = 0$ such that, for all $\gamma' \in \mathcal{G}$, and for all $a > a_0$,*

$$\sum_{\gamma \not\sim \gamma'} e^{-a|\gamma|} \leq F_0(a)\, |\gamma'| \qquad \Box \tag{A.14}$$

**Proof of Theorem A.2:** Let $\{(M_j^{\gamma'},\, E^{M_j^{\gamma'}})\}$, and $\{(P_i^{\gamma'},\, D^{P_i^{\gamma'}})\}$ be the set of mutually disconnected sets of bonds, respectively the set of mutually co-disconnected sets of plaquettes and their colours which make up $\gamma'$. Then

$$\sum_{\gamma \not\sim \gamma'} e^{-a|\gamma|} \leq \sum_i \sum_{\gamma \not\sim (P_i^{\gamma'},\, D_i^{\gamma'})} e^{-a|\gamma|} + \sum_j \sum_{\gamma \not\sim (M_j^{\gamma'},\, E_j^{\gamma'})} e^{-a|\gamma|}$$

$$\leq F_p(a)\sum_i |P_i^{\gamma'}| + F_b(a)\sum_j |M_j^{\gamma'}| \leq F_0(a)|\gamma'|, \tag{A.15}$$

where the second inequality comes from Proposition A.1 and where $F_0 := F_b + F_p$ is differentiable, convex and decreases monotonically to zero ■

As mentioned, this proves Theorem A.1. To prove Proposition A.1 we need two Lemmas and a Corollary.

**Lemma A.1** *There are convex, differentiable, monotonically decreasing functions $g^b$, $g^p$: $(a_0, \infty) \to \mathbb{R}_+$, for some $a_0 \geq 0$, with $\lim_{a\to\infty} g^{b,p}(a) = 0$ such that for all $a \in \mathbb{R}_+$, a large enough,*

$$\sum_{M'\in C_B(M)} e^{-a|M'|} \leq g^b(a)\,|M|, \tag{A.16}$$

*for all $M \in \mathcal{B}_{total}(\mathbb{Z}^{d+1})$ and*

$$\sum_{P'\in C_P(P)} e^{-a|P'|} \leq g^p(a)\,|P|. \tag{A.17}$$

*for all $P \in \mathcal{P}_{total}(\mathbb{Z}^{d+1})$* □

**Lemma A.2** *There are convex, differentiable, and monotonically decreasing functions $f^b$, $f^p$: $(a_0, \infty) \to \mathbb{R}_+$, for some $a_0 \geq 0$, with $\lim_{a\to\infty} f^{b,p}(a) = 0$ such that for all $a \in \mathbb{R}_+$, a large enough,*

$$\sum_{M'\in W_b(P)} e^{-a|M'|} \leq f^b(a)\,|P|, \tag{A.18}$$

*for all $P \in \mathcal{P}_{total}(\mathbb{Z}^{d+1})$ and*

$$\sum_{P'\in W_p(M)} e^{-a|P'|} \leq f^p(a)\,|M|. \tag{A.19}$$

*for all $M \in \mathcal{B}_{total}(\mathbb{Z}^{d+1})$* □

To avoid breaking the stream of the argument we postpone the proof of these Lemmas to the end of this Appendix.

**Corollary A.1** *If Lemma A.2 holds one has:*

$$\sum_{M'\in W_b^{total}(P)} e^{-a|M'|} \leq e^{f^b(a)\,|P|}, \tag{A.20}$$

*$P \in \mathcal{P}_{total}(\mathbb{Z}^{d+1})$, and*

$$\sum_{P'\in W_p^{total}(M)} e^{-a|P'|} \leq e^{f^p(a)\,|M|}, \tag{A.21}$$

*$M \in \mathcal{B}_{total}(\mathbb{Z}^{d+1})$* □

**Important Remark:** To avoid misunderstandings we stress that in (A.20), (A.21) and all below we will always be assuming that in the sums over all subsets $M' \in W_b^{total}(P)$ and, respectively, over all subsets $P' \in W_p^{total}(M)$ the terms corresponding to the empty sets $M' = \emptyset$ and, respectively, $P' = \emptyset$ are being included □

**Proof of Corollary A.1:**

For (A.20) one has, as one easily sees

$$\sum_{M'\in W_b^{total}(P)} e^{-a|M'|} \leq \sum_{m=0}^{\infty} \frac{1}{m!} \left( \sum_{M''\in W_b(P)} e^{-a|M''|} \right)^m \leq e^{f^b(a)\,|P|}, \tag{A.22}$$

where the factor $1/m!$ has been introduced to compensate overcountings and where the last inequality follows from Lemma A.2. We used the fact that the non-empty elements of $W_b^{total}(P)$ are build up by disjoint unions of elements of $W_b(P)$. Note also that $m = 0$ is included in the sum over $m$ because the empty set is included in the sum of the left-hand side, as already remarked. The proof of (A.21) is analogous ■

**Proof of Proposition A.1:**

We prove (A.12), the proof of (A.13) being analogous. To prove (A.12) first note that if $\gamma \not\sim (P, D)$ then either there exists at least one connected subset of bounds $M_0 \in \gamma_g$ with $M_0 \in W_b(P)$ or there exists at least one co-connected subset of plaquettes $P_0 \in \gamma_g$ with $P_0 \in C_p(P)$. Keeping this in mind, one can, after some thought, convince oneself that the following inequality holds:

$$\sum_{\gamma \not\sim (P,D)} e^{-a|\gamma|} \leq J_C(a - \ln(N-1)) + J_W(a - \ln(N-1)), \tag{A.23}$$

where

$$\begin{aligned} J_C(a) \;:=\; & \lim_{I\to\infty} \sum_{P_0\in C_p(P)} e^{-a|P_0|} \sum_{M_1\in W_b^{total}(P_0)} e^{-a|M_1|} \sum_{P_1\in W_p^{total}(M_1)} e^{-a|P_1|} \cdots \\ & \sum_{M_I\in W_b^{total}(P_{I-1})} e^{-a|M_I|} \sum_{P_I\in W_p^{total}(M_I)} e^{-a|P_I|} \end{aligned} \tag{A.24}$$

and

$$\begin{aligned} J_W(a) \;:=\; & \lim_{I\to\infty} \sum_{M_0\in W_b(P)} e^{-a|M_0|} \sum_{P_1\in W_p^{total}(M_0)} e^{-a|P_1|} \sum_{M_1\in W_b^{total}(P_1)} e^{-a|M_1|} \cdots \\ & \sum_{P_I\in W_p^{total}(M_{I-1})} e^{-a|P_I|} \sum_{M_I\in W_b^{total}(P_I)} e^{-a|M_I|}. \end{aligned} \tag{A.25}$$

The idea is the following. If, for instance, there exists a $P_0 \in \gamma_g$ with $P_0 \in C_p(P)$ then, since $\gamma$ is a polymer, there exists $M_1 \in W_b^{total}(P_0)$ contained in $\gamma_g$, $P_1 \in W_p^{total}(M_1)$ contained in $\gamma_g$ and so on. Since all polymers are finite this chain has to break somewhere, what is considered in (A.24) and (A.25) since $M_i = \emptyset$ or $P_j = \emptyset$, $i, j \neq 0$, are allowed to occur in the sums and since $W_{b,p}^{total}(\emptyset) = \emptyset$. The factors $(N-1)^{|M_a|}$ and $(N-1)^{|P_a|}$, which are intrinsically present in (A.24) and (A.25) for $a \to a - \ln(N-1)$, as needed for the left hand side of (A.23) are, as already observed, upper bounds on the number of different colourings associated to each geometric object $M_a$ and $P_a$ appearing in the sums.

Making alternate use of (A.20) and (A.21) one gets:

$$J_C(a) \leq \lim_{I\to\infty} \sum_{P_0\in C_p(P)} e^{-(a - L_I^{bp}(a))|P_0|} \tag{A.26}$$

and

$$J_W(a) \leq \lim_{I\to\infty} \sum_{M_0\in W_b(P)} e^{-(a - L_I^{pb}(a))|M_0|}, \tag{A.27}$$

where $L_I^{bp}$ is defined inductively by $L_1^{bp}(a) = f^b(a - f^p(a))$, $L_{I+1}^{bp}(a) = f^b(a - f^p(L_I^{bp}(a)))$, and analogously for $L_I^{pb}$ with the upper indices $b$ and $p$ interchanged.

Define $L^{bp}(a) = \lim_{I\to\infty} L_I^{bp}(a)$ and $L^{pb}(a) = \lim_{I\to\infty} L_I^{pb}(a)$. These functions satisfy

$$L^{bp}(a) = f^b(a - f^p(L^{bp}(a))), \qquad L^{pb}(a) = f^p(a - f^b(L^{pb}(a))). \tag{A.28}$$

Using Lemma A.1 for (A.26) and Lemma A.2 for (A.27), we get

$$J_C(a) \le g^p(a - L^{bp}(a))|P| \tag{A.29}$$

and

$$J_W(a) \le f^b(a - L^{pb}(a))|P|. \tag{A.30}$$

Hence it is natural to define $F_p(a) := g^p(a' - L^{bp}(a')) + f^b(a' - L^{pb}(a'))$ and, correspondingly, $F_b(a) := g^b(a' - L^{pb}(a')) + f^p(a' - L^{bp}(a'))$, where $a' := a - \ln(N-1)$.

The proof of the Lemma is then finished showing that the functions $g^p(a - L^{bp}(a))$, $f^b(a - L^{pb}(a))$, $g^b(a - L^{pb}(a))$ and $f^p(a - L^{bp}(a))$ are, for $a$ large enough, positive, differentiable, convex and decay monotonically to zero for $a \to \infty$. We establish this separately in Proposition A.2 and Corollary A.2 at the end ■

**Proof of Lemma A.1:**

The proof of this Lemma is a standard piece of the literature of cluster expansions and relies in the solution of the "Königsberger Brückenproblem" (see f.i. [12], Lemma 3.11). Let us show the proof of (A.16). The proof of (A.17) is analogous. Let $b_0$ be an arbitrary bond of $M$. Since there are $|M|$ such bonds we have

$$\sum_{M' \in C_b(M)} e^{-a|M'|} \le |M| \sum_{M' \in C_b(b_0)} e^{-a|M'|} = |M| \sum_{m=1}^{\infty} e^{-am} |C_b^{(m)}(b_0)|, \tag{A.31}$$

where $C_b^{(m)}(b_0) := \{M' \in C_b(b_0)$ such that $|M'| = m\}$. We can find an estimate for $|C_b^{(m)}(b_0)|$ in the following way. Starting from $b_0$ one can move through $M' \in C_b(b_0)$ in a path that meets each bond in $M'$ at most twice. So, one can find a geometry-dependent constant $G_{b,\,d+1}$ (for sets of bonds one can choose, for instance, $G_{b,\,d+1} = 2d+1$) so that $|C_b^{(m)}(b_0)| \le (G_{b,\,d+1})^{2m}$, since $(G_{b,\,d+1})^{2m}$ is the number of connected paths of length $2m$ starting from a fixed point. Returning to (A.31) the proof is completed by choosing $g^b(a) := G_{b,\,d+1}\, e^{-a}/(1 - G_{b,\,d+1}\, e^{-a})$ with $a > \ln G_{b,\,d+1}$. One easily checks that this $g^b$ is convex and decays monotonically to zero. For the proof of (A.17) one has to replace $G_{b,\,d+1}$ by an other constant $G_{p,\,d+1}$ ■

Finest estimates for the general case of $i$-cells in $d+1$ dimensions can be found in [1].

**Proof of Lemma A.2:**

This Lemma in analogous to Proposition A.3.3 in [1] but our proof is a little different, since we were not able to reproduce all estimates used in that proof for the kind of polymers we deal with here. In spite of this our proof seems to be simpler, although our estimates may not be optimal.

Let us prove (A.18), the proof of (A.19) is analogous. We have

$$\sum_{M' \in W_b(P)} e^{-a|M'|} = \sum_{m=1}^{\infty} e^{-am} |W_b^{(m)}(P)|, \tag{A.32}$$

where $W_b^{(m)}(P) := \{M' \in W_b(P)$ so that $|M'| = m\}$. In order to find an estimate for $|W_b^{(m)}(P)|$ we note that if $M' \in W_b^{(m)}(P)$ then there exists at least one $p_0 \in P$ so that $D(p_0)S(p_0) \ne 0$, where $S \in (\mathbb{Z}^{d+1})^2$ is such that $\operatorname{supp} d^* S = M'$ with minimal $\langle S, S\rangle_{(\mathbb{Z}^{d+1})^2}$. All $M' \in \mathcal{B}(\mathbb{Z}^{d+1})$, $|M'| = m$ eventually satisfying such condition for a given $p_0$ are contained in a (d+1)-dimensional cube, $K_{p_0}$, of size $(2m)^{d+1}$ centered at $p_0$. The total number of sets $M' \in \mathcal{B}(\mathbb{Z}^{d+1})$, $|M'| = m$, contained inside of $K_{p_0}$ is bounded by $(2m)^{d+1}(G_{b,\, d+1})^{2m}$, since there are $(2m)^{d+1}$ starting points in $K_{p_0}$ for a path of length $2m$ in $\mathbb{Z}^{d+1}$ and since there are at most $(G_{b,\, d+1})^{2m}$ such paths for a fixed starting point (see proof of Lemma A.1). Hence $|W_b^{(m)}(P)| \leq |P|(2m)^{d+1}(G_{b,\, d+1})^{2m}$, the factor $|P|$ coming from the fact that there are $|P|$ possible choices for $p_0$. Therefore, choosing $a > \ln G_{b,d+1}$,

$$\sum_{M' \in W_b(P)} e^{-a|M'|} \leq |P|\, 2^{d+1} \sum_{m=0}^{\infty} (m)^{d+1} e^{-cm} = |P|\, 2^{d+1} (-1)^{d+1} h_b^{(d+1)}(c), \tag{A.33}$$

where $c = a - \ln(G_{b,\, d+1})$, $h_b(c) = e^{-c}/(1 - e^{-c})$ and $h_b^{(k)}$ is the $k$-th derivative of $h_b$.

Defining $H_{b,\, k}(c) = (-1)^k h_b^{(k)}(c)$ we complete the proof of the Lemma showing that the function $H_{b,\, k}\colon \mathbb{R}_+ \to \mathbb{R}$ is, for all $k \in \mathbb{N}$, positive, convex, monotonically decreasing with $\lim_{c\to\infty} H_{b,\, k}(c) = 0$. But this is clear since $H_{b,\, k}(c) = \sum_{m=1}^{\infty} m^k e^{-cm} > 0$ for all $k$ and by the definition $(H_{b,\, k})' = (-1) H_{b,\, k+1} < 0$ and $(H_{b,\, k})'' = H_{b,\, k+2} > 0$. The fact that $\lim_{c\to\infty} H_{b,\, k}(c) = 0$ follows from

$$H_{b,\, k}(c) = e^{-c} \left(1 + \sum_{m=1}^{\infty} (m+1)^k e^{-cm}\right), \tag{A.34}$$

which implies, using $m + 1 \leq 2m$, $H_{c,\, k}(c) \leq e^{-c}/(1 - 2^k e^{-c})$ for $c$ large enough. The proof of the Lemma is then completed by choosing $f^b = H_{b,\, d+1}$ ■

Let us now complete the details for the proof of Proposition A.1.

**Proposition A.2** *The functions $L^{bp}(a)$ and $L^{pb}(a)$ are positive, differentiable, convex and decay monotonically to zero for $a \to \infty$* □

**Proof:** We will proof the Proposition for $L^{bp}$, the proof for $L^{pb}$ is identical. To simplify the notation we call $L(a) := L^{bp}(a)$ and $h(a) := a - f^p(L^{bp}(a))$. By (A.28), $L = f^b \circ h$.

First note that, since $f^p$ is bounded, $\lim_{a\to\infty} h(a) = \infty$. Hence $\lim_{a\to\infty} L(a) = \lim_{a\to\infty} f^b(h(a)) = 0$. Now $L' = ((f^b)' \circ h) \cdot h'$ and since $h' = 1 - ((f^b)' \circ L) \cdot L'$ one gets

$$L' = \frac{(f^b)' \circ h}{1 + ((f^b)' \circ h) \cdot ((f^p)' \circ L)} < 0 \tag{A.35}$$

since $(f^b)' < 0$ and $(f^p)' < 0$. Analogously $L'' = ((f^b)'' \circ h) \cdot (h')^2 + ((f^b)' \circ h) \cdot h''$ and using the fact that $h'' = -((f^b)'' \circ L) \cdot (L')^2 - ((f^b)' \circ L) \cdot L''$ one gets

$$L'' = \frac{((f^b)'' \circ h) \cdot (h')^2 - ((f^b)' \circ h) \cdot ((f^b)'' \circ L) \cdot (L')^2}{1 + ((f^b)' \circ h) \cdot ((f^p)' \circ L)} > 0 \tag{A.36}$$

since $(f^b)' < 0$, $(f^p)' < 0$, $(f^b)'' > 0$ and $(f^p)'' > 0$ ■

**Corollary A.2** *The functions $g^b(a - L^{pb}(a))$, $f^p(a - L^{bp}(a))$, $g^p(a - L^{bp}(a))$ and $f^b(a - L^{pb}(a))$ are positive, differentiable, convex and decay monotonically to zero for $a \to \infty$* □

**Proof:** We establish this for $G(a) := g^p(a - L^{bp}(a))$. The proof for the other cases is identical. Define $k(a) := a - L^{bp}(a)$. Since $L^{bp}$ is bounded, $\lim_{a\to\infty} G(a) = \lim_{a\to\infty} g^p(k(a)) = 0$. Now $G' = ((g^p)' \circ k) \cdot k' = ((g^p)' \circ k) \cdot (1 - L')$. Since, by the previous Proposition, $L' < 0$ and $(g^p)' < 0$ one concludes $G' < 0$. Analogously one has $G'' = ((g^p)'' \circ k) \cdot (k')^2 - ((g^p)' \circ k) \cdot L'' > 0$, since $L'' > 0$, by the previous Proposition, and $(g^p)'' > 0$, $(g^p)' < 0$.

With this, the proof of Proposition A.1 is now complete ■

## B Classical Duality Transformations and the Proof of Theorem 5.2.

This Theorem has been first proven for the $\mathbb{Z}_2$ case in [8] using polymer expansions.

Let us first study duality transformation for the classical expectations of the $\mathbb{Z}_N$-Higgs model. Let us consider the sets of oriented 1- and 2-cells in $\mathbb{Z}^3$, which we call $l_1$ and $l_2$, respectively. Let us define a geometric duality map $\delta$ between $l_1$ and $l_2$ mapping oriented bonds in oriented plaquettes an oriented plaquettes in oriented bonds as described in Figure 5. These transformations can be described in words in the following way. Considering the cells as points in $\mathbb{Z}^3/2$ the transformation $\delta$ translates the cells by (1/2, 1/2, 1/2) and reverses their orientation. The transformation $\delta$ induces naturally a transformation between 1- and 2-forms. Let $l^1$ and $l^2$ be the linear spaces of 1- and 2-forms on $\mathbb{Z}^3$, respectively, with finite support. Define

$$\begin{aligned} \mathcal{D}: l^1 \to l^2: \quad \mathcal{D}(\alpha)(p) &:= \alpha(\delta(p)), \\ \mathcal{D}: l^2 \to l^1: \quad \mathcal{D}(\beta)(b) &:= \alpha(\delta(b)), \end{aligned} \tag{B.1}$$

for all $\alpha \in l^1$, $\beta \in l^2$ and all $b \in l_1$, $p \in l_2$. Note that $\delta$ and $\mathcal{D}$ are invertible and that $\delta^2$ is a translation by $(1,1,1)$ in $l_1$ and $l_2$.

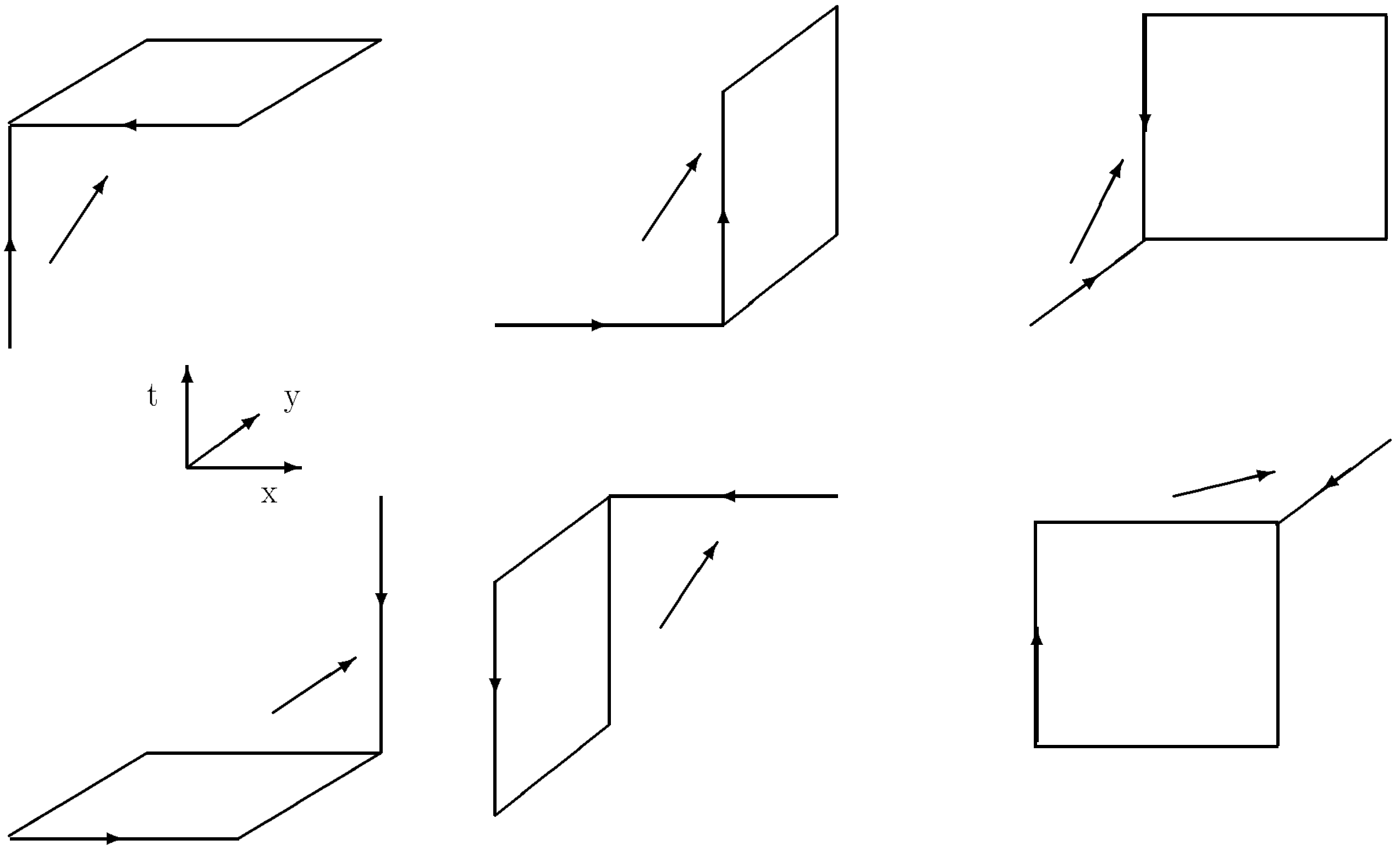

Figure 5: *The geometric duality transformations $\delta$ acting on oriented bonds and plaquettes.*

The following important relations can be established:

$$\begin{aligned} \text{on } l^1\text{:} \quad \mathcal{D} \circ d &= d^* \circ \mathcal{D}, \\ \text{on } l^2\text{:} \quad \mathcal{D} \circ d^* &= d \circ \mathcal{D}. \end{aligned} \tag{B.2}$$

This in particular means that, if $\alpha \in l^1$, $\beta \in l^2$ satisfy $d^*\alpha = 0$ and $d\beta = 0$, then $d(\mathcal{D}^{-1}\alpha) = 0$ and $d^*(\mathcal{D}^{-1}\beta) = 0$. With this we are able to establish the

**Proposition B.1** *With the definitions above and for all $\alpha \in l^1$, $\beta \in l^2$,*

$$\langle B(\alpha,\, \beta)\rangle_V = \langle B(\mathcal{D}^{-1}\beta,\, \mathcal{D}^{-1}\alpha)\rangle'_{V^*} \tag{B.3}$$

*where the prime denotes the previously discussed duality transformation on functions of the couplings, and $V^* = V + (1/2,\, 1/2,\, 1/2)$* □

**Proof.** This Proposition follows directly from (4.18) using also the fact that, for $\alpha \in l^1$, $\beta \in l^2$ with $d^*\alpha = 0$ and $d\beta = 0$, then $[\beta : \alpha] = [\mathcal{D}^{-1}\alpha : \mathcal{D}^{-1}\beta]$, which, in turn, comes from the fact that $\langle \alpha,\, \gamma\rangle_{l_1} = \langle \mathcal{D}\alpha,\, \mathcal{D}\gamma\rangle_{l_2}$, with $\alpha,\, \gamma \in l^1$, from the symmetry of this scalar product and from relations (B.2) ■

If $\gamma \in \underline{l}^1$ is a 1-form defined in $\mathbb{Z}^2$ we denote by $\gamma_n$ the embedding of $\gamma$ in $l^1$ defined in the following way: if $b \in l_1$ is a space-like bond at Euclidean time $n \in \mathbb{Z}$ then $\gamma_n(b) := \gamma(z_b)$, $z_b \in l_1$ being the projection of $b$ on the time-zero hyperplane. Otherwise $\gamma_n(b) = 0$. In words: $\gamma_n$ is a copy of $\gamma$ at time $n$. For such a $\gamma$ and $a \in \mathbb{Z}$ one can also associate an element $\gamma_{a,\, a+1} \in l^2$ in the following way: if $p \in l_2$ is a time-like plaquette spanned by the oriented space-like bonds $b_a$ and $b_{a+1}$, at Euclidean times $a$ and $a+1 \in \mathbb{Z}$, respectively, then $\gamma_{a,\, a+1}(p) := \gamma((z_{b_a}))$ (note that $z_{b_a} = -z_{b_{a+1}}$). Otherwise we define $\gamma_{a,\, a+1}(p) := 0$. Using Definition 5.1 we can establish the following two facts:

$$\begin{aligned} \mathcal{D}^{-1}(\gamma_0) &= (*\gamma)_{-1,0}\,, \\ \mathcal{D}^{-1}(\gamma_{0,1}) &= -(*\gamma)_0\,. \end{aligned} \tag{B.4}$$

Now we are able to complete our task. Without loss we consider an element of $\mathfrak{B}_0$ of the form $A := U_3(\gamma)\beta^{*-1}(U_1(\delta))'$, where $\gamma$ and $\delta$ are 1-forms with finite support in $\mathbb{Z}^2$. For such operators one has $\beta^{*-1} \circ \Delta(A) = \beta^{*-1}(U_1(*\gamma))\alpha_i(U_3(-(*\delta)))$. We have to show that $\omega_0(A')' = \omega_0 \circ \beta^{*-1} \circ \Delta(A)$, i.e., that

$$\langle [A']^{cl}\rangle' = \langle [\beta^{*-1} \circ \Delta(A)]^{cl}\rangle. \tag{B.5}$$

For the classical functions we have $[A']^{cl} = B(-\gamma_0,\, -\delta_{0,1})$ and $[\beta^{*-1} \circ \Delta(A)]^{cl} = B((*\delta)_1,\, -(*\gamma)_{0,1})$. In face of (4.20), (B.3) and (B.4) the proof is completed if the classical expectations have a unique translation invariant thermodynamic limit, what can be proven, for instance, in the convergence region of the polymer and cluster expansions ■

# C Proof of Theorem 6.2.

In order to prove Theorem 6.2 notice first that the equalities

$$\frac{\mu_0(\Phi^E(\underline{V}))}{\omega_0(\Phi^E(\underline{V}))} = \frac{\mu_3(\Phi^M(\underline{\tilde{V}}))}{\omega_3(\Phi^M(\underline{\tilde{V}}))} \tag{C.1}$$

$$\frac{\mu_2(\Phi^M(\underline{\tilde{V}}))}{\omega_2(\Phi^M(\underline{\tilde{V}}))} = \frac{\mu_1(\Phi^E(\underline{V}))}{\omega_1(\Phi^E(\underline{V}))} \tag{C.2}$$

and

$$\frac{\mu_0(\Phi^M(\underline{\tilde{V}}))}{\omega_0(\Phi^M(\underline{\tilde{V}}))} = \frac{\mu_3(\Phi^E(\underline{V}^1))}{\omega_3(\Phi^E(\underline{V}^1))} \tag{C.3}$$

$$\frac{\mu_2(\Phi^E(\underline{V}^1))}{\omega_2(\Phi^E(\underline{V}^1))} = \frac{\mu_1(\Phi^M(\underline{\tilde{V}}))}{\omega_1(\Phi^M(\underline{\tilde{V}}))} \tag{C.4}$$

follow trivialy from the definitions. Above $\underline{V}^1$ is simply $\underline{V}$ translated of $(-1,\,-1)$. Using polymer expansions we will prove that the left hand side of both (C.1) and (C.2) converge to the factor $e^{\frac{-2\pi i n}{N}}$ and that the left hand side of both (C.3) and (C.4) converge to 1.

**Case (C.1).**

This first case has to be analyzed with more detail because its proof differs slightly from that of the simple $\mathbb{Z}_2$ case, as found in [1], due to some additional effects present in the $\mathbb{Z}_N$ case.

First consider $\omega_0(\Phi^E(\underline{V}))$. The classical function associated to $\Phi^E(\underline{V})$ is

$$\prod_{p\in P_{\delta^*\underline{V}}} \frac{g(du(p)-1)}{g(du(p))} \prod_{b\in(\delta^*\underline{V},\,0)} \left(\frac{\mathcal{F}[h](u(b)-1)}{\mathcal{F}[h](u(b))}\right)^{1/2}, \tag{C.5}$$

where $b\in(\delta^*\underline{V},\,n)$ is the set of bonds $\delta^*\underline{V}$ placed at Euclidean time $n$ and $P_{\delta^*\underline{V}}$ is the set of plaquettes spanned by $(\delta^*\underline{V},\,0)$ and $(\delta^*\underline{V},\,1)$. Introducing this function in the expectation values we get

$$\begin{aligned}\omega_0(\Phi^E(\underline{V})) &= \frac{H(0)^{|\delta^*\underline{V}|}}{Z_V^1} \sum_{\substack{D\in V^2\\ d(D-\beta_{\underline{V}})=0}} \sum_{\substack{E\in V^1\\ d^*E=0}} \left[D-\beta_{\underline{V}}:\,E\right]\\ &\times \left[\prod_{p\in\mathrm{supp}D} g((D)(p))\right]\left[\prod_{b\in\mathrm{supp}E\setminus(\delta^*\underline{V},0)} h((E)(b))\right]\left[\prod_{c\in(\delta^*\underline{V},0)} h_S((E)(c))\right],\end{aligned} \tag{C.6}$$

where $\beta_{\underline{V}}$ is the closed two form which takes the value $-1$ on $P_{\delta^*\underline{V}}$ and $h_S(n):=H(n)/H(0)$, with

$$H(n) := \mathcal{F}^{-1}[H_1H_2](n) \tag{C.7}$$

and

$$H_1(m) \;:=\; (\mathcal{F}[h](m-1))^{1/2}, \tag{C.8}$$

$$H_2(m) \;:=\; (\mathcal{F}[h](m))^{1/2}. \tag{C.9}$$

Repeating the steps which led to (4.23) we get

$$\omega_0(\Phi^E(\underline{V})) = H(0)^{|\delta^*\underline{V}|}\exp\left(\sum_{\Gamma\in\mathcal{G}_{clus}} c_\Gamma\left(b^\Gamma_{\emptyset,\beta_{\underline{V}}}\mu^\Gamma_S-\mu^\Gamma\right)\right), \tag{C.10}$$

where

$$\begin{aligned}\mu_S(\gamma) &:= [D^\gamma:\,E^\gamma]\prod_{i=1}^{A_\gamma}\left[\prod_{p\in P_i^\gamma} g(D_i^\gamma(p))\right]\\ &\quad\times\prod_{j=1}^{B_\gamma}\left\{\left[\prod_{b\in M_j^\gamma\setminus(\delta^*\underline{V},0)} h(E_j^\gamma(b))\right]\left[\prod_{c\in(\delta^*\underline{V},0)} h_S(E_j^\gamma(c))\right]\right\},\end{aligned} \tag{C.11}$$

First we have to show that (C.10) makes sense. For that we have to prove that $\|\mu_S\|$ can be chosen small and for this it is enough to prove that $h_S(m)$, for $m\neq 0$, can be chosen small. This is the content of the following Lemma.

**Lemma C.1** *For the function $h_S$ defined above one has*

$$h_S(m) = \frac{\delta_{m,0} + c(m)}{1 + c(0)}, \tag{C.12}$$

*where $c(m)$, $m = 0, \ldots, N-1$, are analytical functions of $h(1), \ldots, h(N-1)$ and converge to zero when $|h| := \max\{|h(1)|, \ldots, |h(N-1)|\}$ goes to zero* □

**Remark:** In the $\mathbb{Z}^2$ case one has $h_S(n) = \delta_{n,0}$ □

**Proof.** Since $h(0) = 1$ one has $(\mathcal{F}[h](n))^{1/2} = N^{-1/2} + a(n)$, where $a(n)$, $n = 0, \ldots, N-1$, are analytical functions of $h(1), \ldots, h(N-1)$ and go to zero for $|h| \to 0$. In this way we can write $H_1(n)H_2(n) = N^{-1} + b(n)$ where $b$ is again analytical and goes to zero for $|h| \to 0$. Therefore, if we compute the Fourier transform of $H_1 H_2$, we get $H(n) = N^{-1/2}(\delta_{n,0} + c(n))$, where $c$ is analytical and converges to zero for $|h| \to 0$ ■

If we now compute $\mu_0(\Phi^E(\underline{V}))$ we get

$$\frac{\mu_0(\Phi^E(\underline{V}))}{\omega_0(\Phi^E(\underline{V}))} = \lim_{r\to\infty} \left[\beta_{\underline{V}} : \alpha_r^n\right] \exp\left(\sum_{\Gamma\in\mathcal{G}_{clus}} c_\Gamma\, \nu(\Gamma, r, \underline{V})\right), \tag{C.13}$$

where

$$\nu(\Gamma, r, \underline{V}) := a^\Gamma_{\alpha_r^n} b^\Gamma_{\beta_{\underline{V}}} \mu^\Gamma_S - a^\Gamma_{\alpha_r^n}\mu^\Gamma - b^\Gamma_{\beta_{\underline{V}}}\mu^\Gamma_S + \mu^\Gamma. \tag{C.14}$$

Above we used a simplified notation and called $a_\alpha$ for $a_{\emptyset,\alpha}$ and $b_\beta$ for $b_{\emptyset,\beta}$.

For $r$ large enough one can conclude after a careful inspection that the only clusters for which $\nu(\Gamma, r, \underline{V})$ is non-zero are those which simultaneously have polymers with a non-trivial winding number with $\alpha_r^n$ and polymers which have either a non-trivial winding number with $\beta_{\underline{V}}$ or have bonds contained in $(\delta^*\underline{V}, 0)$ or both. Since such $\Gamma$'s are clusters their size has to be at least dist $(B^r, (\delta^*\underline{V}, 0))$. By (A.3) their contribution disappears after taking the limits $r \to \infty$ and $\underline{V} \uparrow \mathbb{Z}^d$, in this order. The factor $\left[\beta_{\underline{V}} : \alpha_r^n\right]$ remains and equals $e^{\frac{-2\pi i n}{N}}$. This proves (6.37) ■

**Case (C.2).**

This case is simpler. Using the fact that $\omega_2 = \omega_0'$ and using (6.48) we write the left hand side of (C.2) as

$$\lim_{r\to\infty} \left[-\beta_r^n : \alpha_{\underline{V}}\right] \exp\left(\sum_{\Gamma\in\mathcal{G}_{clus}} c_\Gamma \left(a^\Gamma_{\alpha_{\underline{V}}} b^\Gamma_{-\beta_r^n} - a^\Gamma_{\alpha_{\underline{V}}} - b^\Gamma_{-\beta_r^n} + 1\right) {\mu'}^\Gamma\right), \tag{C.15}$$

where $\alpha_{\underline{V}}$ is a closed 1-form and takes values $\pm 1$ on its support $*\delta^*\underline{V}$ (which is a closed loop). By an analogous argumentation to the previous case one sees that the sum over clusters converges to zero and we get the final factor from $\left[-\beta_r^n : \alpha_{\underline{V}}\right]$ ■

**Case (C.3).**

For the left hand side of (C.3) we have

$$\lim_{r\to\infty} \exp\left(\sum_{\Gamma\in\mathcal{G}_{clus}} c_\Gamma \left(a^\Gamma_{\alpha_{\underline{V}}} a^\Gamma_{\alpha_r^n} - a^\Gamma_{\alpha_{\underline{V}}} - a^\Gamma_{\alpha_r^n} + 1\right) \mu^\Gamma\right), \tag{C.16}$$

which analogously converges to 1 ■

**Case (C.4).**

For this case we have to deal with

$$\lim_{r\to\infty} \exp\left(\sum_{\Gamma\in\mathcal{G}_{clus}} c_\Gamma\,\zeta(\Gamma,\, r,\, \underline{V})\right), \tag{C.17}$$

where

$$\zeta(\Gamma,\, r,\, \underline{V}) := b^\Gamma_{-\beta^n_r} b^\Gamma_{\beta_{\underline{V}}} {\mu'}^\Gamma_S - b^\Gamma_{-\beta^n_r}{\mu'}^\Gamma - b^\Gamma_{\beta_{\underline{V}}}{\mu'}^\Gamma_S + {\mu'}^\Gamma, \tag{C.18}$$

which analogously converges to 1 ■

# D Finishing the Proof of Proposition 8.2.

We want to show that

$$\frac{\left\|\pi_{\mu_0}\left(A_{p+1}(\underline{R}_{\underline{x}\to\underline{0}})A_{p'+1}(\underline{L}_{\underline{0}\to\underline{x}})\right)\phi_{\mu_0}\right\|}{\left\|\pi_{\mu_0}\left(A_{p}(\underline{R}_{\underline{x}\to\underline{0}})A_{p'}(\underline{L}_{\underline{0}\to\underline{x}})\right)\phi_{\mu_0}\right\|} \tag{D.1}$$

converges to one when the limits $p'$, $p\to\infty$ are taken. There are many ways to show this fact using cluster expansions. The one we present is perhaps the quickest. In terms of cluster expansions the expression above is give by the limit $r\to\infty$ of

$\exp\frac{1}{2}\sum_\Gamma c_\Gamma \mu^\Gamma$

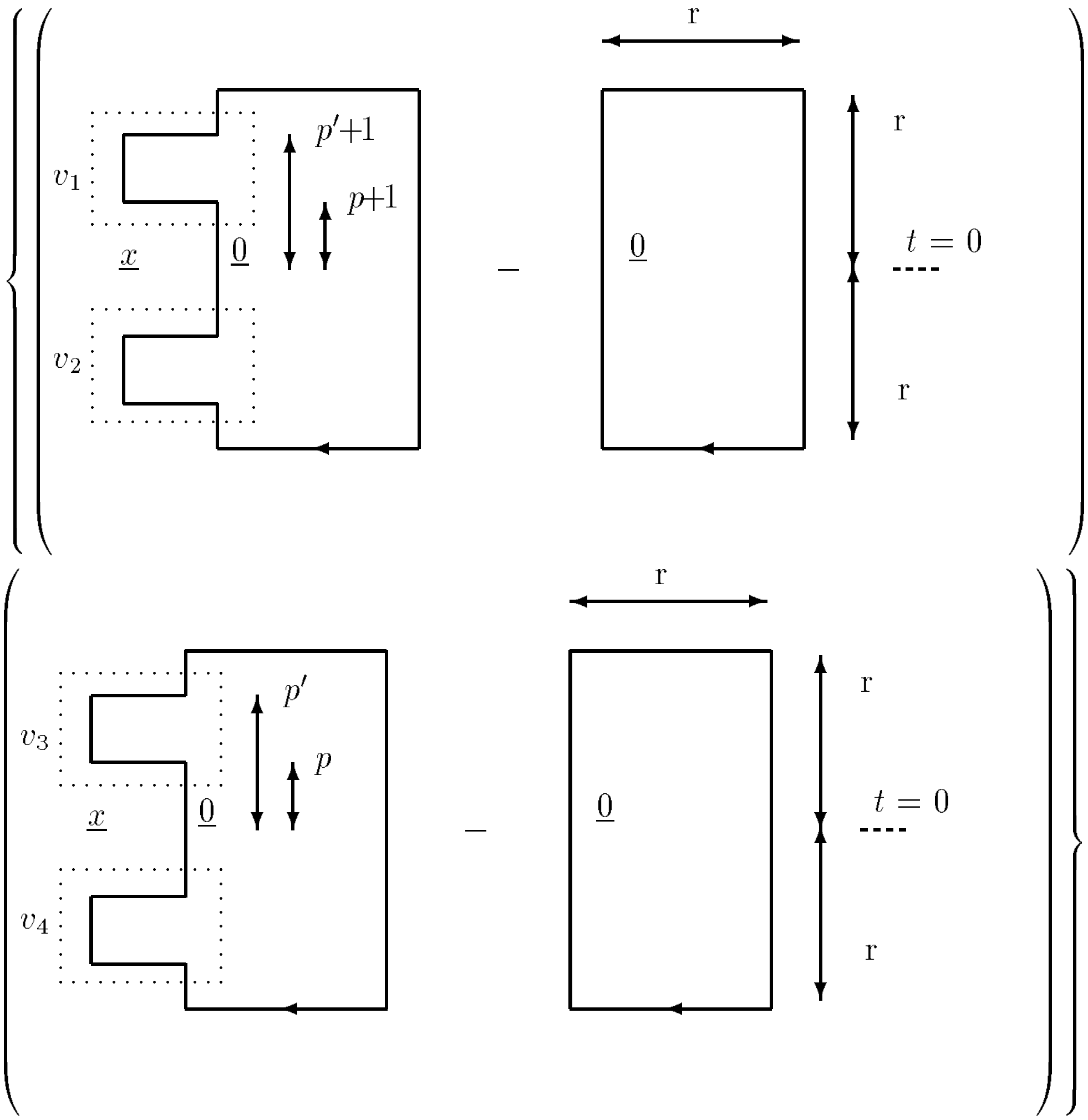

, (D.2)

where in the figure above we represented schematically the terms corresponding to winding numbers of cluster with respect to four loops. The limit $r$, presents no difficulties. As for the limits $p'$ and $p \to \infty$, we argument as follows. The contributions of the clusters incompatible with the boxes $v_1$ and $v_2$ in the first and second loops are canceled by the corresponding ones incompatible with the boxes $v_3 = v_1 - (1, 0, 0)$ and $v_4 = v_2 + (1, 0, 0)$ of the third and fourth loops, except for some cluster which are simultaneously incompatible with, say, $v_1$ and $v_2$ and for this reason have sizes larger than $p$. Other clusters are either canceled exactly or have sizes larger than $p$. Hence, the expression above converges quickly to 1 ■

# References

[1] K. Fredenhagen and M. Marcu. "Charged states in $\mathbb{Z}_2$ gauge theories". *Commun. Math. Phys.*, **92**: 81–119, (1983).

[2] K. Fredenhagen, M. Gaberdiel, and S. M. Rüger. "Scattering States of Plektons (Particles with Braid Group Statistics) in 2+1 Dimensional Quantum Field Theory ". *University of Hamburg. Preprint.*, , (1992 ).

[3] J.C.A. Barata and K. Fredenhagen. "Charged particles in $\mathbb{Z}_2$ gauge theories". *Commun. Math. Phys.*, **113**: 403–417, (1987).

[4] J.C.A. Barata. "Scattering states of charged particles in the $\mathbb{Z}_2$ gauge theories ". *Commun. Math. Phys.*, **138**: 175–191, (1991).

[5] J.C.A. Barata and K. Fredenhagen. "Particle scattering for Euclidean lattice field theories". *Commun. Math. Phys.*, **138**: 507–519, (1991).

[6] R. Savit. "Topological exitations in U(1)-invariant theories.". *Phys. Rev. Lett.*, **39**: 55–58, (1977).

[7] M. E. Peskin. "Mandelstam-'tHooft duality in abelian lattice models.". *Ann. Phys.*, **133**: 122–152, (1978).

[8] Frank Gaebler. Quasiteilchen mit anomaler Statistik in zwei- und dreidimensionalen Gittertherien. Diplomarbeit, (1990). Freie Universität Berlin.

[9] J.C.A. Barata and F. Nill. In Preparation. .

[10] Ola Bratteli and Derek W. Robinson. *"Operator Algebras and Quantum Statistical Mechanics I". $C^*$ and $W^*$-Algebras. Symmetry Groups. Decomposition of States.* Springer Verlag, (1979).

[11] Walter Rudin. *"Functional Analysis".* McGraw-Hill International Editions, second edition, (1991).

[12] Erhard Seiler. *"Gauge Theory as a Problem of Constuctive Quantum Field Theory and Statistical Mechanics". Lecture Notes in Physics, 159.* Springer Verlag, (1982).

[13] H.G. Evertz, K. Jansen, H.A. Kastrup, K. Fredenhagen, and M. Marcu. "Proof of universal perimeter law behavior of the Wegner-Wilson loop in nonabelian lattice gauge theories with Higgs fields. ". *Phys. Lett.*, **194B**: 277, (1987).